\documentclass[aps,twocolumn,showpacs,preprintnumbers,nofootinbib,prd,superscriptaddress,groupedaddress,10pt]{revtex4-1}

\makeatletter
\def\l@subsubsection#1#2{}
\def\l@subsubsubsection#1#2{}
\makeatother

\usepackage{graphicx,amssymb,amsmath,amsthm,amsfonts,epsfig,epsf,fixmath}
\usepackage[usenames]{color}
\usepackage{epstopdf}
\usepackage{mathtools}
\usepackage{relsize}
\usepackage{placeins} 
\usepackage{booktabs} 
\usepackage{siunitx} 
\DeclareSIUnit \parsec {pc} 
\usepackage{aas_macros}
\usepackage{bm}
\usepackage{dcolumn}
\usepackage{latexsym}
\usepackage{rotating}
\usepackage{longtable} 

\usepackage{enumerate}
\usepackage{tensor,multirow}
\usepackage{url}
\usepackage[linktocpage]{hyperref}

\def\nn{\nonumber}

\newcommand{\vett}[1]{\underline{#1}}

\newcommand{\dd}{\mathrm{d}}

\newcommand{\totder}[2]{\frac{\mathrm{d} #1}{\mathrm{d} #2}}
\newcommand{\modulo}[1]{\left|#1\right|}

\newcommand{\media}[1]{\left\langle#1\right\rangle}

\newcommand{\nswsh}{\!\prescript{}{-2}{S^{\hat{a}\hat{\omega}}_{\ell m}}} 
\newcommand{\rswsh}{S^{\hat{a}\hat{\omega}}_{\ell m}} 
\newcommand{\Rin}{R^{\textup{in}}_{\ell m \hat{\omega}}} 
\newcommand{\Rup}{R^{\textup{up}}_{\ell m\hat{\omega}}} 
\newcommand{\Xin}{X^{\textup{in}}_{\ell m\hat{\omega}}} 
\newcommand{\Xup}{X^{\textup{up}}_{\ell m \hat{\omega}}} 

\begin{document}
\title{Extreme mass ratio inspirals with spinning secondary:\\ a detailed study of equatorial circular motion}
\author{
Gabriel Andres Piovano$^1$,
Andrea Maselli$^1$,
Paolo Pani$^{1}$}

\affiliation{$^{1}$ Dipartimento di Fisica, ``Sapienza" Università di Roma \& Sezione INFN Roma1, Piazzale Aldo Moro 5, 
00185, Roma, Italy}

\begin{abstract} 
Extreme mass-ratio inspirals detectable by the future Laser Interferometer Space Antenna provide a unique way to test 
general relativity and fundamental physics. Motivated by this possibility, here we study in detail the 
EMRI dynamics in the presence of a spinning secondary, collecting and extending various results that appeared in 
previous work and also providing useful intermediate steps and new relations for the first time. We present the results 
of a frequency-domain code that computes gravitational-wave fluxes and the adiabatic orbital evolution for the case of 
circular, equatorial orbits with (anti)aligned spins. The spin of the secondary starts affecting the gravitational-wave 
phase at the same post-adiabatic order as the leading-order self-force terms and introduces a detectable dephasing, 
which can be used to measure it at $5-25\%$ level, depending on individual spins. In a companion 
paper we discuss the implication of this effect for tests of the Kerr bound.
\end{abstract}
\maketitle
\tableofcontents
\section{Introduction}
Extreme mass-ratio inspirals~(EMRIs) are among the most interesting gravitational-wave~(GW) sources for the future 
space-based Laser~Interferometer~Space~Antenna~(LISA)~\cite{Audley:2017drz} and for evolved 
concepts thereof~\cite{Baibhav:2019rsa}. 
An EMRI consists of a stellar-size compact object (henceforth dubbed as secondary) orbiting a supermassive object 
(henceforth dubbed as primary). The mass ratio of the binary is 
$q = \mu/M\in (10^{-7} - 10^{-4})$ and the secondary makes ${\cal O}(1/q)$ cycles before plunging. This provides a 
unique opportunity to map the spacetime of the primary and to study radiation-reaction effects that govern the 
evolution of the orbit.

While parameter estimation still faces challenging open problems~\cite{Babak:2017tow,Chua:2019wgs}, in principle an 
EMRI detection with LISA can provide exquisite measurements of the properties of the binary~\cite{Babak:2017tow}.
In addition, EMRIs are unique probes of fundamental physics~\cite{Barack:2018yly,Barausse:2020rsu}. Probing both the 
conservative and the dissipative sector of the dynamics, they allow for novel tests of 
gravity~\cite{Sopuerta:2009iy,Yunes:2011aa,Pani:2011xj,Barausse:2016eii,Chamberlain:2017fjl,Cardoso:2018zhm} and of the 
nature of supermassive objects~\cite{Barack:2006pq,Pani:2010em,Babak:2017tow,Pani:2019cyc,Datta:2019epe}.

With these motivations in mind, in this work we provide a detailed study of the EMRI dynamics in the presence of a 
spinning secondary. The spin of the secondary starts affecting the gravitational phase to the first order in the post-adiabatic expansion, 
being thus comparable to the leading-order post-adiabatic self-force effects
(which come from the  conservative first-order and dissipative second-order in the mass ratio parts of the self-force)~\cite{Pound:2015tma,Barack:2018yvs,Dolan:2013roa,Burko:2015sqa,Warburton:2017sxk, Akcay:2019bvk}. 

EMRI detection and parameter estimation require accurate first post-adiabatic models of the 
waveforms~\cite{Pound:2015tma,Barack:2018yvs,Akcay:2019bvk}. Therefore, no EMRI inspiral and waveform model is complete without including 
the spin of the secondary, which motivates several work on this topic. 

Earlier work in perturbation theory mostly focused on the effect of the spin on unbound 
orbits~\cite{Mino:1995fm,Saijo:1998mn,Tominaga:2000cs}, and the spin of the 
secondary was taken to be unrealistically large in order to maximize its effect and compensate for the mass-ratio 
suppression. One of the first work to consider dissipative spin effects on bound orbits is Ref.~\cite{Tanaka:1996ht}, 
which estimated post-Newtonian terms for the fluxes by expanding the Teukolsky equation (see also 
Ref.~\cite{Nagar:2019wrt} for a more recent analysis).
A more recent work~\cite{Dolan:2013roa} considered the precession of a gyroscope in Schwarzschild spacetime induced by 
the conservative self-torque of the particle. The effects of conservative spin-curvature coupling and self-force were 
studied in Ref.~\cite{Burko:2003rv,Burko:2015sqa} for circular orbits in Schwarzschild, and later on in 
Ref.~\cite{Warburton:2017sxk} for generic orbits.
The GW fluxes for circular orbits in Schwarzschild and Kerr spacetimes were computed accurately using a 
time-domain code~\cite{Harms:2015ixa,Harms:2016ctx, Lukes-Gerakopoulos:2017vkj}, comparing also some of the most used 
choices for the supplementary spin conditions discussed below.
Recently, Ref.~\cite{Akcay:2019bvk} considered spin dissipative effects with a spinning test particle and 
derived new flux-balance laws relating the asymptotic fluxes of energy and angular momentum to the adiabatic changes of 
the orbital parameters, focusing on the case of circular orbits around a Schwarzschild and secondary spin perpendicular 
to the orbital plane.

An estimate of the conservative contributions on the phase induced by the secondary spin was provided 
in Ref.~\cite{Yunes:2010zj} using effective-one-body models, while recently Ref.~\cite{Chen:2019hac} calculated 
the gravitational fluxes including the spin-induced quadrupole in the case of a near extremal Kerr BH.
However, to the best of our knowledge, none of the previous work went on to compute explicitly the adiabatic evolution 
to the leading order and the corresponding spin-correction to the GW phase in a Kerr spacetime, which is crucial to 
estimate the detectability of the secondary spin. In this work we present a detailed study in this direction for 
circular, equatorial orbits around a Kerr BH and (anti)aligned spins.

The plan of the paper is as follows. Section~\ref{sec:moments} is devoted to an introduction of the motion of a 
spinning test particle in curved spacetime. In Sec.~\ref{sec:orbits} the problem is specialized to the case of a 
primary Kerr metric and, in particular, to circular, equatorial orbits with (anti)aligned spins. The adiabatic 
approximation used to evolve the orbit and to compute the dephasing is discussed in Sec.~\ref{sec:radreaction}.
Section~\ref{sec:num} is devoted to a brief discussion of the numerical methods used to solve the problem. Results 
are presented in Sec.~\ref{sec:results}. Future work is discussed in the conclusion, Sec.~\ref{sec:conclusion}. In 
Appendices~\ref{app:SN} and \ref{app:Teu source term} we provide some details on the Sasaki-Nakamura~(SN) equation and 
on the Teukolsky source term for a spinning particle, collecting and extending various results that appeared in 
previous work and also providing useful intermediate steps and new relations for the first time. Finally, a comparison 
with the GW fluxes computed in previous work is presented in Appendix~\ref{app:comparison}.

In a companion paper we discuss how measurements of the spin of the secondary can be used to devise 
model-independent tests of the Kerr bound, i.e. the fact that spinning black holes~(BHs) in general relativity cannot 
spin above a critical value of the angular momentum~\cite{Piovano:2020ooe}.

Throughout this work we use geometric units, $G=c=1$, and define the Riemann tensor as 
\begin{equation}
{R_{\mu\nu\sigma}}^\delta \omega_\delta =2 \nabla_{[\mu}\nabla_{\nu]}\omega_\sigma \ ,
\end{equation}
where $ \nabla_\mu$ is the covariant derivative and $\omega_\delta$ an arbitrary 1-form, while the square brackets 
denote the antisymmetrization. This is the same notation adopted in the package \textsc{xAct}~\cite{xAct} of the 
software \textsc{Mathematica}, which we used for all the tensor computations.  The metric signature is $(-,+,+,+)$.

\section{Multipole moments and EMRI dynamics} \label{sec:moments}

The dynamical evolution of an EMRI can be suitably studied in the framework of 
perturbation theory, in which a small (secondary) object perturbs the background 
metric of a larger (primary) BH. If the size of the small 
body is considerably smaller than the typical scale of the binary, set by the curvature 
radius of the central object, its stress-energy tensor $T^{\mu\nu}$ allows for a multipolar 
expansion within the so-called {\it gravitational skeletonization}~\cite{Tulczyjew:1959,Dixon:1964NCim, Dixon:1970I,Dixon:1970II}. 
Retaining only the first two multipoles is equivalent to consider the secondary as a 
spinning particle and to neglect tidal interactions, which are encoded 
in higher multipoles. 

For a given worldline $ X^\alpha(\tau)$, specified by the secondary proper time $\tau$, the multipole 
moments in general relativity have the following integral representation~\cite{Kyrian:2007zz}
\begin{equation}
\int_{x^0=const} T^{\mu \nu}\delta x^{\alpha_1}\cdots \delta x^{\alpha_n} \sqrt{-g}\,\dd^3 x \ ,
\end{equation}
where $\delta x^{\alpha}= x^\alpha -X^\alpha$ is the deviation from $ X^\alpha(\tau)$, defined 
inside the world-tube of the body, and $g = \det(g_{\mu\nu})$ is the determinant of the 
metric $g_{\mu\nu}$. 
Hereafter we consider the {\it pole-dipole} approximation, by neglecting all moments of the secondary higher than the 
first two: the linear momentum $p^\mu$, and the spin-dipole described by the skew-symmetric 
tensor $S^{\mu \nu}$:
\begin{align}
p^\alpha &= \int_{x^0=const} \sqrt{-g}\dd^3 x T^{\alpha 0} \ , \label{mom1}\\
S^{\alpha\beta}(X^\alpha) &= \int_{x^0=const} \sqrt{-g}\dd^3 x (\delta x^\alpha T^{\beta0} - \delta x^\beta T^{\alpha 
0})  \ .\label{mom2}
\end{align}
The integrals~\eqref{mom1}-\eqref{mom2} are computed choosing a coordinate 
frame such that $\delta x^0 = 0$, while $\delta x^i$ lie inside the integration region.  
We refer the reader to Refs.~\cite{Tanaka:1996ht,Dixon:1964NCim,Dixon:1978} for a covariant representation of the 
multipole moments and for a detailed discussion on their properties.

The covariant conservation of the energy-momentum tensor, $\nabla_\mu T^{\mu\nu}=0$, leads to 
the Mathisson- Papapetrou-Dixon (MPD) equations of motion for the spinning test body. 
These equations were first obtained by Mathisson in linearized theory of gravity~\cite{Mathisson:1937zz}, 
and then by Papapetrou in full general relativity~\cite{Papapetrou:1951pa, Corinaldesi:1951pb}. A covariant 
formulation was obtained by Tulczyjew~\cite{Tulczyjew:1959} and Dixon~\cite{Dixon:1964NCim,Dixon:1970I,Dixon:1970II}, 
who also included the higher-order multipole moments of the secondary. A modern derivation is given in 
Ref.~\cite{Steinhoff:2009tk}. 
The MPD equations of motion read:
\begin{align}
\totder{X^\mu}{\zeta}& = v^\mu \ ,\\
\nabla_{\vec{v}}p^\mu &=-\frac{1}{2}{R^{\mu}}_{\nu\alpha\beta}v^\nu S^{\alpha \beta}\ , \label{eq:2MPD}\\ 
\nabla_{\vec{v}}S^{\mu\nu} &= 2p^{[\mu}v^{\nu]} \ , \label{eq:3MPD}\\
\mathfrak{m}&\equiv -p_\mu v^\mu \ ,  \label{eq:4MPD}
\end{align}
where $ \nabla_{\vec{v}} \equiv v^\mu \nabla_\mu$, $v^\mu$ is the tangent vector to the representative worldline, 
and $\zeta$ is an affine parameter that can be different from the proper time $\tau$. Thus, the tangent 
vector $v^\mu$ does not need to be the 4-velocity of a physical observer. The timelike condition $v^2 \equiv v^\mu 
v_\mu 
<0$ is not a priori guaranteed by the MPD equations, i.e., $v^2$ is not necessarily an integral of motion. 
The mass $\mathfrak{m}$ is the so-called monopole rest-mass, which is related to the 
energy of the particle as measured in the center of mass frame. The total or dynamical rest mass of the object is given 
by
\begin{equation}
\mu^2 =- p^\sigma p_\sigma \ ,
\end{equation}
and represents the mass measured in a reference frame where the spatial components 
of $p^\mu$ vanish. Neither $\mathfrak{m}$ nor $\mu$ are necessarily constants of motion~\cite{Semerak:1999qc}. 
The spin parameter $S$ is defined as
\begin{equation}
S^2 \equiv \frac{1}{2}S^{\mu\nu} S_{\mu\nu} \ ,
\end{equation}
which is also not a priori conserved. The 4-velocity and the linear momentum are not aligned 
since 
\begin{equation}
p^\mu =\frac{1}{v^2} (\mathfrak{m} v^{\mu} - v_\sigma \nabla_{\vec{v}}S^{\mu\sigma}) \ .
\end{equation}

The system of MPD equations is undetermined, since there are $18$ dynamical variables 
$ \{ X^\mu ,  v^\mu , p^\mu , S^{\mu\nu} \}$ (note that $S^{\mu\nu}$ is skew-symmetric) and only $15$ 
equations of motion. 
One therefore needs to specify $3$ additional constraints to close the system 
of equations. These constraints are given by choosing
a spin-supplementary condition, which fixes the reference worldline with respect to 
which the moments are computed. We choose as a reference worldline the body's
center of mass. However, in general relativity the center of mass of a spinning body is 
observer-dependent, thus it is necessary to specify a reference frame by fixing, for example, 
the spin-supplementary condition covariantly as\footnote{There are several possible physical spin-supplementary 
conditions, at least in the pole-dipole approximation. See for example Ref.~\cite{Kyrian:2007zz} 
for a summary of the most common choices used in the literature.} 
\begin{equation}
S^{\mu\nu}V_\nu = 0 \ ,
\end{equation}
and by choosing $V^\nu$ as the 4-velocity of a physical observer. The representative 
worldline $X^\mu(\zeta)$ identifies then the center of mass measured by an observer with 
timelike 4-velocity $V^\nu$ (for more details see~\cite{Costa:2011zn,Costa:2014nta}) 

Hereafter we choose the Tulczyjew-Dixon condition:
\begin{equation}
S^{\mu\nu}p_\nu=0 \ ,
\end{equation}
which corresponds to $V^\mu \equiv p^\mu$, i.e. one requires that the center of mass is 
measured in the frame where $p^i=0$. This spin condition fixes a unique worldline, 
and gives a relation between the 4-velocity $v^\mu$ and the linear momentum $p^\mu$: 
\begin{equation}
v^\mu = \frac{\mathfrak{m}}{\mu^2}\bigg(p^\mu + \frac{2 S^{\mu \nu}R_{\nu\rho\sigma \!\lambda}p^\rho S^{\sigma 
\!\lambda}}{4 \mu^2+ R_{\alpha\beta\gamma\delta}S^{\alpha\beta}S^{\gamma\delta}}\bigg) \ . \label{eq:vprel}
\end{equation}
Moreover, as a consequence of the Tulczyjew-Dixon spin-supplementary condition, the mass $\mu$ and the spin $S$ become
constants of motion, unlike the mass term $\mathfrak{m}$. To fix the latter, we first need to 
choose an affine parameter $\zeta$ for the MPD equations. One possible 
choice is setting $\zeta$ equal to the proper time $\tau$, which guarantees that $v^\mu v_\mu = -1$ 
throughout the dynamics. Imposing $v^\mu v_\mu = -1$ automatically fixes $\mathfrak{m}$. 
Another possibility, first proposed in~\cite{Ehlers:1977}, (see also~\cite{Lukes-Gerakopoulos:2017cru,Witzany:2018ahb}) 
consists in rescaling $\zeta$ such that 
\begin{equation}
p^\mu v_\mu = -\mu \implies \mu = \mathfrak{m} = const  \ ,\label{eq:evolparcond}
\end{equation}
which makes $\mathfrak{m}$ constant. In this case however we need to check that $v^\mu v_\mu < 0$ during the 
orbital evolution. This choice of the affine parameter will be labeled with $\zeta \equiv \lambda$, to 
differentiate it from the generic affine parameter $\zeta$.
It has been numerically shown that, by 
imposing the same initial conditions, $\lambda$ and $\tau$ are equivalent and 
lead to the same worldline~\cite{Lukes-Gerakopoulos:2017cru}. In the next sections 
we will also check that the condition $v^\mu v_\mu < 0$ is always satisfied for all configurations, and that it is 
equivalent to impose $v^\mu v_\mu = -1$ 
and to require that $\mathfrak{m} \in \mathbb{R}$. 
Finally, the conservation of the mass parameter $\mu$ in the Tulczyjew-Dixon spin-supplementary condition 
guarantees that the normalization $\mu^2 = - p^\mu p_\mu$ holds during the dynamical 
evolution.

Plugging Eq.~\eqref{eq:vprel} into Eq.~\eqref{eq:3MPD}, it is easy to see that
\begin{equation}
 \nabla_{\vec{v}}S^{\mu\nu} = {\cal O}(q)\,. \label{constantS}
\end{equation}
Thus, the spin tensor is parallel-transported along the worldline to leading order in the mass ratio.

The freedom in the choice of the spin-supplementary condition reflects the physical requirement 
that in classical theories particles with intrinsic angular momentum must 
have a finite size, and that any point of the body  can be used to fix the 
representative worldline.  Given $R$ the size of the rotating object, 
it has been shown that $R \geq S/\mu $ where $S/\mu$ is the M\o ller radius~\cite{Moller:1949}. 
Hence, assuming $R=S/\mu$ and denoting with 
$\modulo{{R}_{\mu\nu\rho\sigma}}$ the 
magnitude of the Riemann tensor, the MPD equations are valid as long as the condition 
$
\modulo{{R}_{\mu\nu\rho\sigma}}^{-1}\gg (S/\mu)^{\!2}
$
is satisfied, i.e if the size of the spinning secondary is much smaller than the curvature radius of the primary. For 
a Kerr spacetime, the Kretschmann scalar is $48 M^2/r^6 $ on the equatorial plane, so $\modulo{{R}_{\mu\nu\rho\sigma}} 
\approx M/r^3$. Thus, the validity condition of the MPD equations 
for a Kerr background becomes
\begin{equation}
\left(\frac{r}{M}\right)^3 \gg  \left(\frac{S}{\mu M} \right)^{\!2}  \ . \label{eq:MPDval}
\end{equation}

In the following it will be useful to define the dimensionless spin parameter $\sigma$ 
as
\begin{equation}
 \sigma \coloneqq \frac{S}{\mu M} = \chi q \ , \label{def:sigma}
\end{equation}
where $\chi= S/\mu^2$ is the reduced spin of the secondary. Regardless of the nature of the secondary, in 
EMRIs it is expected $|\chi|\ll 1/q$, which implies $|\sigma|\ll1$. This also shows that Eq.~\eqref{eq:MPDval} is 
always 
satisfied in the EMRI limit.

\section{Orbital motion}\label{sec:orbits}
In this section we review the orbital motion of a spinning test particle in the Kerr metric, focusing on the case of 
circular, equatorial orbits and (anti)aligned spins. Along the way we present some useful intermediate steps and novel 
relations that, to the best of our knowledge, have not been presented anywhere else.

The background spacetime is described by the Kerr metric in Boyer-Lindquist coordinates,
\begin{align}
ds^2=&-dt^2+\Sigma(\Delta^{-1}dr^2+d\theta^2)+(r^2+a^2)\sin\theta^2 d\phi^2\nonumber\\
&+2Mr/\Sigma(a\sin\theta^2-dt)^2\ ,
\end{align}
where $\Delta=r^2-2Mr+a^2$, $\Sigma=r^2+a^2\cos^2\theta$, and $a$ 
is the spin parameter such that $\vert a\vert \leq M$. Without loss of generality, we assume that the specific spin $a$ 
of the primary is aligned to the $z$-axis, namely $a\geq0$. The spin $S$ of the secondary is positive (negative) when 
it is align (antialigned) with the primary spin.

The computations in this section are valid for a generic spin parameter $\sigma$, although later on we will be 
interested mostly in the case $\sigma\ll1$ which is relevant for EMRIs.

\subsection{Field equations in the tetrad formalism and constants of motion}
To describe the orbital motion it is convenient to introduce 
the following orthonormal tetrad frame (in Boyer-Lindquist coordinates) 
\begin{align}
e^{(0)}_\mu &= \left(\sqrt{\frac{\Delta}{\Sigma}},0,0,-a \sin^2\theta \sqrt{\frac{\Delta}{\Sigma}} \right) \ , \\
e^{(1)}_\mu &= \left(0, \sqrt{\frac{\Sigma}{\Delta}},0,0 \right) \ , \\
e^{(2)}_\mu &= \left( 0,0,\sqrt{\Sigma},0 \right) \ , \\
e^{(3)}_\mu &= \left(-\frac{a}{\sqrt{\Sigma}}\sin\theta,0,0,\frac{r^2+ a^2}{\sqrt{\Sigma}}\sin\theta \right) \ .
\end{align}
We use the notation $e^{(a)}_\mu = \big(e^{(a)}_t,\, e^{(a)}_r,\, e^{(a)}_\theta, \,e^{(a)}_\phi \big)$, with 
the Latin indices for the 
tetrad components, which are raised/lowered using the metric $\eta^{ab}=\mathrm{diag}(-1,1,1,1)$. 

The equations of motion then read
\begin{align}
&\totder{}{\lambda}p^{(a)}= {\omega_{(b)(c)}}^{(a)} v^{(b)} p^{(c)} - 
\frac{1}{2}{R^{(a)}}_{(b)(c)(d)}v^{(b)}S^{(c)(d)}\label{eq:totdertetrad1} \ ,\\
&\totder{}{\lambda}S^{(a)(b)} =-2 v^{(e)} {\omega_{(e)(c)}}^{\mathlarger{[}(a)}S^{(b)\mathlarger{]}(c)} + 2 
p^{\mathlarger{[}(a)}v^{(b)\mathlarger{]}} \ ,
\end{align}
where $p^{(a)}=p^\mu e^{(a)}_\mu$ and so on, whereas
${\omega_{(a)(b)}}^{(c)}\equiv e^{\mu}_{(a)} e^{\nu}_{(b)} \nabla_{\mu} e^{(c)}_\nu$ are 
the Ricci rotation coefficients \cite{Mino:1995fm}.

The timelike and spacelike Killing vector fields of the Kerr spacetime ($\xi^\mu = (1,0,0,0)$ and $\Xi^\mu = 
(0,0,0,1)$, respectively), can be written in the tetrad frame as
\begin{align}
\xi^\mu &= \sqrt{\frac{\Delta}{\Sigma}}\,  e^\mu_{(0)} -  \frac{a \sin\theta }{\sqrt{\Sigma}}\, e^\mu_{(3)} \ ,  
\\
\Xi^\mu &= -a \sin^2\theta \sqrt{\frac{\Delta}{\Sigma}}\, e^\mu_{(0)} +  \frac{(r^2 + a^2) \sin \theta}{ 
\sqrt{\Sigma}}\, e^\mu_{(3)} \ .
\end{align}
For a generic Killing field $\kappa^\mu$ of the background spacetime there exists a first integral of motion
\begin{equation}
C_\kappa = p_\mu \kappa^\mu -\frac{1}{2}\nabla_\nu \kappa_{\mu}S^{\mu\nu} \ , \label{Ckappa}
\end{equation}
which is conserved also when higher multipoles are included~\cite{Ehlers:1977}.
The conserved quantities $C_\xi \equiv E$ and $C_\Xi \equiv J_z$ are associated with $\xi^\mu$ and $\Xi^\mu$, 
respectively~\cite{Saijo:1998mn}.

It is convenient to introduce the spin vector
\begin{equation}
s^{(a)} \equiv -\frac{1}{2}\epsilon^{(a)(b)(c)(d)}u_{(b)}S_{(c)(d)} \ ,
\end{equation}
where $\epsilon_{(a)(b)(c)(d)}$ is the antisymmetric Levi-Civita tensor ($\epsilon_{(0)(1)(2)(3)}=1$) 
and $u^{(a)} = p^{(a)}/\mu$. The spin tensor can be recast in the following form
\begin{equation}
S^{(a)(b)} \equiv \epsilon^{(a)(b)(c)(d)}u_{(c)}s_{(d)} \ .
\end{equation}

\subsection{Equations of motion on the equatorial plane}
When the orbit is equatorial, and neglecting radiation-reaction effects, it can be shown that if the spin vector 
is parallel to 
the $z$-axis, i.e. $s^\mu = s^\theta \delta^\mu_\theta$, the spinning particle is constrained 
on the equatorial plane. In fact, suppose we set $s^\mu = s^\theta \delta^\mu_\theta$ as initial condition. 
By construction, $s^\mu p_\mu = 0$, which implies $ p^\theta = 0$ and $S^{\mu\theta} = 0$. 
Thus, using the equations of motion~\eqref{eq:3MPD}:
\begin{equation}
\nabla_{\vec{v}}S^{\mu\theta} = 0 \implies p^\mu v^\theta - p^\theta v^\mu = 0  \implies  p^\mu v^\theta = 0\ ,
\end{equation}
which implies the only nontrivial solution $v^\theta = 0$.
One also needs to prove that $\theta = \pi/2$ is a solution of the equations of motion. 
From Eq.~\eqref{eq:2MPD}, we have
\begin{equation}
\nabla_{\vec{v}}p^\theta = 0 \implies  0 =-\frac{1}{2}{R^{\theta}}_{\nu\alpha\beta}v^\nu S^{\alpha \beta} \propto \cos 
\theta\ ,
\end{equation}
which shows that $\theta = \pi/2$ is a solution. If $\theta = \pi/2$ at $\lambda =0$, then the 
initial condition  $s^\mu = s^\theta \delta^\mu_\theta$ guarantees that $\theta = \pi/2$ for any value of 
the evolution parameter $\lambda$. Note that this property does not depend on the spin-supplementary condition.

Hereafter, in order to simplify the notation, we introduce the hatted dimensionless quantities as $\hat{a}= a/M$ and 
$\hat{r} = r/M$.  We also set $s^{(2)}  \equiv - S$, such that for $S>0$ (resp. $S<0$) the spin is parallel (resp. 
antiparallel) to the $z$-axis\footnote{In spherical coordinates on the equatorial plane, 
$\partial_\theta$ and $\partial_z$ are anti-aligned, therefore $s^{(2)}= r s^\theta <0$ means that the spin is aligned 
to $\partial_z$, and so to the spin of the primary.
}

Using Eqs.~\eqref{eq:vprel},\eqref{eq:evolparcond} and the normalization $u^{(a)}u_{(a)}=-1$, it is possible to write 
the 
velocities $v^{(a)}$ in terms of the normalized momenta $u^{(a)}$
\begin{align}
&v^{(0)} = \frac{1}{N} \bigg(1- \frac{\sigma^2}{\hat{r}^3}\bigg) u^{(0)} \label{eq:v2u1}\ ,\\
&v^{(1)} = \frac{1}{N}\bigg(1- \frac{\sigma^2}{\hat{r}^3}\bigg) u^{(1)}\label{eq:v2u2}\ ,\\
&v^{(3)} = \frac{1}{N}\bigg(1+ \frac{2 \sigma^2}{\hat{r}^3}\bigg) u^{(3)}\label{eq:v2u3} \ ,
\end{align}
with $N =1- \frac{\sigma^2}{\hat{r}^3}\left[1+3 \big(u^{(3)}\big)^2\right]$.
Likewise, the conserved quantities can be written as~\cite{Saijo:1998mn}
\begin{align}
&\hat{E} =  \frac{\sqrt{\Delta}}{\hat{r}} u^{(0)}+ \frac{\hat{a}\hat{r} + \sigma}{\hat{r}^2}u^{(3)}\label{eq:energy2u}\ 
,\\
&\hat{J_z} = \frac{\sqrt{\Delta}}{\hat{r}}(\hat{a}+\sigma) u^{(0)}+ \left[\frac{\hat{r}^2 + \hat{a}^2}{\hat{r}} + 
\frac{\hat{a}\sigma}{\hat{r}^2} (1+\hat{r})\right]u^{(3)}\label{eq:Jz2u}\ ,
\end{align}
where
$\hat{E} = E/\mu$ and $\hat{J}_z= J_z/(\mu M)$.
Since we assumed $a\geq0$, the orbit is prograde and retrograde 
for $\hat{J}_z >0$ and $\hat{J}_z <0$, respectively. 
At infinity\footnote{Or, equivalently, in the weak-field and slow-motion regime (see Appendix~B 
of Ref.~\cite{Steinhoff:2012rw} for details).} 
the constant of motion $J_z$ can be interpreted as the total angular momentum on the $z$-axis, i.e. the sum 
$J_z \approx L_z + S$ of the orbital angular momentum $L_z$ and of the spin $S$ of the secondary.

The above relations can be inverted to obtain $u^{(0)}$ and $u^{(3)}$ in terms of $\hat{E} $ and $\hat{J_z}$:
\begin{align}
 u^{(0)} &=- \frac{\hat{E} \hat{r}^3 + ( \hat{E} \hat{a} -  \hat{J_z}) \sigma +  \hat{r} \hat{a}[\hat{J_z} - \hat{E}(a 
+ 
\sigma)]}{\Sigma_\sigma\sqrt{\Delta}}\label{eq:u2constmotion1}\ , \\
u^{(3)}  &=  \frac{\hat{r}[\hat{J_z} -\hat{E}(\hat{a} + \sigma )]}{\Sigma_\sigma}\label{eq:u2constmotion2}\ ,
\end{align}
where
\begin{equation}
\Sigma_\sigma = \hat{r}^2\left(1 - \frac{\sigma^2}{\hat{r}^3} \right) >0\ ,
\end{equation}
which is positive due to the constraint~\eqref{eq:MPDval}. Using 
Eqs.~\eqref{eq:u2constmotion1}-\eqref{eq:u2constmotion2} 
and the relations between the velocities $v^{(a)}$ and the normalized momenta 
$u^{(a)}$ [Eqs.~\eqref{eq:v2u1}-\eqref{eq:v2u3}], 
we can write the equations of motion in Boyer-Lindquist coordinates as (see also Ref.~\cite{Saijo:1998mn})
\begin{align}
&\Sigma_\sigma \Lambda_\sigma\totder{\hat{t}}{\hat{\lambda}} =\hat{a} \bigg(1 + \frac{3 \sigma^2}{\hat{r} 
\Sigma_\sigma} \bigg)[\hat{J_z} - \hat{E}(\hat{a} +  \sigma) ] + \frac{\hat{r}^2+\hat{a}^2}{\Delta} P_\sigma \ ,\\
&(\Sigma_\sigma \Lambda_\sigma)^2\bigg(\totder{\hat{r}}{\hat{\lambda}}\bigg)^{\!2} = R^2_\sigma \ , \\
&\Sigma_\sigma \Lambda_\sigma\totder{\phi}{\hat{\lambda}}= \bigg(1 + \frac{3 \sigma^2}{\hat{r} \Sigma_\sigma} 
\bigg)[\hat{J_z} - \hat{E}(\hat{a} + \sigma) ] + \frac{\hat{a}}{\Delta} P_\sigma\ ,
\end{align}
where
\begin{align}
& \Lambda_\sigma = 1 - \frac{3\sigma^2 \hat{r}[-(\hat{a} + \sigma)\hat{E}+\hat{J_z}]^2}{\Sigma^3_\sigma} \ ,\\
&R_\sigma = P^2_\sigma -\Delta \bigg( \frac{\Sigma^2_\sigma}{\hat{r}^2} + [-(\hat{a}+\sigma)\hat{E}+\hat{J_z}]^2 \bigg) 
\ ,\\
&P_\sigma= \bigg[(\hat{r}^2+\hat{a}^2) + \frac{\hat{a}\sigma}{\hat{r}} (\hat{r}+1) \bigg]\hat{E} -  \bigg[ \hat{a} 
+\frac{\sigma}{\hat{r}} \bigg]\hat{J_z} \ ,
\end{align}
and $\frac{1}{\hat{r}^2}\Sigma_\sigma \Lambda_\sigma = N$.

As previously discussed, condition~\eqref{eq:evolparcond} does not necessarily imply $v^{(a)}v_{(a)}<0$ and the latter 
condition must be checked during the dynamics. The norm of $v^{(a)}$ reads
\[
v^{(a)}v_{(a)}=\frac{-\hat{r}^6 + 3 \sigma^2 (u^{(3)})^2\big(2\hat{r}^3 + \sigma^2 \big) + 2\sigma^2\hat{r}^3 - 
\sigma^4}{(\hat{r}^3 N)^2}\ ,
\]
and the constraint $v^{(a)}v_{(a)}<0$ leads to
\begin{equation}
\Lambda_\sigma >\frac{\hat{r}^3+2\sigma^2}{2\hat{r}^3+\sigma^2} \label{eq:velconstraint}\ .
\end{equation}
Equation~\eqref{eq:velconstraint} shows that $\Lambda_\sigma $ must be positive definite, which implies $N>0$. 
Moreover, for realistic values of $\sigma$ (recall that $|\sigma|\ll1$ when $|\chi|\ll 1/q$, see 
Eq.~\eqref{def:sigma}) the constraint~\eqref{eq:velconstraint} reduces to
\begin{equation}
\Lambda_\sigma \gtrsim \frac{1}{2} \quad \text{ for} \quad \sigma \ll1
\end{equation}
and, since $\hat{E}$ and $\hat{J}_z$ are usually ${\cal O}(1)$ during the dynamics,  
$\Lambda_\sigma \approx 1 \text{ for }  \sigma \ll1 $. Thus Eq.~\eqref{eq:velconstraint} is 
always satisfied for bound equatorial EMRIs. 
Finally, we note that choosing the proper time of the object as evolution parameter, the 
condition $v^{(a)}v_{(a)}=-1$ fixes the kinematical mass $\mathfrak{m}$ as
\begin{equation}
\mathfrak{m}(\hat{r})=\frac{\hat{r}^3 N}{\sqrt{\hat{r}^6 - 3 \sigma^2 (u^{(3)})^2\big(2\hat{r}^3 + \sigma^2 \big) - 
2\sigma^2\hat{r}^3 + \sigma^4}} \ .
\end{equation}
Imposing that $\mathfrak{m}(\hat{r})$ is a real number gives again the constraint~\eqref{eq:velconstraint}. 

\subsection{Effective potential, ISCO, and orbital frequency}
%
For circular orbits, there are two additional constraints on the motion: one enforces zero radial velocity, the other 
requires zero radial acceleration. The condition $v^r = 0$ implies $v^{(1)}=0$ and, together with Eq.~\eqref{eq:v2u2} 
yields $p^{(1)} =0$, whereas zero radial acceleration requires $\totder{}{\lambda}p^{(1)} = 0$. Imposing these constraints  is 
equivalent to ask the orbital radius to be the local minimum of an effective potential. 
For a spinning particle moving on the equatorial plane of a Kerr BH, the effective potential 
depends on the spin-supplementary condition (see Refs.~\cite{Harms:2016ctx, Lukes-Gerakopoulos:2017vkj} 
for the form of the effective potentials for some common choices of the spin-supplementary conditions). 
Following Ref.~\cite{Jefremov:2015gza} we use 
\begin{equation}
V_\sigma(\hat{r}) = \frac{1}{\hat{r}^4}(\alpha_\sigma \hat{E}^2 -2\beta_\sigma \hat{E} + \gamma_\sigma) \ ,
\end{equation}
where
\begin{align}
\alpha_\sigma &= \left[\hat{r}^2+\hat{a}^2 + \frac{\hat{a}\sigma(\hat{r}+1)}{\hat{r}} \right]^2 - \Delta 
(\hat{a}+\sigma)^2\ , \\
\beta_\sigma &= \bigg[\bigg(\hat{a}+\frac{\sigma}{\hat{r}}\bigg)\bigg(\hat{r}^2+\hat{a}^2 + 
\frac{\hat{a}\sigma(\hat{r}+1)}{\hat{r}}\bigg) - \Delta (\hat{a}+\sigma) \bigg] \hat{J}_z\ ,\\
\gamma_\sigma &= \bigg(\hat{a}+\frac{\sigma}{\hat{r}}\bigg)^{\! \! 2} \hat{J}_z^2 - \Delta \bigg[\hat{r}^2 \bigg(1 - 
\frac{\sigma^2}{\hat{r}^3}\bigg)^{\! \! 2} + \hat{J}_z^2 \bigg]\ .
\end{align}
The effective potential reduces to the standard one for a nonspinning particle in Kerr when 
$\sigma=0$.
The condition for a circular orbit with radius  $\hat{r}_0$ translates to
\begin{alignat*}{2}
V_\sigma(\hat{r}_0) = 0 & \quad \quad \ , \left.\totder{V_\sigma}{\hat{r}} \right \rvert_{\hat{r}=\hat{r}_0}=0\ ,
\end{alignat*}
and stability of such orbits against radial perturbations requires $\left. \frac{\dd^2 V_\sigma}{\dd 
\hat{r}^2}\right\rvert_{\hat{r}=\hat{r}_0}<0$,
although the orbit might still be unstable under perturbation in the $\theta$ direction~\cite{Suzuki:1997by}. The 
innermost stable circular orbit~(ISCO) is obtained by imposing $\left. \frac{\dd^2 V_\sigma}{\dd 
\hat{r}^2}\right\rvert_{\hat{r}=\hat{r}_0}=0$.

In order to compute the GW fluxes, we also need the orbital frequency of a 
circular equatorial orbit as measured by an observer located at infinity, 
\[
\widehat{\Omega} = M\Omega  = \totder{\phi}{\hat{t}} 
= \frac{\hat{a} v^{(0)} + \sqrt{\Delta} v^{(3)}}{(\hat{r}^2+\hat{a}^2)v^{(0)}+\hat{a}\sqrt{\Delta}v^{(3)}} \ .
\]
In terms of the momenta $\hat{\Omega}$ is given by
\begin{equation}
\widehat{\Omega}=  \frac{\hat{a}(\hat{r}^3 - \sigma^2)u^{(0)} + \sqrt{\Delta} (\hat{r}^3+2\sigma^2) 
u^{(3)}}{(\hat{r}^2+\hat{a}^2)(\hat{r}^3-\sigma^2)u^{(0)}+\hat{a}\sqrt{\Delta}(\hat{r}^3+2\sigma^2) u^{(3)}}\ ,
\end{equation}
where $u^{(0)}$ and $u^{(3)}$ are given in terms of $\hat{r}$ by solving $\totder{}{\lambda}p^{(1)} = 0$: 
\begin{alignat}{2}
u^{(0)}= \frac{1}{\sqrt{1 - U_\mp^2}} \ , &\qquad u^{(3)}= \frac{U_\mp}{\sqrt{1 - U_\mp^2}} \ , \label{eq:uproretro}
\end{alignat}
where~\cite{Tanaka:1996ht}
\begin{equation}
U _\mp =\frac{u^{(3)}}{u^{(0)}} = -\frac{2 \hat{a}\hat{r}^3+3\sigma \hat{r}^2 + \hat{a}\sigma^2\mp 
{\cal D}}{2\sqrt{\Delta}(\hat{r}^3 + 2\sigma^2)} \ , \label{eq:u32u0}
\end{equation}
with
\begin{equation}
 {\cal D}=\sqrt{4 \hat{r}^7 
+ 12 \hat{a}\sigma\hat{r}^5 +13\sigma^2 \hat{r}^4 + 6\hat{a}\sigma^3\hat{r}^2 - 8\sigma^4\hat{r}+ 9\hat{a}^2\sigma^4}\ ,
\end{equation}
and the $\mp$ sign corresponding to co-rotating and counter-rotating orbits, respectively. 
Note that the argument of the square root is not positive definitive for generic values of $\sigma$. Nevertheless, 
for $\sigma\ll1$, it is easy to see that Eq.~\eqref{eq:u32u0} is always real.
Using Eq.~\eqref{eq:uproretro}, the orbital frequency $\widehat{\Omega}$  can be recast as
\begin{equation}
\widehat{\Omega} = \frac{(2 \hat{a} + 3\sigma)\hat{r}^3 +3(2\hat{a}^2\sigma+\hat{a}\sigma^2 
)\hat{r}+4\hat{a}\sigma^2\mp \hat{r} {\cal D}}{2 (\hat{a}^2 + 3 \hat{a}\sigma + 
\sigma^2) \hat{r}^3 + 6\sigma(\hat{a}+\sigma) \hat{a}^2 \hat{r}+4\hat{a}^2\sigma^2 -2 \hat{r}^6} \ . \label{eq:omegaofr}
\end{equation}
This formula agrees with the one shown in Ref.~\cite{Harms:2015ixa}. Plugging Eq.~\eqref{eq:uproretro} into 
Eqs.~\eqref{eq:energy2u}-\eqref{eq:Jz2u} finally yields the first integrals $\hat{E}$ and $\hat{J}_z$ 
for a spinning object in circular equatorial orbit in the Kerr spacetime:
\begin{align}
\hat{E} &=\frac{\hat{r} \sqrt{\Delta} + (\hat{a}\hat{r} + \sigma) U_\mp }{\hat{r}^2 \sqrt{1 - U_\mp^2}} \ 
,\label{eq:energyofr}\\
\hat{J}_z &=\frac{\hat{r} \sqrt{\Delta} (\hat{a} + \sigma) +  [\hat{r}^3 + \hat{r}\hat{a} (\hat{a} + \sigma) + 
\hat{a}\sigma] U_\mp}{\hat{r}^2 \sqrt{1 - U_\mp^2}} \ .\label{eq:jzofr}
\end{align}
The minus and plus sign in Eq.~\eqref{eq:omegaofr}-\eqref{eq:jzofr} correspond to prograde and 
retrograde orbits, respectively.
Expressions~\eqref{eq:energyofr} and~\eqref{eq:jzofr} will be useful when studying the adiabatic evolution of 
the orbit.

\begin{figure}[!htpb]
\includegraphics[width=0.45\textwidth]{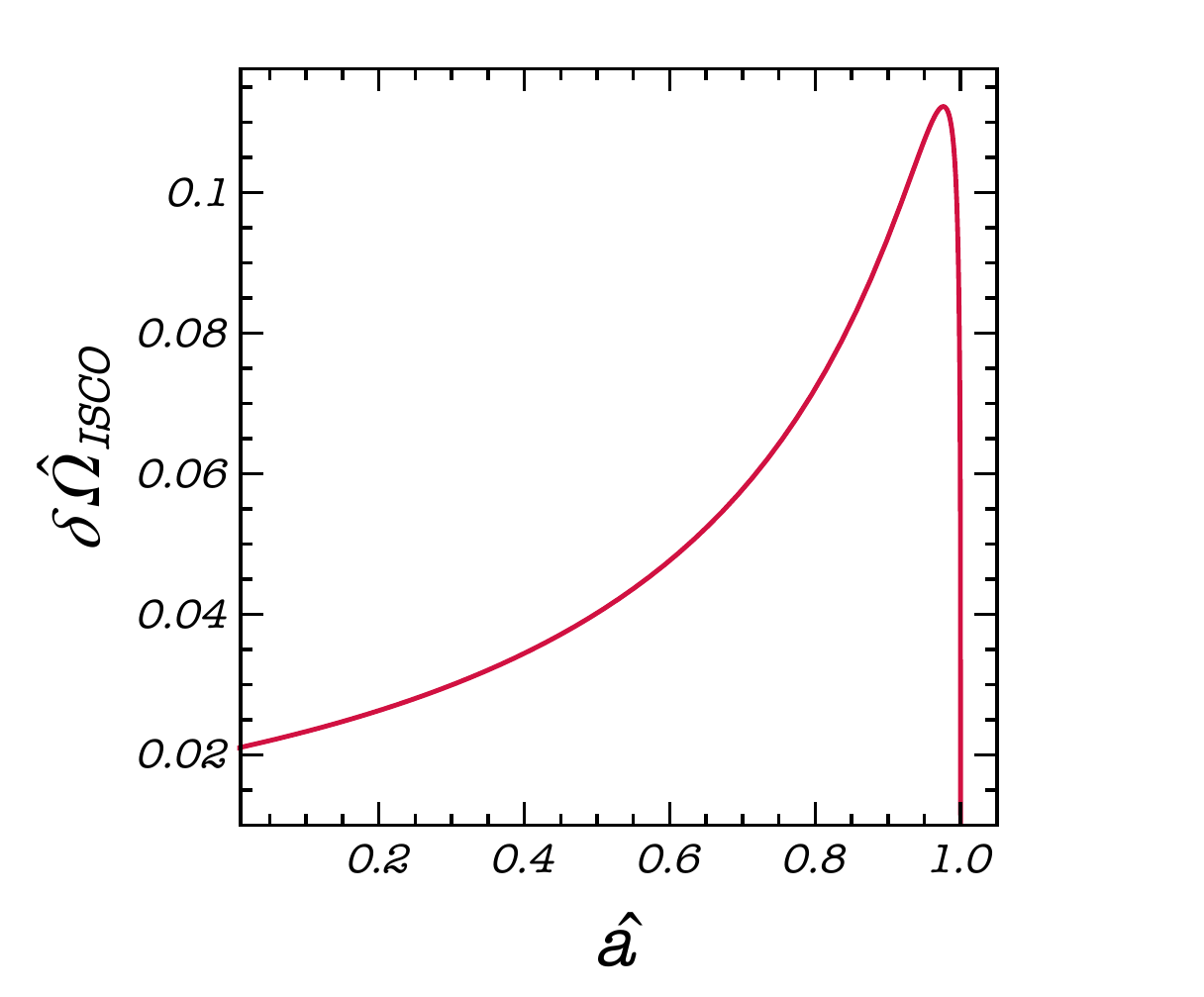}
\caption{Spin correction to the orbital frequency at the ISCO as a function of 
$\hat a$ for prograde orbits (upper sign Eq.~\eqref{eq:freqcorrISCO})}
\label{fig:dOmegaISCO} 
\end{figure}

Furthermore, the above quantities can be used to derive analytical expressions for the ISCO location and frequency to 
${\cal O}(\sigma)$ (see also Ref.~\cite{Jefremov:2015gza}). The orbital frequency can be written as
\begin{equation}
\widehat{ \Omega}(\hat r) = \widehat{\Omega}^0(\hat r) + \sigma \delta\widehat\Omega(\hat r) + {\cal O}(\sigma^2) \ , \label{eq:Omega}
\end{equation}
where $\widehat{\Omega}^0(\hat r)=1/(\hat a\pm\hat r^{3/2})$ is the orbital frequency of a nonspinning particle around Kerr, 
and
\begin{equation}
 \delta\widehat\Omega(\hat r)= -\frac{3}{2}\frac{\sqrt{\hat r}\mp \hat a}{\sqrt{\hat r}(\hat r^{3/2}\pm a)^2}\,.
\end{equation}
The ISCO location can be expanded in the same way and its leading-order spin correction reads
\begin{equation}
 \delta \hat r_{\rm ISCO}= \frac{4\hat a}{\hat r_{\rm ISCO}^0} \mp \frac{4}{\sqrt{\hat r_{\rm ISCO}^0}}\,,
\end{equation}
where $\hat r_{\rm ISCO}^0$ is the (normalized) ISCO location of the Kerr metric for a nonspinning secondary, which is 
solution to $\hat r^2-6 \hat r+ 8\hat a \hat r^{1/2}-3\hat a^2=0$ (its analytical 
expression as a function of $\hat a$ can be found in Ref.~\cite{1972ApJ...178..347B}). Using the above results, the 
leading-order spin correction to the ISCO orbital frequency is
\begin{equation}
 \delta \widehat\Omega_{\rm ISCO}= \frac{9}{2}\left(\frac{\sqrt{\hat r_{\rm ISCO}^0}\mp\hat a}{\sqrt{\hat r_{\rm 
ISCO}^0}\left((\hat r_{\rm ISCO}^0)^{3/2} \pm\hat a\right)^2}\right)\,. \label{eq:freqcorrISCO}
\end{equation}
This quantity is shown in Fig.~\ref{fig:dOmegaISCO} as a function of $\hat a$ for prograde orbits (upper sign Eq.~\eqref{eq:freqcorrISCO}). Note that $\delta \widehat\Omega_{\rm 
ISCO}>0$ for any $\hat a$ (being zero in the extremal case), i.e., if the spin of the secondary is aligned to that of 
the primary the orbital frequency at the ISCO is higher. 

\section{Radiation-reaction effects and balance laws}\label{sec:radreaction}

We study radiation-reaction effects within the \textit{adiabatic approximation}, 
assuming that the emission timescale is much longer than orbital period, namely 
\begin{equation}
 \frac{2\pi}{\widehat{\Omega}} \ll \hat{r}\modulo{\totder{\hat{r}}{\hat{t}}}^{-1}\,. \label{adiabcondition}
\end{equation}
In this approximation, changes to the mass terms $\mu$ and $M$ and to the spin 
$\hat{a}$ are smaller than the leading-order dissipative terms~\cite{Hinderer:2008dm}. The change to the primary mass 
and spin due to GW absorption at the horizon formally enter at the next-to-leading order, although with a small 
coefficient~\cite{Hughes:2018qxz}. 

Thus, for a \emph{nonspinning} object on an equatorial orbit around a Kerr BH 
\begin{equation}
\totder{E}{t} = \Omega  \totder{L_z}{t}\,.
\end{equation}
In the adiabatic approximation, the following balance equations hold:
\begin{alignat}{1}
\left(\totder{E}{t}\right) _{\!\!\text{GW}} =-\media{\totder{E}{t}}\,, \quad & \quad 
\left(\totder{L_z}{t}\right)_{\!\!\text{GW}}=-\media{\totder{L_z}{t}}\,,
\end{alignat}
where the brackets denote time-averaging over a time length much longer than the time evolution of the orbital 
parameters but shorter than the radiation time scales. The gravitational energy and angular momentum luminosities 
include both the contribution at infinity and at the event horizon, and are calculated by averaging over several 
wavelengths. 
Equation~\eqref{adiabcondition} breaks down at the onset of the inspiral/plunge transition region, 
where the adiabatic approximation is no longer valid (see Ref.~\cite{Ori:2000zn} and Refs.~\cite{Burke:2019yek, 
Compere:2019cqe} for a recent discussion on this topic).  
Nonetheless, the difference between the ISCO frequency and the transition frequency scales as $q^{2/5}\ll1$.
Thus, for a typical EMRI, Eq.~\eqref{adiabcondition} is valid for almost all the inspiral prior to plunge.

For a spinning particle in Kerr, there is an extra degree of freedom related to the spin of the small object. 
In general the evolution of the constants of motion can also depend on the secondary spin evolution. However, it was 
recently shown that the evolution of the $E$ and $J_z$ are formally the same as those above to first order in 
$\sigma$~\cite{Akcay:2019bvk}.
On the other hand, the evolution of the spin tensor $S_{\mu\nu}$ depends on \emph{local} metric perturbations and not 
only on asymptotic fluxes~\cite{Akcay:2019bvk}. This evolution determines that of the particle $4$-velocity through 
Eq.~\eqref{Ckappa}. However, as shown in Eq.~\eqref{constantS}, the spin tensor evolves at ${\cal O}(q)$ and it 
affects the particle acceleration to higher order in the mass ratio.
Likewise, the effect of the secondary spin on the adiabatic changes to $M$ and $\hat a$ is subleading.
Thus --~for what concerns the leading-order spin corrections to the dynamics~-- the evolution of the binary
masses and spins can be neglected.

It remains to prove that the equation
\begin{equation}
\totder{\hat{E}}{\hat{t}} = \widehat{\Omega} \totder{\hat{J}_z}{\hat{t}} \label{eq:balance_eq1}
\end{equation}
holds for a spinning object with the above assumptions.
Using the chain rule, Eq.~\eqref{eq:balance_eq1} is equivalent to
\begin{equation}
  \widehat{\Omega} 
  = \frac{\partial \hat{E}}{\partial \hat{r}} 
\bigg(\frac{\partial \hat{J}_z}{\partial \hat{r}}\bigg)^{\!-1} \,.
\end{equation}
and by plugging this into Eqs.~\eqref{eq:omegaofr}--\eqref{eq:jzofr}, it is straightforward to see 
that the previous relation is satisfied in our case for any value of the spin. This is the generalization of Eq.~(20) 
in  Ref.~\cite{Kennefick:1998ab}, which derived an equivalent formula in the case of a non-spinning secondary. In 
Ref.~\cite{Tanaka:1996ht}, the authors considered circular orbits for a spinning particle moving slightly off 
the equatorial plane by a quantity ${\cal O}(\sigma)$, and they showed in a similar manner that 
Eq.~\eqref{eq:balance_eq1} is valid to ${\cal O}(\sigma)$. 

Noteworthy, the above argument assumes that circular orbits for a spinning particle remains circular under 
radiation reaction, i.e. that Eq.~\eqref{eq:balance_eq1} remains valid throughout the adiabatic inspiral. In other 
words, one needs to prove that an initial circular orbit for a spinning particle does not become slightly eccentric 
during inspiral due to backreaction effects, following the same procedure of
Refs.~\cite{Kennefick:1998ab,Kennefick:1995za} in the case of a nonspinning secondary. We leave the analysis of this 
important issue for future work. Here we just note that, under the assumption that the secondary  spin remains constant, 
it is self-consistent to use Eq.~\eqref{eq:balance_eq1}, as also shown in Ref.~\cite{Tanaka:1996ht}.

\subsection{GW fluxes in the Teukolsky formalism}\label{sec:teuk}

We use the Teukolsky formalism to compute the gravitational wave flux at infinity. 
Metric perturbations of the Kerr background are decomposed using 
the Newman-Penrose tetrad basis, that allows to isolate the nontrivial  degrees of freedom of the Riemann tensor. 
At infinity, the two GW polarizations are both encoded in the $\Psi_4$ Weyl scalar:
\begin{equation}
\Psi_4 (r \rightarrow \infty) =  \frac{1}{2} \frac{\partial^2}{\partial\hat{t}^2}\big(h_+ - ih_\times\big) \ .
\label{eq:Psi4_GW_amp}
\end{equation}
In the Fourier space, 
\begin{equation}
\Psi_4= \rho^4 \displaystyle\sum_{\ell=2}^{\infty}\displaystyle\sum_{m=-\ell}^{\ell}\int_{-\infty}^\infty \dd 
\hat{\omega}  
R_{\ell m\hat{\omega}}(\hat{r})\prescript{}{-2}{S^{\hat{a}\hat{\omega}}_{\ell m}(\theta)} e^{i (m\phi -\hat{\omega} 
\hat{t})} \ , \label{eq:Psi4}
\end{equation}
where $\rho = [\hat{r} - i \hat{a} \cos\theta]^{-1}$, and the $s=-2$ spin-weighted orthonormal 
spheroidal harmonics $\nswsh(\theta)$ and radial  function $R(\hat{r})$ obey two 
decoupled ordinary differential equations. For the angular component:
\begin{eqnarray}
&&\left.\Bigg[\frac{1}{\sin\theta}\totder{}{\theta}\left(\sin\theta\totder{}{\theta} \right) - \hat{a}^2 \hat{\omega}^2 
\sin^2\theta - \left( \frac{m - 2 \cos\theta}{\sin\theta}\right)^{\!\!2} \right.\nn\\
&+&\left. 4 \hat{\omega} \cos\theta- 2 
+2\hat{a}m\hat{\omega} 
+\lambda_{\ell m\hat{\omega}} \right.\Bigg]\nswsh(\theta) = 0  \ ,\label{eq:swsweq}
\end{eqnarray}
where $\lambda_{\ell m\hat{\omega}} = E_{\ell m \hat{\omega}} - 2m\hat{a}\hat{\omega} +\hat{a}^2\hat{\omega}^2 - 2 $. 
The eigenvalues and the eigenfunctions satisfy the following identities: $\lambda_{\ell m -\hat{\omega}} = 
\lambda_{\ell-m \hat{\omega}}$ and
\begin{equation}
 \prescript{}{-2}{S^{-\hat{a}\hat{\omega}}_{\ell -m}}(\theta) = (-1)^l \!\!\, 
\prescript{}{-2}{S^{\hat{a}\hat{\omega}}_{\ell 
m}}(\pi-\theta) \ ,
\end{equation}
while $\nswsh(\theta) e^{im\phi}$ reduces to the spin-weighted spherical harmonics for $\hat a=0$ or $\hat{\omega}=0$. 
We have employed the numerical routines provided by the BH Perturbation Toolkit~\cite{BHPToolkit} to 
compute $\lambda_{\ell m\hat{\omega}}$, the spin-weighted spheroidal harmonics, and their derivatives.

The radial Teukolsky equation is given by
\begin{equation}
\Delta^2 \totder{}{\hat{r}}\left(\frac{1}{\Delta}\totder{R_{\ell m\omega}}{\hat{r}}\right) - V(\hat{r}) R_{\ell m 
\hat{\omega}}(\hat{r}) = 
\mathcal{T}_{\ell m\hat{\omega}} \ , \label{eq:radialTeueq}
\end{equation}
where the source term $\mathcal{T}_{\ell m\hat{\omega}}$ is discussed below and the potential $V(\hat{r})$ reads
\begin{align}
V(\hat{r}) &= -\frac{K^2 + 4i(\hat{r}-1)K}{\Delta} + 8i \hat{\omega} \hat{r} + \lambda_{\ell m \hat{\omega}} \ 
,\\
K &= (\hat{r}^2 + \hat{a}^2)\hat{\omega} -\hat{a}m \ .
\end{align}
The homogeneous Teukolsky equation admits two linearly independent solutions, $\Rin$ and $\Rup$, with the following 
asymptotic values at horizon $\hat{r}_+$ and at infinity:
\begin{align}
\Rin \sim
\begin{cases}
B^{\textup{tran}}_{\ell m \hat{\omega}}\Delta^2 e^{-i \hat{\kappa} \hat{r}^\ast} \quad  &\hat{r} \to 
\hat{r}_+ \ ,\\[0.5mm]
B^{\textup{out}}_{\ell m \hat{\omega}} \hat{r}^3  e^{i \hat{\omega} \hat{r}^{\ast}} +B^{\textup{in}}_{\ell m 
\hat{\omega}}\frac{1}{\hat{r}}  e^{-i \hat{\omega}
\hat{r}^\ast} \, \quad &\hat{r} \to \infty  \ ,\label{eq:inBCTeu}
\end{cases}
\end{align}
\begin{align}
\Rup \sim 
\begin{cases}
D^{\textup{out}}_{\ell m\hat{\omega}} \hat{r}^3  e^{i \hat{\kappa} \hat{r}^\ast} + D^{\textup{in}}_{\ell 
m\hat{\omega}}\Delta^2  e^{- i \hat{\kappa}
\hat{r}^\ast} \quad   &\hat{r}\to \hat{r}_+ \ , \\[0.5mm]
D^{\textup{tran}}_{\ell m \hat{\omega}} \hat{r}^3  e^{i \omega \hat{r}^\ast}  \quad &\hat{r} \to \infty  \ 
, \label{eq:upBCTeu}
\end{cases}
\end{align}
where $\hat{\kappa} = \hat{\omega} - m \hat{\omega}_+$,  $\hat{r}_\pm = 1 \pm \sqrt{1-\hat{a}^2}$, 
$\hat{\omega}_+ = \hat{a}/(2 \hat{r}_+)$, and being $\hat{r}^\ast$ the tortoise coordinate of the Kerr metric,
\begin{equation}
\hat{r}^\ast = \hat{r} + \frac{2 \hat{r}_+}{\hat{r}_+ - \hat{r}_-}\ln\Big(\frac{\hat{r} -\hat{r}_+}{2}\Big) - 
\frac{2 r_-}{r_+ -  
\hat{r}_-}\ln\Big(\frac{\hat{r} - \hat{r}_-}{2}\Big)\ .
\end{equation}
The radial Teukolsky equation can be solved through the Green function 
method~\cite{Mino:1997bx}. The solution with the correct asymptotics reads
\begin{align}
R_{\ell m \hat{\omega}}(\hat{r}) &= \frac{1}{W_{\hat{r}}} \left\{\Rup(\hat{r}) \int_{\hat{r}_+}^{\hat{r}} \dd 
\hat{r}'\frac{\Rin(\hat{r}') \mathcal{T}_{\ell m \hat{\omega}}(\hat{r}')}{\Delta^2} \right.\nonumber\\
&\left.+\Rin(\hat{r}) \int_{\hat{r}}^{\infty} \dd \hat{r}'\frac{\Rup(\hat{r}') \mathcal{T}_{\ell m 
\hat{\omega}}(\hat{r}')}{\Delta^2}  \right\} \ ,
\end{align}
with the constant Wronskian given by
\begin{equation}
W_{\hat{r}} \equiv\!\left(\! {\Rin}\totder{\Rup}{\hat{r}^*} - {\Rup}{}\totder{\Rin}{\hat{r}^*} 
\!\right) \!= 2i \hat{\omega} 
B^{\textup{in}}_{\ell m\hat{\omega}} D^{\textup{tran}}_{\ell m \hat{\omega}} \ .
\end{equation}
The solution is purely outgoing at infinity and purely ingoing at the horizon:
\begin{align}
R_{\ell m \hat{\omega}}(\hat{r} \to \hat{r}_+) &  = Z^\infty_{\ell m \hat{\omega}} \Delta^2 e^{-i \hat{\kappa} 
\hat{r}^\ast} \ ,  \\
R_{\ell m\hat{\omega}}(\hat{r}\to \infty) & = Z^H_{\ell m \hat{\omega}} \hat{r}^3 e^{i \hat{\omega} \hat{r}^\ast} 
\ ,
\end{align}
with
\begin{align}
Z^\infty_{\ell m \hat{\omega}}  &= C^\infty_{\ell m\hat{\omega}} \int_{\hat{r}_+}^{\infty} \dd 
\hat{r}'\frac{\Rup(\hat{r}') 
}{\Delta^2}\mathcal{T}_{\ell m \hat{\omega}}(r') \ , \label{eq:infamp}\\
Z^H_{\ell m \hat{\omega}} &=C^H_{\ell m \hat{\omega}}\int_{\hat{r}_+}^{\infty} \dd 
\hat{r}'\frac{\Rin(\hat{r}')}{\Delta^2} \mathcal{T}_{\ell m\hat{\omega}}(\hat{r}') \ , \label{eq:horamp}
\end{align}
and 
\begin{equation}
C^H_{\ell m \hat{\omega}} =  \frac{1}{2i \hat{\omega} B^{\textup{in}}_{\ell m\omega}} \ , \qquad
C^\infty_{\ell m\hat{\omega}} = \frac{B^{\textup{tran}}_{\ell m \hat{\omega}}}{2i \hat{\omega} B^{\textup{in}}_{\ell 
m\hat{\omega}} 
D^{\textup{tran}}_{\ell m\hat{\omega}}} \ . \label{DBconst}
\end{equation}
The amplitudes $Z^H_{\ell m \hat{\omega}}$ and $Z^\infty_{\ell m \hat{\omega}}$ fully determine the 
asymptotic GW fluxes at infinity and at the horizon.  
The factors $B^{\textup{tran}}_{\ell m \hat{\omega}}$ and $D^{\textup{tran}}_{\ell m \hat{\omega}}$ are 
arbitrary, but it is convenient to fix their values as shown in Appendix~\ref{app:SN}.
As discussed in Sec.~\ref{sec:num}, we compute $\Rin$ and $\Rup$ using two different methods: the 
Mano Suzuki Takasugi~(MST) method~\cite{Mano:1996vt,Fujita:2004rb,Fujita:2009us} and by solving the SN equation (see 
Appendix~\ref{app:SN}). These methods agree with each others within the numerical 
accuracy.

The source term $\mathcal{T}_{\ell m\hat{\omega}}$ of the radial Teukolsky equation is rather cumbersome, even 
for nonspinning bodies. For generic bound orbits, the source term is given by
\begin{equation}
Z^{H,\infty}_{\ell m \hat{\omega}}=C^{H,\infty}_{\ell m \hat{\omega}}\! \int\limits_{-\infty}^{\infty} \! \! \dd 
\hat{t} \, e^{i(\hat{\omega} \hat{t} - m\phi(\hat{t}))}\mathcal{I}^{H,\infty}\big[\hat{r}(\hat{t}),\theta(\hat{t})\big]
\ ,\label{ZHinf}
\end{equation}
where $\mathcal{I}^{H,\infty}\big[\hat{r}(\hat{t}),\theta(\hat{t})\big]$ is
\begin{align}
&\mathcal{I}^{H,\infty}\big[\hat{r}(\hat{t}),\theta(\hat{t})\big] =  \left[\!A_0  - (A_1 + B_1) \totder{}{\hat{r}} + 
\right. \nonumber\\
&\left. \left.+ (A_2+B_2) \frac{\dd^2}{\dd \hat{r}^2}  - B_3 \frac{\dd^3 }{\dd \hat{r}^3} 
\right]\!R^{\textup{in},\textup{up}}_{\ell m \hat{\omega}} \right\rvert_{\theta = \theta(\hat{t}) , \hat{r} = 
\hat{r}(\hat{t})} \ .
\end{align}
Related technical details as well as the explicit form of this term are given in 
Appendix~\eqref{app:Teu source term} [e.g., Eq.~\eqref{eq:genTeuamp}].

At infinity, Eqs.~\eqref{eq:Psi4} and \eqref{eq:horamp} lead to the 
gravitational-wave signal
\begin{equation}
h_+ -i h_\times \sim -\frac{2}{\hat{r}} \displaystyle\sum_{\ell m} \int\limits_{-\infty}^{\infty}\!\frac{\dd 
\hat{\omega}}{\hat{\omega}^2}Z^H_{\ell m \hat{\omega}} e^{i\hat{\omega}( \hat{r}^\ast - \hat{t})}\!\nswsh(\vartheta) 
e^{i m\varphi} \ , \label{waveform}
\end{equation}
where $\vartheta$ is the angle between the observer's line of sight and the spin axis of the primary 
(here aligned with the $z$-axis), while $\varphi \equiv \phi(\hat{t}=0)$.

For a circular equatorial orbit, the form of the source term greatly simplifies
and, since $\phi(\hat{t})=\widehat{\Omega}\hat{t}$, Eq.~\eqref{ZHinf} 
reduces to
\begin{equation}
Z^{H,\infty}_{\ell m \hat{\omega}}= \delta(\hat{\omega} - m \widehat{\Omega}) \mathcal{A}^{H,\infty}_{\ell m 
\hat{\omega}} \ ,
\end{equation}
with $\mathcal{A}^{H,\infty}_{\ell m \hat{\omega}} =  2\pi C^{H,\infty}_{\ell m 
\hat{\omega}}\mathcal{I}^{H,\infty}(\hat{r}_0,\pi/2) $ computed for a specific orbital radius $\hat{r}_0$.
In this case the waveform~\eqref{waveform} reduces to
\begin{equation}
h_+ -i h_\times \sim -\frac{2}{\hat{r}} \displaystyle\sum_{\ell m}\frac{\mathcal{A}^H_{\ell m \hat{\omega}}}{(m 
\widehat{\Omega})^2} e^{im \widehat{\Omega}(\hat{r}^\ast - \hat{t})}\!\nswsh(\vartheta) e^{i m\varphi} \ 
,\label{eq:waveform}
\end{equation}
and the GW energy fluxes are given by
\begin{align}
\left( \frac{\dd \hat{E}}{\dd \hat{A} \dd \hat{t}} \right)^{\!\infty}_{\!\text{GW}} &= 
\frac{1}{16\pi}\media{(\dot{h}_+)^2 + (\dot{h}_\times)^2}_{\text{GW}}\\
&= \frac{1}{4\pi \hat{r}^2}\displaystyle\sum_{\ell m}\frac{\modulo{\mathcal{A}^H_{\ell m 
\hat{\omega}}}^2}{(m\widehat{\Omega})^2}\modulo{\nswsh(\vartheta)}^2 \ ,
\end{align}
where the angle brackets here denote averaging over several wavelengths. 
Using the waveform~\eqref{eq:waveform} and the normalization condition of the spin-weighted 
spheroidal harmonics, the gravitational luminosities are obtained by 
integrating the fluxes over the solid angle, which yields:
\begin{align}
\bigg(\frac{\dd \hat{E}}{ \dd \hat{t}} \bigg)^{\!\infty}_{\!\text{GW}} &= 
\displaystyle\sum_{\ell=2}^{\infty}\displaystyle\sum_{m=1}^{\ell}\frac{\modulo{\mathcal{A}^H_{\ell 
m\hat{\omega}}}^2}{2\pi(m\widehat{\Omega})^2} \ ,
  \label{eq:energyfluxinf}\\
\bigg(\frac{\dd \hat{J}_z}{ \dd \hat{t}} \bigg)^{\!\infty}_{\!\text{GW}} &= 
\displaystyle\sum_{\ell=2}^{\infty}\displaystyle\sum_{m=1}^{\ell} 
\frac{m\modulo{\mathcal{A}^H_{\ell m \hat{\omega}}}^2}{2\pi(m\widehat{\Omega})^3} \ ,
\end{align}
where the sum over $m$ goes for $m=1, \dots ,\ell $ since $Z^{H,\infty}_{\ell-m-\hat{\omega}} = 
(-1)^{\ell}\bar{Z}^{H,\infty}_{\ell m\hat{\omega}} $ and the bar denotes complex conjugation.

Similarly, the GW luminosities at the horizon read~\cite{Hughes:1999bq}
\begin{align}
\bigg( \frac{\dd \hat{E}}{ \dd \hat{t}} \bigg)^{\!H}_{\!\text{GW}} &= 
\displaystyle\sum_{\ell=2}^{\infty}\displaystyle\sum_{m=1}^{\ell}\alpha_{\ell 
m}\frac{\modulo{\mathcal{A}^{\infty}_{\ell 
m\hat{\omega}}}^2}{2\pi(m\Omega)^2} \ , \label{eq:energyfluxhor} \\
\bigg(\frac{\dd \hat{J}_z}{ \dd \hat{t}} \bigg)^{\!H}_{\!\text{GW}} &= 
\displaystyle\sum_{\ell=2}^{\infty}\displaystyle\sum_{m=1}^{\ell}\alpha_{\ell m} 
\frac{m\modulo{\mathcal{A}^{\infty}_{\ell m\hat{\omega}} }^2}{2\pi(m \widehat{\Omega})^3} \ ,
\end{align}
where
\[
\alpha_{\ell m} = \frac{256(2\hat{r}_+)^5\hat{\kappa}(\hat{\kappa}^2 + 4\epsilon^2)(\hat{\kappa}^2 + 
16\epsilon^2)(m\widehat{\Omega})^3}{\modulo{C_{\ell m}}^2}
\]
with $\epsilon = \sqrt{1-\hat{a}^2}/(4\hat{r}_+)$, and
\begin{align}
\modulo{C_{\ell m}}^2 &= [(\lambda_{\ell m \widehat{\Omega}} + 2)^2 + 4\hat{a}(m\widehat{\Omega}) - 
4\hat{a}^2(m\widehat{\Omega})^2]\nn\\
&\times[\lambda_{\ell m\widehat{\Omega}}^2 + 36m\hat{a}(m\widehat{\Omega}) 
- 36\hat{a}^2 (m\widehat{\Omega})^2]\nn\\
 &+ (2\lambda_{lm\widehat{\Omega}}+3)[96\hat{a}^2(m\widehat{\Omega})^2-48m\hat{a}(m\widehat{\Omega})]\nn\\
 &+144(m\widehat{\Omega})^2(1-\hat{a}^2) \ .
\end{align}
%
\subsection{Orbital evolution and GW phase}\label{sec:dephasing}

To compute the overall orbital phase $\Phi$ accumulated during the EMRI, it is necessary to 
calculate the total energy luminosities (from now on also called ``fluxes'', with a slightly abuse of 
terminology):
\begin{align}
\mathcal{F} 
&= \frac{1}{q}\Bigg[ \bigg( \frac{\dd \hat{E}}{ \dd \hat{t}} \bigg)^{\!\!H}_{\!\!\text{GW}} + \bigg( \frac{\dd 
\hat{E}}{ \dd \hat{t}} \bigg)^{\!\!\infty}_{\!\!\text{GW}} \Bigg] \ .
\end{align}
All fluxes were calculated in normalized units, and they were rescaled by the mass ratio $q$. $\mathcal{F}_{\ell m}$ 
denotes the flux for the harmonic indexes $l$ and $m$. We remind that $\hat{E} =  E/\mu$. Since $\dot 
E\propto q^2$ to the leading order, the normalized flux $\mathcal{F}$ does not depend on $q$.

With the fluxes $\mathcal{F}$ at hand, it is possible to calculate the adiabatic evolution of the orbital radius 
$\hat{r}(\hat t)$ and phase $\Phi(\hat t)$ due to radiation losses as follows: 
\begin{alignat}{2}
\totder{\hat{r}}{\hat t} &= -q\mathcal{F}(\hat{r}) \bigg( \totder{\hat{E}}{\hat{r}}\bigg)^{\!-1} 
\qquad 
\totder{\Phi}{\hat t} &= 
\widehat{\Omega}(\hat{r}(\hat{t}))\,, \label{eqdephasing}
\end{alignat}
with $\hat{E}$ given by Eq.~\eqref{eq:energyofr}. 

Finally, for the dominant mode, the GW phase is related to the orbital phase by $\Phi_{\rm GW}=2\Phi$.

\section{Numerical methods}\label{sec:num}
The solutions $\Rin$ and $\Rup$ to the homogeneous Teukolsky equation were calculated in two different ways:
\begin{itemize}
\item through the MST method~\cite{Fujita:2004rb,Fujita:2009us}, as implemented in the \textsc{Mathematica} packages of 
the BH Perturbation Toolkit~\cite{BHPToolkit}.
\item by first solving the SN equation and then transforming the obtained 
solution to $\Rin$ and $\Rup$ (see Appendix~\ref{app:SN}).
\end{itemize}
Both methods require arbitrary precision arithmetic, and the MST method is usually faster and more 
accurate than solving directly the SN equation. Unfortunately, the implementation of the MST 
method of~\cite{BHPToolkit} has one limitation: the precision of 
$\Rin$ and $\Rup$ crucially depends on the gravitational frequency $m \widehat{\Omega}$. As $m \widehat{\Omega}$ 
increases, the precision of the input parameters should drastically increase as well, in order for the computed $\Rin$ 
and $\Rup$ to have enough significant figures. Thus, the MST method tends to become slower for large values of $\ell $ 
and when $\hat{r}$ approaches the ISCO\footnote{For instance, let us consider a nonspinning particle at the ISCO for a 
Kerr BH with $\hat{a} =0.9$: for $\ell=m=2$, with $35$ figures in input, $\mathcal{F}$ is returned with $18$ figures, 
while for $\ell=m=20$, using $90$ figures in input returns fluxes with only $9$ figures of precision. The SN 
method, albeit generally slower, does not has the same issue; the precision of the fluxes 
in output is not affected by the gravitational frequency. }.

We, therefore, took the best of the two methods and implemented both in a \textsc{Mathematica} code. We checked 
that the methods agree with each other within numerical accuracy in the entire parameter space.

Our algorithm is the following: 
\begin{itemize}
\item Choose the parameters $\hat{a}$ and $\chi$;
\item Loop on the harmonic index $\ell$, starting with $\ell=2$ until $\ell_{\rm max}$. We typically used 
$\ell_{\rm max}=20$, see discussion below;
\item If $\ell\leq 8$, loop on the index $ m = 1, \dots , \ell$ starting with $m=1$. For larger values of $\ell$, 
we only considered the $m=\ell$ and $m=\ell-1$, since the others are negligibly small\footnote{When $\ell>8$, we 
compare the flux for $m=\ell$ with the flux for $ m=\ell-i$ at the ISCO. When
\[
\frac{\mathcal{F}_{\ell \ell-i}}{\modulo{\mathcal{F}_{\ell \ell} - \mathcal{F}_{\ell \ell-i}}}< 10^{-6}
\]
for a certain $i =1, \dots ,\ell -1$, we truncate the $m$ series.};
\item Loop on the values of an array of orbital radii $\hat{r}$, starting from $\hat{r}_\textup{start}$. The starting 
point $\hat{r}_\textup{start}$ is calculated in such a way that all the spinning test objects start the inspiral with 
the same frequency of a nonspinning object (i.e $\chi=0$) at the reference value $\hat{r} = 10.1$;
\item Compute the energy fluxes $\mathcal{F}$, using the MST method as implemented in~\cite{BHPToolkit} to obtain 
$\Rin$ and $\Rup$. 
\item The above point is performed within a certain precision threshold. If the MST method fails to give the fluxes 
with prescribed precision (for increasing number of figures in the input parameters; the number 
depends on $\ell$), switch to the SN method. To solve the SN equation, we employed the boundary conditions described in 
Appendix~\ref{app:SN BCs}, keeping $10$ and $13$ terms for 
the series at the horizon and infinity, respectively. 
\item Stop the $\hat{r}$ loop at the ISCO. Interpolate the fluxes in the range $\hat 
r\in(\hat{r}_\textup{ISCO},\hat{r}_\textup{start})$;
\item Using the interpolated fluxes, solve Eq.~\eqref{eqdephasing} to compute the orbital phase.
\end{itemize}

All the fluxes were calculated for prograde stable orbits. The parameters chosen for the numerical simulations are the 
following:
\begin{itemize} 
\item $\hat{a} = (0,\,0.1,\,0.2,\dots \,0.9, \,0.95, \,0.97, \,0.990, \,0.995)$
\item $\chi\in(-2,2)$ with steps $\delta \chi = 0.2$
\item $\mu =30 M_{\odot}$ and $M =10^6 M_{\odot}$, hence $q=3\times10^{-5}$.
\end{itemize}

To estimate the maximum truncation errors of our code, we computed the fluxes at the ISCO for a spinning particle with $\chi=2$
for $\ell=21$ and $\ell=22$ and compared with the corresponding fluxes summed up $\ell_{\rm max}=20$. Choosing 
$\chi=2$ as a reference is just for convenience: the truncation error is practically independent 
of the spin of the secondary, but it is greatly affected by $\hat{a}$ and by the orbital radius. 
In Table~\ref{tab:trunc_err} we report the 
fractional truncation error $\Delta^\text{tr}(\mathcal{F})$ obtained by comparing, for $\chi=2$ and $q=3\times10^{-5}$, 
the fluxes at the ISCO truncated at $\ell=20$ with the fluxes including the $\ell=21$ and $\ell=22$ contributions. 

\begin{table}[!htpb]
\begin{center}
 \begin{tabular}{c|c}
\hline 
\hline 
$\hat{a}$ & $\Delta^\text{tr}(\mathcal{F})$ \\
\hline 
$0$     & $3.5 \times 10^{-11}$ \\
$0.3$   & $4.5 \times 10^{-10}$ \\
$0.5$   & $3.7 \times 10^{-9}$  \\
$0.8$   & $3.4 \times 10^{-7}$  \\
$0.9$   & $3.8 \times 10^{-6}$  \\
$0.97$  & $6.1 \times 10^{-5}$  \\
$0.995$ & $5.0 \times 10^{-4}$  \\
\hline 
\hline 
\end{tabular}
\end{center}
\caption{Fractional truncation error $\Delta^\text{tr}(\mathcal{F})$, obtained by taking $\chi=2$ and 
$q=3\times10^{-5}$ as reference. The error were estimated at the ISCO by comparing the fluxes truncated at $\ell_{\rm 
max}=20$ with the ones truncated at $\ell_{\rm max}=22$.}
\label{tab:trunc_err}
\end{table}

In Appendix~\ref{app:comparison}, we compare our results for the fluxes with previous work, overall finding 
excellent agreement.

\section{Results}\label{sec:results}

\subsection{Spin corrections to fluxes and GW phase}
\begin{figure*}[!htpb]
\includegraphics[width=0.99\textwidth]{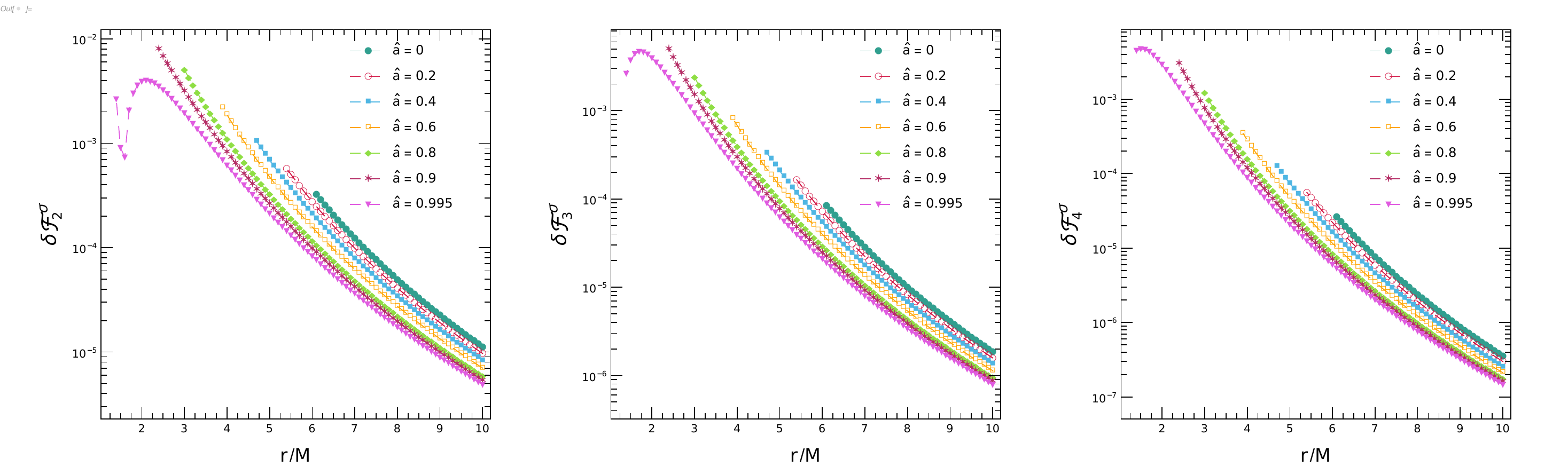}
\includegraphics[width=0.99\textwidth]{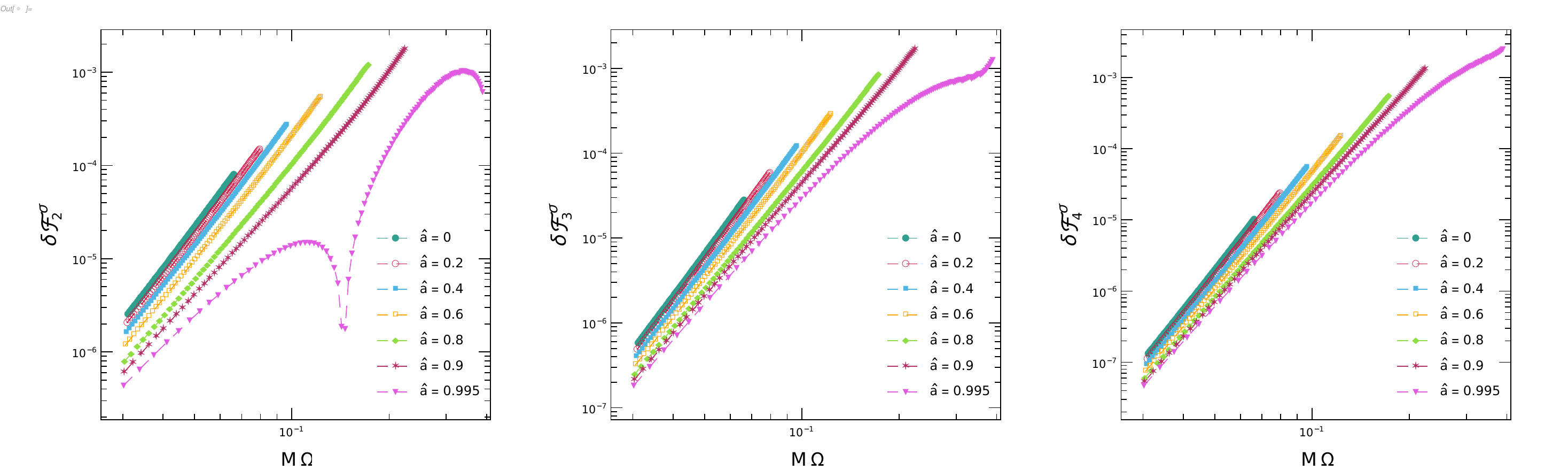}
\caption{Top panels: The spin-correction coefficient $\delta{\cal F}^\sigma_\ell$ [see Eqs.~\eqref{eq:flux_exp} and 
~\eqref{eq:flux_exp2}] as a function of the 
orbital radius (up to the ISCO) for different values of the spin $\hat{a}$ of the primary and for $\ell=2,3,4$ (from 
left to right), summing up to all values of $m$ such that $|m|\leq \ell$. Bottom panel: the same but for the fluxes as 
a function of the orbital frequency.
An analogous plot for the total 
spin-correction $\delta \mathcal{F}^\sigma=\sum_{\ell=2}^{20}\delta{\cal F}^\sigma_\ell$ is presented in a companion 
paper~\cite{Piovano:2020ooe}. Data for the fluxes are available online~\cite{webpage} and on the BH 
Perturbation Toolkit webpage~\cite{BHPToolkit}.
Note that, for nearly-extremal primary ($\hat a \gtrsim 0.99$), 
$\delta{\cal F}^\sigma_2$ is nonmonotonic near the ISCO, although near extremality $\ell=2$ is not the 
dominant spin correction to the flux~\cite{Gralla:2016qfw} and the total correction $\delta {\cal F}^\sigma$ is 
monotonic~\cite{Piovano:2020ooe}.
}
\label{fig:fluxes_corr_over_r} 
\end{figure*}

Due to the small mass ratio, the GW fluxes $\mathcal{F}$ can be expanded at fixed orbital radius $\hat{r}$as 
\begin{equation}
\mathcal{F}(\hat{r},\sigma) = \mathcal{F}^0(\hat{r}) + \sigma \delta \mathcal{F}^\sigma(\hat{r}) + {\cal O}(\sigma^2) \ , \label{eq:flux_exp}
\end{equation}
where $\mathcal{F}^0$ are the fluxes for a nonspinning secondary around a Kerr primary and $\delta \mathcal{F}^\sigma$ 
are the linear spin corrections. The coefficients $\delta \mathcal{F}^\sigma$  were obtained by fitting the fluxes 
$\mathcal{F}$ with a cubic polynomial in $\sigma$ and then retaining only the linear terms. Such 
fitting procedure was repeated for each value of $\hat{r}$ 
at which we computed the fluxes. The top panels of Fig.~\ref{fig:fluxes_corr_over_r} show the linear spin corrections
\begin{equation}
 \delta{\cal F}^\sigma_\ell=\sum_{m=-\ell}^{\ell}\delta {\cal F}_{\ell m}^\sigma\ , \label{eq:flux_exp2}
\end{equation}
for $\ell=2,3,4$ and summing up to all values of $m$ such that 
$|m|\leq \ell$. An analogous plot for the total flux, $\delta \mathcal{F}^\sigma=\sum_{\ell=2}\delta{\cal 
F}^\sigma_\ell$ (summing up to $\ell=20$) is presented in Ref.~\cite{Piovano:2020ooe}.

In the bottom panels of Fig.~\ref{fig:fluxes_corr_over_r} we also show $\delta{\cal F}^\sigma$ for fixed values 
of the orbital frequency instead of $\hat{r}$, since the latter is a gauge dependent quantity.
To this aim, for a given primary spin $\hat{a}$, we considered an evenly spaced grid of frequencies, 
with the same number of points for all the values of $\sigma$, such that 
\begin{equation}
\widehat\Omega(i) = \widehat\Omega_{\text{start}} + (i-1) \delta \widehat\Omega\ ,\quad i=1,...100\ ,
\end{equation}
where $\delta \widehat\Omega = (\widehat\Omega_{\text{ISCO}}-\widehat\Omega_{\text{start}})/100$. 
$\widehat\Omega_{\text{ISCO}}$ and $\widehat\Omega_{\text{start}}$ are the orbital frequency at the ISCO 
and at $\hat{r}_{\text{start}} = 10.1$ for a nonspinning particle, respectively. To compare the fluxes at equal 
frequencies, $\widehat\Omega_{\text{ISCO}}$ was not included in the grid.  At fixed spins, 
it is then possible to find a map between $\widehat\Omega$ and the orbital radius $\hat r$, 
which allows to recast Eq.~\eqref{eq:flux_exp} as
\begin{equation}
\mathcal{F}(\widehat{\Omega},\sigma) = \mathcal{F}^0(\widehat{\Omega}) + \sigma \delta
\mathcal{F}^\sigma(\widehat{\Omega}) + {\cal O}(\sigma^2) \,.
\end{equation}

Having computed the fluxes, we can now proceed to determine the adiabatic orbital evolution and the 
orbital phase by solving Eqs.~\eqref{eqdephasing}. We consider an inspiral starting at $\hat{r}=\hat{r}_{\rm start}$.
Ideally, one would like to evolve the inspiral up to the ISCO. However, since the latter depends on 
$\sigma$, so it does the duration of the inspiral, also for a fixed value of $\hat a$. It 
would therefore be complicated to compare the phase evolution for different spins of the secondary. 
Thus, we chose\footnote{A more rigorous choice is to determine the 
end of the evolution for each binary as the onset of the transition region 
where the adiabatic approximation breaks down~\cite{Ori:2000zn,Burke:2019yek, 
Compere:2019cqe}. However, since the latter depends on the 
secondary spin, a choice of a reference time $t_\text{ref}$ equal for all values of $\sigma$ would still 
be required.} to evolve the inspiral up to a reference end time $t_\text{ref}= t_\text{end} - 1/2 \text{ 
day}$, 
where $ t_\text{end}$ is the time to reach the ISCO for a nonspinning secondary for a given value of $\hat a$.
The offset of $1/2 \text{ day}$ is chosen so that the evolution stops before the ISCO for any value of  $\hat a$ and $\chi$. 

Throughout the inspiral, the phase $\Phi(t)$ can be written as 
\begin{equation}
\Phi (t) = \Phi^0(t) +\frac{ \sigma}{q} \delta \Phi^\sigma(t) +  {\cal O}(\sigma^2/q)\,, 
\label{eq:totphase}
\end{equation}
where $\Phi^0(t)$ is the phase for a nonspinning secondary and $ \delta \Phi^\sigma(t)$ is the 
change due to the ${\cal O}(\sigma)$ contribution. Note that, since $\sigma =q\chi$, the linear 
spin correction is independent of $q$ to the 
leading order, and it is therefore suppressed by a factor $q$ relative to $\Phi^0(t)={\cal O}(1/q)$.
The coefficients $\delta \Phi^\sigma(t)$ were obtained by interpolating  $\Phi(t) -\Phi^0(t)$ with a 
cubic polynomial in $\chi$ as follows
\begin{equation}
 \Phi(t) -\Phi^0(t) = a_0 +\chi a_1 + q \chi^2  a_2 + q^2 \chi^3 a_3\,,
\end{equation}
where $a_i$ are the fit coefficients, with $a_0\approx0$. 
The reported values of $a_1 \equiv \delta \Phi^\sigma(t)$ are robust against the truncation order of the fit.

\begin{figure}[!htpb]
 \includegraphics[width=0.45\textwidth]{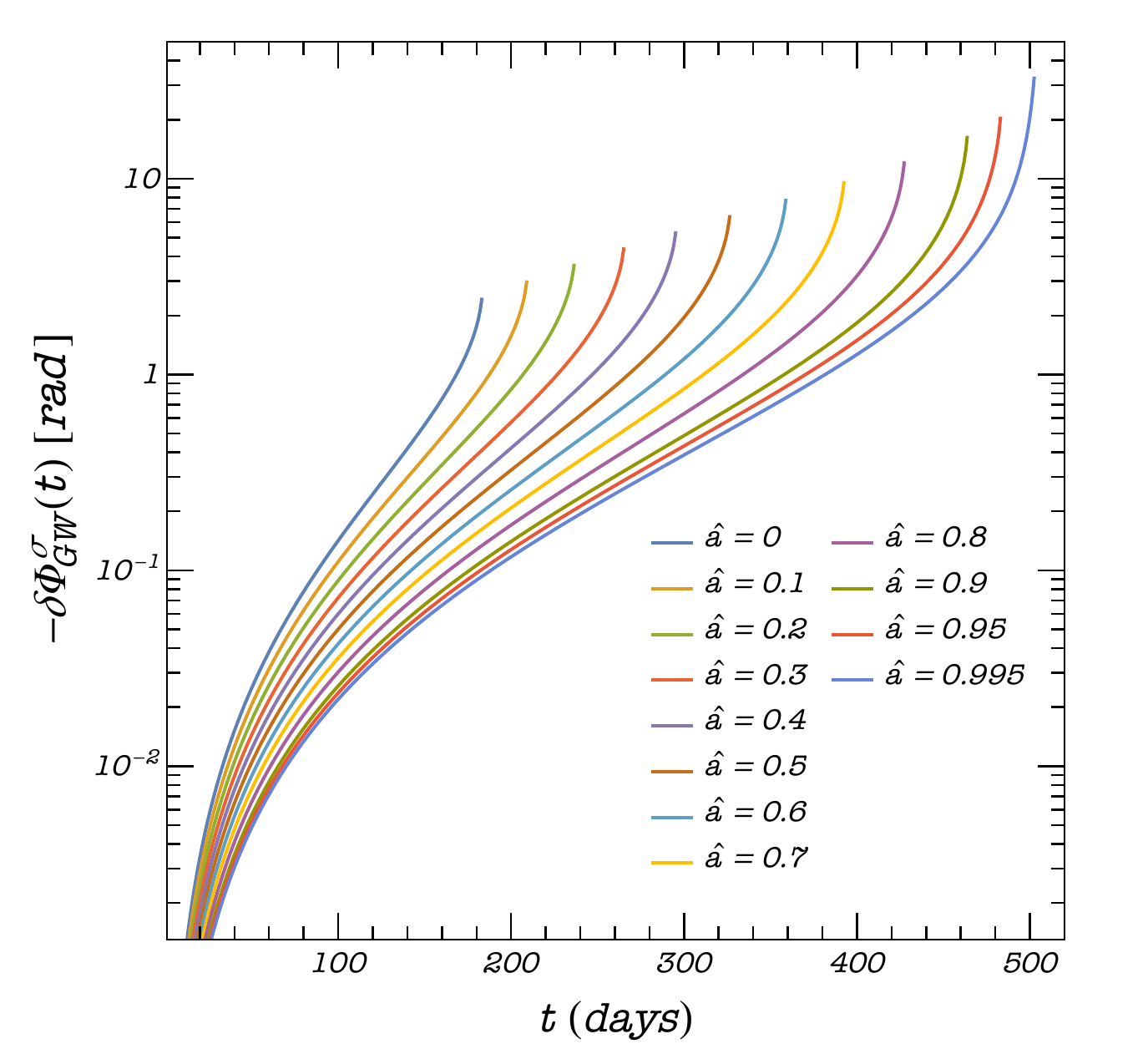}
 \caption{Time evolution of the linear spin corrections to the GW phase $\delta \Phi_\text{GW}^\sigma(t)$ 
for different values of $\hat{a}$.}
 \label{fig:phase_t_ev} 
\end{figure}

\begin{figure}[!htpb]
\includegraphics[width=0.46\textwidth]{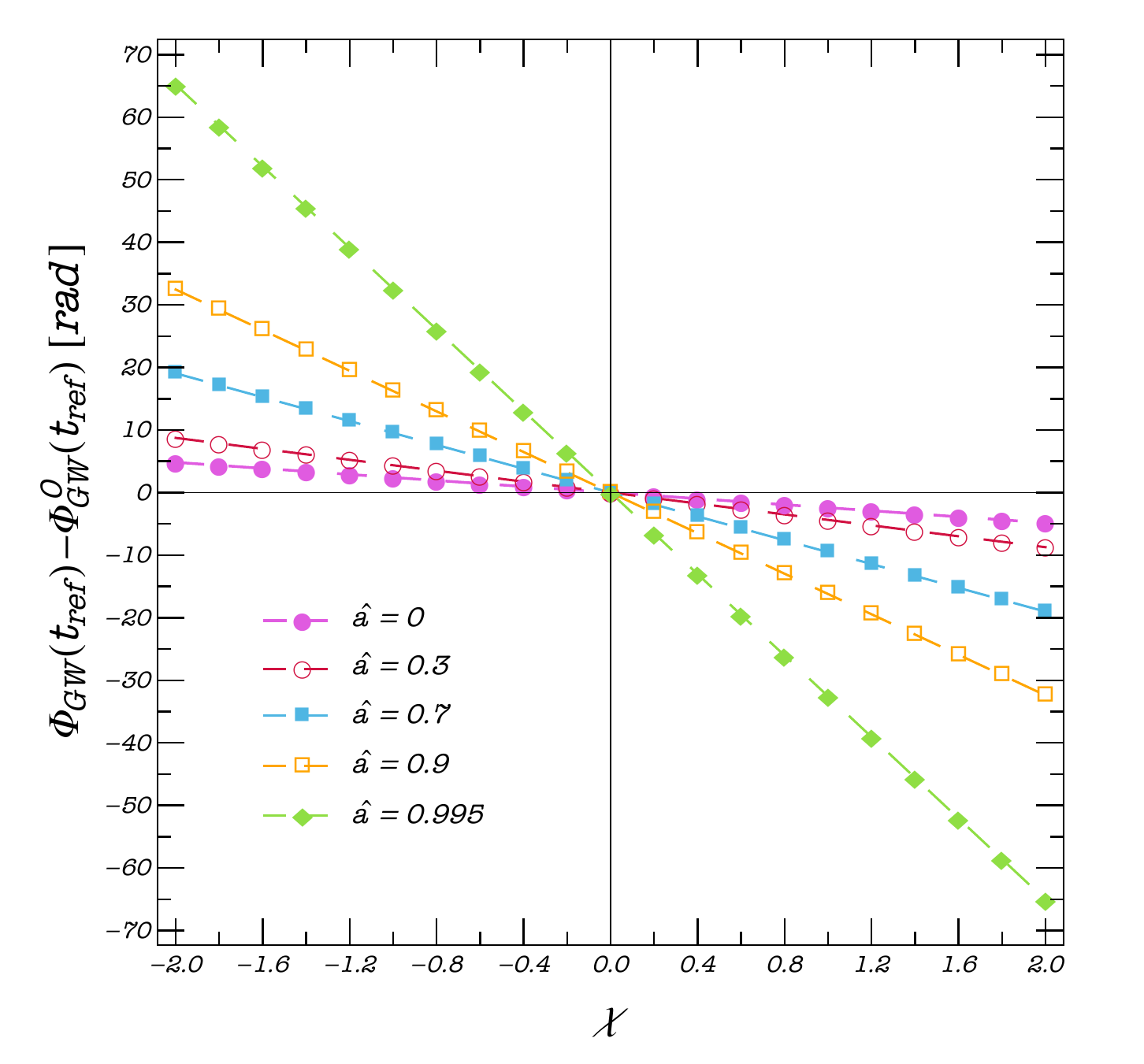}
\caption{ Phase difference $\Phi_\text{GW}(t_\text{ref}) -\Phi_\text{GW}^0(t_\text{ref})$  
between a spinning and nonspinning secondary as a function of $\chi$, calculated at $t_\text{ref}= t_\text{end} - 1/2 
\text{ day}$, where $ t_\text{end}$ is the time to reach the ISCO for a nonspinning secondary. Note that the curves are 
linear to an excellent accuracy, showing that $\Phi_\text{GW}(t_\text{ref}) -\Phi_\text{GW}^0(t_\text{ref})\propto 
\chi$.}
 \label{fig:figphasediff} 
\end{figure}

The orbital phase $\Phi(t)$ is then related 
to the GW phase of the dominant mode by $\Phi_{\rm GW}(t)=2\Phi(t)$. 
The GW phase as a function of time is shown in Fig.~\ref{fig:phase_t_ev} for various values 
of $\hat a$.
Figure~\ref{fig:figphasediff} also shows the phase difference $\Phi_\text{GW}(t_\text{ref}) 
-\Phi^0_\text{GW}(t_\text{ref})$ computed at $t_\text{ref}$ as a function of the spin $\chi$, showing that it is 
linear to excellent accuracy. Although we only present the range $|\chi|\leq 2$, the phase difference is 
linear provided $|\sigma|\ll1$, i.e. $|\chi|\ll 1/q$, as expected.

\begin{table}[!htpb]
\begin{tabular}{c|c|c}
\hline
\hline
 $\hat{a}$ & $\delta \Phi^\sigma_{\rm GW}(t_\text{ref}) [\text{rad}]$ & $\Delta\chi$ \\
\hline 
0   & -2.416 & -0.414  \\
0.1 & -2.962 & -0.338 \\
0.2 & -3.606 & -0.277\\
0.3 & -4.367 & -0.229\\
0.4 & -5.277 & -0.189\\
0.5 & -6.379 & -0.157 \\
0.6 & -7.748 & -0.129\\
0.7 & -9.522 & -0.105\\
0.8 & -12.013 & -0.0832\\
0.9 & -16.215 & -0.0617\\
0.95 &-20.328 & -0.0492\\
0.97 & -23.271 & -0.0430 \\
0.990 &-29.201 & -0.0342\\
0.995 & -32.570 & -0.0307\\
\hline
\hline
\end{tabular}
\caption{Spin corrections to the phase $\delta \Phi^\sigma_{\rm GW}(t_\text{ref})$ and its inverse (which gives the 
resolution on a measurement of $\chi$ according to criterion~\eqref{criterion} with $\alpha=1$) for different values of 
$\hat{a}$. 
}
\label{tab:spin_phase_corr}
\end{table}

The values of $\delta \Phi^\sigma_{\rm GW}(t_\text{ref}) $ (i.e., the slope of the lines shown in 
Fig.~\ref{fig:figphasediff}) for different values of 
$\hat{a}$ are given in Table~\ref{tab:spin_phase_corr} and plotted in Ref.~\cite{Piovano:2020ooe}. 
We fitted these data with two different fits.
The first one is
\begin{equation}
 \delta \Phi^\sigma_{\rm GW}(t_{\rm ref}) = \sum^3_{i=0} b_i (1-\hat a^2)^{i/2} + b_4\hat a\ ,\label{fit1}
\end{equation}
where $b_0=38.44, b_1=-90.36, b_2=99.43, b_3=-44.95, b_4=1.91$. This fit is accurate within 
$5\%$ in the whole range $\hat a\in[0,0.995]$, with better accuracy at large $\hat a$.  
The second fit is
\begin{equation}
 \delta \Phi^\sigma_{\rm GW}(t_{\rm ref}) = \left\{\begin{array}{lll}
                                                   \sum_{i=0}^3 d_i \hat a^i &\quad& \hat a\leq0.7 \\
                                                   \sum_{i=0}^3 e_i (1-\hat a^2)^{i/2}&\quad& 0.7\leq \hat a<0.995
                                                  \end{array}
                                                  \right. \label{fit2}\ ,
\end{equation}
where $d_0=-2.40$, $d_1=-5.70$, $d_2=0.13$, $d_3=-9.25$, and $e_0=-41.42$, $e_1/e_0=-2.49$, $e_2/e_0=3.30$, 
$e_3/e_0=-2.47$. This piecewise fit is accurate within $1\%$ in the whole range $\hat a\in[0,0.995]$.

Finally, we note that the order of magnitude of our dephasing is consistent with previous results that used 
approximated waveforms. In particular, our dephasing is compatible with the results of 
Refs.~\cite{Barack:2006pq,Huerta:2011kt} that used ``kludge'' waveforms, and it 
agrees within a factor $\approx 2$, with the results of Ref.~\cite{Yunes:2010zj}, which used effective-one-body 
waveforms to model the EMRI signal. 

\subsection{Minimum resolvable spin of the secondary}

In a companion paper~\cite{Piovano:2020ooe} we briefly discussed how the above results can be used to place a constraint
on the spin of the secondary in a model-independent fashion, i.e. without assuming any property of the secondary other 
than its mass and spin. Here we take the opportunity to extend that discussion.

Measuring the binary parameters from an EMRI signal is a challenging and open 
problem~\cite{Huerta:2011kt,Babak:2017tow,Chua:2019wgs}, 
which requires developing accurate waveform models, performing a statistical analysis that can 
account for correlations among the waveform parameters, and also taking into account that the 
EMRI events in LISA might overlap with several (possibly louder) simultaneous signals from supermassive BH 
coalescences and other sources~\cite{Audley:2017drz,Chua:2019wgs,LISADataChallenge}.

Postponing a data-analysis study for a follow-up work, here we estimate the minimum 
resolvable $\chi$ by computing the uncertainty on $\chi$ which would lead to a total GW dephasing $\approx 1\,{\rm 
rad}$. A larger dephasing would substantially impact a matched-filter search, leading to a significant loss of detected 
events and potentially to systematics in the parameter estimation~\cite{Lindblom:2008cm}. 

\begin{figure}[t]
\centering
\includegraphics[width=0.48\textwidth]{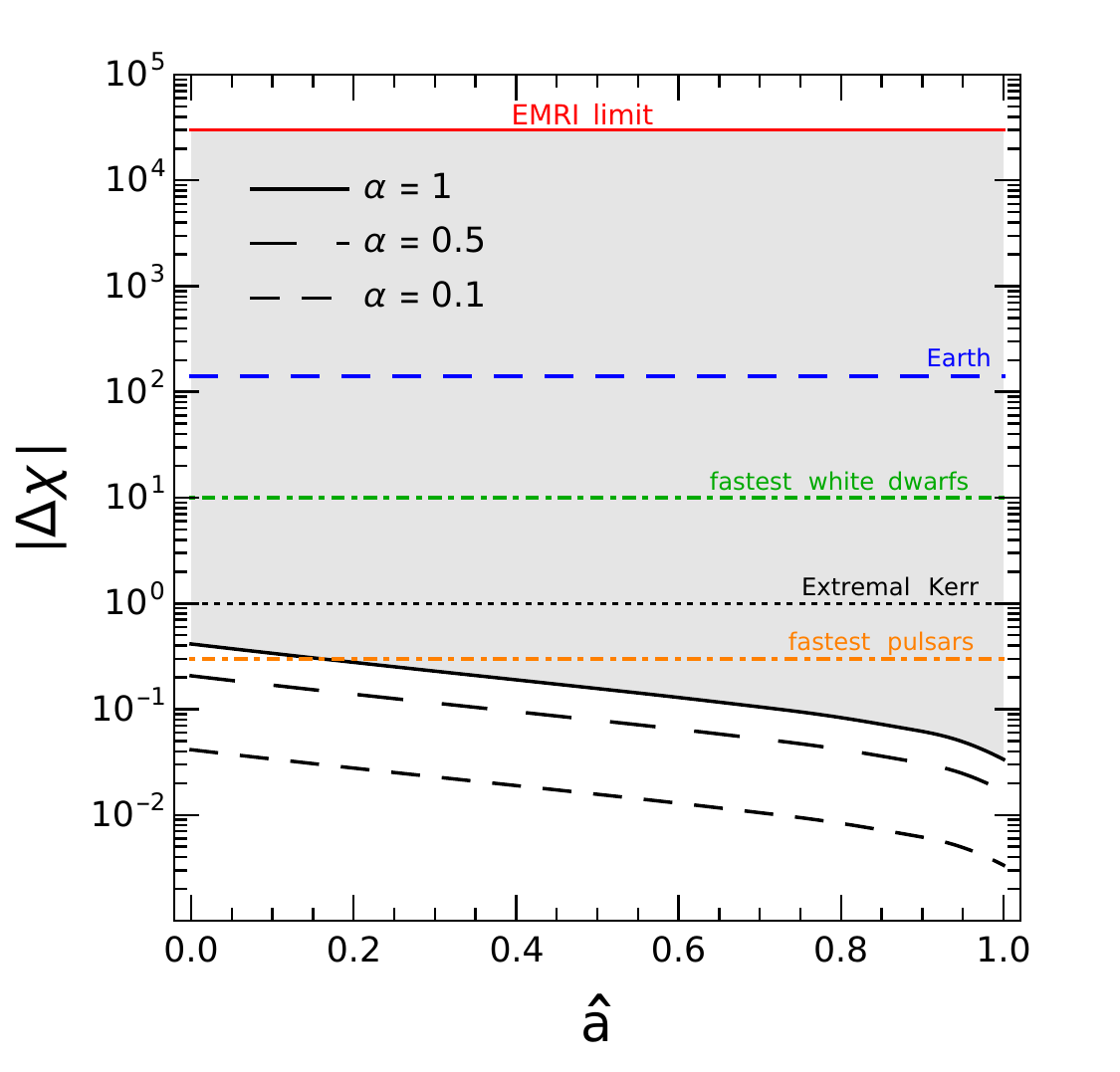} 
\caption{Resolution $|\Delta\chi|$ on a GW measurement of the spin of the EMRI secondary obtained saturating the 
criterion~\eqref{criterion}. A measured GW dephasing at the level of $\alpha\,{\rm rad}$ would probe the region 
above each curve. As a reference, we mark with horizontal lines some typical values of $\chi$ for astrophysical 
objects. Our analysis is valid for $\chi\ll 1/q\approx 3\times 10^4$ (continuous horizontal red line). 
}
\label{fig:exclusion} 
\end{figure}

Let us then suppose that the EMRI masses, the spin of the primary BH $\hat{a}$, and the other waveform parameters 
except $\chi$ are known\footnote{The primary mass and spin and the secondary mass are the parameters that can 
be better constrained in an EMRI~\cite{Barack:2006pq,Huerta:2011kt,Babak:2017tow}.}, i.e. we consider two waveforms 
which differ only by the value of the spin of the secondary, 
$\chi_A$ and $\chi_B$, respectively.
The minimum difference $\Delta \chi=\chi_B-\chi_A$ 
which would lead to a difference in phase larger than $\alpha\,{\rm radiant}$ is~\cite{Piovano:2020ooe}
\begin{equation}
 |\Delta\chi|> \frac{\alpha}{|\delta\Phi^\sigma_{\rm GW}|}\ .\label{criterion}
\end{equation}
The critical value is shown in the last column of Table~\ref{tab:spin_phase_corr} as a function of the primary spin 
$\hat a$ and assuming the $1$-radiant condition, i.e. $\alpha=1$. 
Based on previous analysis in a similar context~\cite{Datta:2019epe}, we expect that more stringent constraints would 
arise by computing the mismatch $\mathcal{M}$ between two waveforms and requiring $\mathcal{M}\gtrsim
1/(2\rho^2)$~\cite{Flanagan:1997kp,Lindblom:2008cm} where $\rho$ is the signal-to-noise ratio of the EMRI signal.
This would suggest using $\alpha<1$ for our estimates, although we shall adopt the more standard and conservative 
requirement and use $\alpha=1$.

Figure~\ref{fig:exclusion} shows the minimum resolution $ |\Delta\chi|$ [obtained saturating Eq.~\eqref{criterion}] as a 
function of the primary spin. For each chosen value of $\alpha$, the area above the corresponding 
curve identifies binary configurations producing a measurable dephasing according to our simplified analysis.
In other words, the spin $\chi$ of a secondary can be measured with 
a relative error $\Delta\chi/\chi$.

It is interesting to compare such resolution with typical values of $\chi$ for known 
astrophysical objects. 
If the secondary is a Kerr BH, then $|\chi|\leq 1$. For the fastest millisecond pulsars, 
$\chi\approx 0.3$, although fast spinning pulsars are all in strongly-accreting binary systems, whereas 
isolated pulsars are expected to spin more slowly.
However, $\chi$ can be much larger than unity for other objects. For example, a ball of radius $\SI{1}{\centi \metre}$ 
and mass $\SI{1}{\kilo \gram}$ making one rotation per second has $\chi \approx 1 \times 10^{17}$. Astrophysical 
objects do not reach such extreme values, but can have $\chi\gg1$~\cite{Hartl:2002ig}. For example, $\chi\approx 140$ 
for Earth, and $\chi\approx10$ for the fastest white dwarfs in accreting binary systems.
The above reference values are shown in Fig.~\ref{fig:exclusion} by horizontal lines. 

Note that $ |\Delta\chi|<1$ in all cases, and therefore our simplified analysis suggests that the spin of a rapidly 
spinning Kerr secondary could be measured with an accuracy greater than $100\%$.

\subsection{Model-independent constraints on ``superspinars''}
%
Compact dark objects which exceed the Kerr bound $|\chi|\leq1$ (so-called 
``superspinars'') were suggested to arise generically in high-energy 
modifications to general relativity such as string theories~\cite{Gimon:2007ur}. Our results of 
Fig.~\ref{fig:exclusion} 
show that the typical resolution on $\chi$ achievable with an EMRI detection can be used to rule out (or detect) 
superspinars in a large region of the parameter space~\cite{Piovano:2020ooe}. For example, if $\chi\approx\hat 
a\approx (0.5-0.7)$, a measurement with absolute error $\Delta\chi$ would exclude $\chi>1$ at 
$3\sigma$ confidence level.
This is particularly interesting in light of the fact that no theoretical upper bound is expected for superspinars, 
besides, possibly, those coming from the ergoregion 
instability~\cite{Pani:2010jz,Maggio:2017ivp,Maggio:2018ivz,Roy:2019uuy}. A measurement of $\chi$ at the level 
reported above can thus potentially probe a vast region of the parameter space for superspinars~\cite{Piovano:2020ooe}.

In principle, a putative EMRI measurement of $|\chi|>1$ could still be degenerate with the secondary being a 
neutron star or a white dwarf. 
Given the theoretical upper bound on the maximum mass of such objects, an EMRI measurement 
of $\mu$ larger than $3M_\odot$ (resp. $\sim 1.4M_\odot$) would exclude a standard origin for the 
superspinar, as a neutron star (resp. a white dwarf). Similarly, no compact object spinning above the Kerr 
bound is know with $\mu\ll M_\odot$.

Moreover, even within the allowed, narrow, mass ranges, 
\emph{isolated} compact stars feature spins smaller than the Kerr bound. Fast rotating 
neutron stars or white dwarfs are expected to evolve in accreting systems. For example, 
the fastest spinning white dwarf to date has $\chi\approx 10$, but it is strongly accreting from a binary 
companion~\cite{1997A&A...317..815B}. Interestingly, all the observed fast rotating neutron stars
\footnote{Including the fastest known pulsar PSR J1748-2446ad with $\chi\approx0.3$~\cite{Hessels:2006ze}. As a 
reference, out of $340$ observations of millisecond pulsars in 
the ATNF Pulsar Database~\cite{Manchester:2004bp}, $\langle \chi\rangle = 0.11\pm0.04$, suggesting that $|\chi|>1$ 
would be very unlikely.}
rotate consistently below their theoretical maximum set by the mass shedding limit. While no solid 
explanation does exist to bridge this gap, EMRIs can provide a new  window to discover neutron stars 
spinning close to the mass-shedding limit. Finally, less compact objects, such as brown dwarfs, might also have 
spin larger then the Kerr bound, but can be easily distinguishable from exotic superspinars, as 
they are tidally disrupted much before reaching the ISCO\footnote{As a reference, the critical 
tidal-disruption 
radius is of the order $R_t\sim M q^{2/3}/C$, where $C=\mu/R$ is the compactness of the secondary with radius $R$. For 
a typical brown dwarf $C\sim 10^{-6}$, and $R_t \sim 100 M$ for $q\sim 10^{-6}$. In general, objects less 
compact than white dwarfs are tidally disrupted at low frequency and can be distinguished on this ground.}.

Finally, in the context of our study one could wonder whether it is theoretically consistent to study a 
secondary superspinar around a primary Kerr BH. This is indeed the case in two scenarios (see 
Ref.~\cite{Cardoso:2019rvt} for a 
review): a) if superspinars arise within general relativity in the presence of exotic matter fields, in such case both 
Kerr BHs and superspinars can co-exist in the spectrum of solutions of the theory; b) if superspinars arise in
high-energy modified theories of gravity such as string theories, as originally proposed~\cite{Gimon:2007ur}. In the 
latter case it is natural to expect that high-energy corrections which are relevant for the secondary might be 
negligible for the primary. Indeed, in an effective-field-theory approach high-energy corrections to general relativity 
modify the Einstein-Hilbert action with the inclusion of higher-order curvature terms of the 
form~\cite{Berti:2015itd,Barack:2018yly}
\begin{equation}
 R+...+ \beta (R_{abcd})^n +... \,, \quad n>1 
\end{equation}
where $R$ is the Ricci scalar, $R_{abcd}$ schematically denotes terms that depend on the Riemann tensor, and $\beta$ 
is a coupling constant with dimensions of a $(\text{length})^{2(n-1)}$. In these theories relative corrections to the metric 
of a compact object of size $\sim L$ are of the order of~\cite{Barausse:2014tra}
\begin{equation}
 \frac{\beta}{L^{2(n-1)}}\,,
\end{equation}
or some power thereof. Thus, the difference between the high-curvature corrections of the secondary relative to those 
of the primary scales as
\begin{equation}
 \sim \frac{M^{2(n-1)}}{\mu^{2(n-1)}} = q^{2(1-n)}\gg1\,.
\end{equation}
This heuristically shows the obvious fact that in an EMRI the secondary is much more affected by the high-curvature 
corrections than the primary, especially for high-order terms (i.e., higher values of $n$).

In certain high-curvature corrections to general relativity, the secondary might also be charged under 
new fundamental fields, in which case there is also extra emission (in particular there could be dipolar, $\ell=1$, 
fluxes)~\cite{Pani:2011xj,Cardoso:2018zhm,Maselli:2020zgv}.

\section{Conclusion and future work}\label{sec:conclusion} 

We have studied the GW fluxes and the adiabatic evolution of a spinning point particle in circular, 
equatorial motion around the Kerr background and with spin (anti)aligned to that of the central 
BH. Our results for the fluxes agree with those previously appeared in the literature, whereas the computation of the 
GW phase in Kerr spacetime is novel .

Since the EMRI dynamics does not depend on the nature of the secondary but only on its multiple moments, 
the GW signal can be used to derive model-independent constraints on the secondary, for example to measure the spin 
of a Kerr secondary, or to distinguish whether 
the secondary is a fast spinning BH or a slowly-spinning neutron star, or also whether the secondary 
satisfies the Kerr bound or is a superspinar~\cite{Piovano:2020ooe}. 

This work represents a first step in the analysis of the impact of the secondary spin on EMRI's evolution, in parallel 
with recent work along related directions. 
Future work will include extensions to generic orbits (e.g., along the lines of Ref.~\cite{Witzany:2019nml}), 
misaligned spins (which introduce precession~\cite{Tanaka:1996ht,Bini:2006pc,Dolan:2013roa,Ruangsri:2015cvg}), 
and the development of data analysis approaches~\cite{Chua:2019wgs} to assess the detectability of such effects. 
In particular, it is important to assess the role of parameter correlations in the measurement of small effects 
such as the spin of the secondary, as discussed in Ref.~\cite{Huerta:2011kt}.
A complete account of dissipative effects in the case of a spinning secondary would also require to consider the spin 
evolution due to self-force effects, which is a more challenging problem, especially for generic 
orbits~\cite{Akcay:2019bvk}. Moreover, an important extension of this work is to include the contribution of the 
conservative first-order self-force on the equations of motion~\cite{Burko:2003rv,Burko:2015sqa,Warburton:2017sxk} and 
study how this affects the ${\cal O}(\sigma)$ in the GW signal.

Another interesting extension is to include the quadrupole moment of the 
secondary~\cite{Hinderer:2013uwa,Steinhoff:2012rw,Bini:2014xyr}. Compared to the spin, this effect is suppressed by a 
further power of the mass ratio and is probably negligible for EMRI detection with LISA, although a rigorous study 
is required to assess whether neglecting this term can affect parameter estimation for the loudest events.
Furthermore, since the quadrupole moment of a Kerr BH is uniquely determined in terms of its mass a spin, measuring the 
quadrupole of the secondary would allow for model-independent tests of the BH no-hair theorem.

Finally, more theoretical related work includes nonintegrability and chaotic motion for generic values of the 
spin~\cite{Zelenka:2019nyp,Lukes-Gerakopoulos:2016udm,Hartl:2002ig}, although these effects might require extremely 
high values for the spin of the secondary and should not be directly relevant for the phenomenology of EMRI signals 
detectable with LISA.

\begin{acknowledgments}
We thank Richard Brito for useful discussion and Niels Warburton for reading the draft and providing valuable 
suggestions. G.A.P. would like to thank Viktor Skoup\'{y} for pointing out a typo in Table~\ref{tab:newtonfluxa0}.
This work makes use of the Black Hole Perturbation Toolkit and \textsc{xAct} \textsc{Mathematica} package. 
P.P. acknowledges financial support provided under the European Union's H2020 ERC, Starting 
Grant agreement no.~DarkGRA--757480, and under the MIUR PRIN and FARE programmes (GW-NEXT, CUP:~B84I20000100001).
The authors would like to acknowledge networking support by the COST 
Action CA16104 and support from the Amaldi Research Center funded by 
the MIUR program "Dipartimento di Eccellenza" (CUP: B81I18001170001).

\end{acknowledgments}


\appendix

\section{Sasaki-Nakamura equation}\label{app:SN}
In this and in the following appendix we provide further technical details on the formalisms that we 
use in Secs.~\ref{sec:radreaction}-\ref{sec:num} to compute the GW fluxes. 

The homogeneous Teukolsky equation is an example of stiff differential problem, 
with the solutions \eqref{eq:inBCTeu}-\eqref{eq:upBCTeu} rapidly diverging at infinity due 
to the long-range character of the potential. High accuracy solutions require therefore 
time-consuming numerical integrations.
A substantial improvement in this direction has been achieved by Sasaki and Nakamura, 
finding a suitable transformation which maps the homogeneous Teukolsky equation to an 
equivalent form with a short-range potential that is easier to solve numerically \cite{Sasaki:1981sx}. 
The SN equation is given by (we remind that hatted quantities are dimensionless)
\begin{equation}
\bigg[f(\hat{r})^2\frac{\dd^2}{\dd \hat{r}^2}+f(\hat{r}) \bigg(\totder{f(\hat{r})}{\hat{r}} - F(\hat{r}) \bigg) 
\totder{}{\hat{r}} -U(\hat{r})\bigg] 
X_{\ell m \hat{\omega}}=0 \ , \label{eq:SNeq}
\end{equation}
with $f(\hat{r})= \totder{\hat{r}}{\hat{r}^*}= \frac{\Delta}{\hat{r}^2+\hat{a}^2}$.
The coefficient $F(\hat{r})$ is defined as
\begin{equation}
F(\hat{r})= \frac{\eta(\hat{r})_{,\hat{r}}}{\eta(\hat{r})}\frac{\Delta}{\hat{r}^2 + \hat{a}^2} \ ,
\end{equation}
where ${}_{,\hat{r}}$ denotes the derivative with respect to $\hat{r}$ and 
\begin{equation}
\eta(\hat{r}) = c_0 + \frac{c_1}{\hat{r}} + \frac{c_2}{\hat{r}^2} + \frac{c_3}{\hat{r}^3} + \frac{c_4}{\hat{r}^4} \ ,
\end{equation}
with
\begin{align}
c_0 &= - 12i \hat{\omega}  + \lambda_{\ell m\hat{\omega}}(\lambda_{\ell m\hat{\omega} }+2) - 12\hat{a}\hat{\omega}( 
\hat{a}\hat{\omega} - m) \ , \label{c0}\\
c_1 &= 8 i \hat{a} [3\hat{a} \hat{\omega} - \lambda_{\ell m\hat{\omega}}(\hat{a}\hat{\omega}-m)] \ , \\
c_2 &=-24 i \hat{a} (\hat{a}\hat{\omega} - m) + 12\hat{a}^2[1-2(\hat{a}\hat{\omega} -m)^2] \ ,  \\
c_3 &= 24i \hat{a}^3(\hat{a}\hat{\omega}-m)-24 \hat{a}^2 \ ,   \\
c_4 &= 12\hat{a}^4 \ .
\end{align}
The function $U(\hat{r})$ in Eq.~\eqref{eq:SNeq} reads
\begin{equation}
U(\hat{r}) = \frac{\Delta U_1(\hat{r})}{(\hat{r}^2+\hat{a}^2)^2}+G(\hat{r})^2 + \frac{\Delta 
G(\hat{r})_{,\hat{r}}}{\hat{r}^2+\hat{a}^2}  - F(\hat{r})G(\hat{r}) \ ,
\end{equation}
where
\begin{align}
G(\hat{r}) &= - \frac{2(\hat{r}-1)}{\hat{r}^2+\hat{a}^2} + \frac{\hat{r}\Delta}{(\hat{r}^2+\hat{a}^2)^2} \ ,\\
U_1(\hat{r}) &= V(\hat{r}) +\frac{\Delta^2}{\beta}\Big[\Big(2 \alpha + \frac{\beta_{,\hat{r}}}{\Delta} 
\Big)_{\!\!,\hat{r}} -\frac{\eta(\hat{r})_{,\hat{r}}}{\eta(\hat{r})}\Big(\alpha + \frac{\beta_{,\hat{r}}}{\Delta}\Big) 
\Big]  \ ,\\
\alpha &= -i K(\hat{r}) \frac{\beta}{\Delta^2}  + 3i K(\hat{r})_{,\hat{r}} + \lambda_{\ell m \hat{\omega}} + 
\frac{6\Delta}{\hat{r}^2} \ ,\\
\beta &=2 \Delta\Big[ -i K(\hat{r}) + \hat{r} - 1 - \frac{2\Delta}{\hat{r}} \Big]  \ .
\end{align}
The two functions $K(\hat{r})$ and $V(\hat{r})$ are the same introduced 
for the Teukolsky radial equation~\eqref{eq:radialTeueq}.

The SN equation admits two linearly independent solutions, $\Xin$ and $\Xup$, which behave asymptotically as
\begin{equation}
\Xin
\sim
\begin{cases}
e^{-i \hat{\kappa} \hat{r}^\ast} \quad  &\hat{r} \to \hat{r}_+  \\
A^{\textup{out}}_{\ell m \hat{\omega}}  e^{i \hat{\omega} \hat{r}^\ast} + A^{\textup{in}}_{\ell m 
\hat{\omega}} e^{-i \hat{\omega} \hat{r}^\ast} \quad 
&\hat{r} \to \infty  \label{eq:inBCSN}
\end{cases}\,,
\end{equation}
\begin{equation}
\Xup\sim
\begin{cases}
 C^{\textup{out}}_{\ell m \hat{\omega}} e^{i \hat{\kappa} \hat{r}^{\ast}} + C^{\textup{in}}_{\ell m 
\hat{\omega}}e^{- i \hat{\kappa} \hat{r}^{\ast}} \quad \,  &r\to r_+   \label{eq:upBCSN} \\
e^{i \hat{\omega} \hat{r}^{\ast}}  \, \quad &\hat{r} \to \infty  
\end{cases}\,.
\end{equation}
The solutions of the Teukolsky and SN equations are related by:
\begin{align}
R^{\textup{in},\textup{up}}_{\ell m \hat{\omega}}(\hat{r}) &=\frac{1}{\eta}\bigg[\bigg(\alpha 
+\frac{\beta_{,\hat{r}}}{\Delta}\bigg)Y^{\textup{in},\textup{up}}_{\ell m \hat{\omega}} - 
\frac{\beta}{\Delta}{Y^{\textup{in},\textup{up}}_{\ell m \hat{\omega}}}_{\!\!, \hat{r}}\bigg] 
\label{eq:fromSNtoTeu} \ ,\\
Y^{\textup{in},\textup{up}}_{\ell m \hat{\omega}}& =\frac{\Delta}{\sqrt{\hat{r}^2+\hat{a}^2}}  
X^{\textup{in},\textup{up}}_{\ell m \hat{\omega}} \ .
\end{align}
With the above normalization of the solutions $\Xin$ $\Xup$, these transformations allow to fix the arbitrary 
constants $D^{\textup{tran}}_{\ell m \hat{\omega}}$ and $B^{\textup{tran}}_{\ell 
m\omega} $ [cf. Eq.~\eqref{DBconst}] as~\cite{Mino:1997bx}: 
\begin{equation}
D^{\textup{tran}}_{\ell m \hat{\omega}}= - \frac{4\hat{\omega}^2}{c_0} \ , \qquad \quad B^{\textup{tran}}_{\ell m 
\hat{\omega}}  = \frac{1}{d_{\ell m\hat{\omega}}} \ ,
\end{equation}
where
\begin{align}
&d_{\ell m\hat{\omega}} = 4\sqrt{2 \hat{r}_+} [(2 - 6i \hat{\omega} - 4\hat{\omega}^2)\hat{r}_+^2 + (3i \hat{a} m 
-4\nonumber  \\
& + 4 \hat{a}\hat{\omega}m + 6i\hat{\omega})\hat{r}_+ -\hat{a}^2m^2 -3iam +2] \ ,
\end{align}
and the coefficient $c_0$ is given in Eq.~\eqref{c0}.

The numerical values of $\Xin$ (resp. $\Xup$) are obtained by integrating 
Eq.~\eqref{eq:SNeq} from $\hat{r}_+$ (resp. infinity) up to infinity (resp. $\hat{r}_+$) 
using the boundary conditions~\eqref{eq:inBCSN} (resp. \eqref{eq:upBCSN}).
In this work we have derived the boundary conditions for the 
homogeneous SN equation in terms of 
explicit recursion relations which can be truncated at arbitrary order (see Sec.~\ref{app:SN BCs}).
We finally transform back $\Xin , \Xup$ to the Teukolsky solutions using Eq.~\eqref{eq:fromSNtoTeu}. The amplitude 
$B^{\textup{in}}_{\ell m \hat{\omega}}$ can be obtained from the Wronskian $W_{\hat{r}}$ at a given orbital separation.

\subsection{Boundary conditions for the SN equation in terms of recursion relations}\label{app:SN BCs}
We have derived accurate boundary conditions 
by looking for series expansions of the master equation 
at the outer horizon $\hat{r}_+$ and at infinity. To this aim we 
have studied the singularities on the real axis 
of Eq.~\eqref{eq:SNeq}, which can be recast in the form
\begin{equation}
\Delta^2\frac{d^2X_{\ell m\hat{\omega}}}{d\hat{r}^2} +\Delta \overline{F}(\hat{r})\frac{dX_{\ell 
m\hat{\omega}}}{d\hat{r}} + \overline{U}(\hat{r})X_{\ell m\hat{\omega}}=0 \ ,
\end{equation}
where
\begin{align}
\overline{F}(\hat{r}) &= (\hat{r}^2+\hat{a}^2) \bigg(\totder{f(\hat{r})}{\hat{r}} - F(\hat{r}) \bigg) \ ,\\
\overline{U}(\hat{r}) &= -  (\hat{r}^2+\hat{a}^2)^2 U(\hat{r}) \ .
\end{align}
Moreover
\begin{align}
F(\hat{r}_\pm) &= 0 \ , & \quad F(\hat{r}) &\xrightarrow[\hat{r} \to \infty]{ } 0 \ , \\
U(\hat{r}_+) &= - \hat{\kappa}^2  \ , &\quad  U(\hat{r}) &\xrightarrow[\hat{r} \to \infty]{ } -\hat{\omega}^2 \ .
\end{align}
Since the functions $\overline{F}(\hat{r}) $ and $\overline{U}(\hat{r}) $ are analytic 
on the positive real axis, it turns out that 
the Eq.~\eqref{eq:SNeq} has three singularities: two at the horizons $\hat{r}=\hat{r}_-$  and $\hat{r}=\hat{r}_+$, both 
of which are regular singularities, and one at $\hat{r}=\infty$ which is an irregular singularity of rank $1$.
By Fuchs theorem, the solutions of the SN equation around 
$\hat{r}_+$ can be written as Frobenius series, with radius of 
convergence
\begin{equation}
\hat{r}_+ - \hat{r}_ -= 2 \sqrt{1-\hat{a}^2} \ .
\end{equation}
For $\hat{r}=\infty$ or $\hat{a}=1$ (for which $\hat{r}_+=\hat{r}_-$) the boundary conditions 
can be written in terms of asymptotic expansions.

\subsubsection{Boundary condition at the horizon}
To compute the boundary conditions at the outer horizon $\hat{r}_+$, it is 
convenient to recast the SN equation as
\begin{equation}
(\hat{r}-\hat{r}_+)^2\frac{d^2X_{\ell m\hat{\omega}}}{d\hat{r}^2} + (\hat{r}-\hat{r}_+) p_H(\hat{r})\frac{dX_{\ell 
m\omega}}{d\hat{r}} + q_H(\hat{r})X_{\ell m \hat{\omega}}=0
\end{equation}
where 
\begin{align}
p_H(\hat{r}) &=\bigg( \frac{\hat{r}^2+\hat{a}^2}{\hat{r}-\hat{r}_-}\bigg) \bigg[\totder{f(\hat{r})}{\hat{r}} - 
F(\hat{r}) \bigg] \ ,\\
q_H(\hat{r}) &= - \bigg(\frac{\hat{r}^2+\hat{a}^2}{\hat{r}-\hat{r}_-}\bigg)^{\!2} U(\hat{r}) \ .
\end{align}
Following the Frobenius method we look for a power series solution of 
the form
\begin{equation}
X_{\ell m \hat{\omega}} = (\hat{r}-\hat{r}_+)^d \displaystyle\sum_{n=0}^{\infty} a_n (\hat{r}-\hat{r}_+)^n \ ,
\end{equation}
where $d$ is one of the solutions of the indicial equation
\begin{equation}
I(d) = d(d-1) + p_H(\hat{r}_+) d + q_H(\hat{r}_+) =0\ .
\end{equation}
For Eq.~\eqref{eq:SNeq}, the latter corresponds to 
\begin{equation}
I(d) = d^2 +  \kappa^2\left(\frac{2\hat{r}_+}{\hat{r}_+ - \hat{r}_-}\right)^{\!\!2} = 0 \ , \quad \ \hat{\kappa} = 
\hat{\omega} - 
\frac{m \hat{a}}{2 \hat{r}_+} \ .
\end{equation}
Given $(d_1,d_2)$ two solutions of the above equation, 
their difference $d_1 -d_2$ is neither zero nor an integer. We have 
therefore two linearly independent solutions such that
\begin{equation}
X_{\ell m \hat{\omega}}=\exp\bigg\{\pm i \hat{\kappa} \frac{2 \hat{r}_+}{\hat{r}_+ - \hat{r}_-} \log(\hat{r} - 
\hat{r}_+)\bigg\} \displaystyle \sum_{n=0}^{\infty} 
a_n (r-r_+)^n \label{eq:BCHor} \ .
\end{equation}
The recursion relation for the coefficients $a_n$ is (setting $a_0=1$)
\begin{equation}
a_n = - \frac{1}{I(d+n)}  \displaystyle\sum_{k=0}^{n-1}\frac{(k+d)p_H^{(n-k)}(\hat{r}_+) +q_H^{(n-k)}(r_+)}{(n-k)!} a_k 
\ ,
\end{equation}
where $p_H^{(k)}(\hat{r}_+) $ and $q_H^{(k)}(\hat{r}_+) $ are the $k$-th 
derivatives of the coefficients $p_H(\hat{r})$ and $q_H(\hat{r})$ with respect 
to $\hat{r}$, and calculated at $\hat{r}_+$.
For $\hat{a}\leq 0.9$, the boundary conditions at the horizon have been 
calculated at $\hat{r}_\text{in} = \hat{r}_+ + \epsilon$ with 
$\epsilon = 10^{-3}$, while for higher spins we have fixed $\epsilon = 10^{-5}$.
To increase precision, we truncate compute the series coefficients up to $n=10$.

\subsubsection{Boundary condition at infinity}

Ordinary differential equations with irregular singularities of rank $1$, like the SN equation, admit general 
expressions for asymptotic 
expansions around such singularities (see Refs.~\cite{Olver:1994:AEC,Olver:1997:ASL} and especially 
Ref.~\cite{Olver:1974asymptotics} for more details). To 
calculate the boundary conditions at infinity we rewrite the SN equation as
\begin{equation}
\frac{d^2X_{\ell m\hat{\omega}}}{d\hat{r}^2} + p_\infty(\hat{r})\frac{dX_{\ell m \hat{\omega}}}{d\hat{r}} + 
q_\infty(\hat{r}) X_{\ell m \hat{\omega}}=0\ ,
\end{equation}
where 
\begin{align}
p_\infty(\hat{r}) &= \frac{(\hat{r}^2+\hat{a}^2)}{\Delta} \bigg[\totder{f(\hat{r}) }{\hat{r}}- F(\hat{r}) \bigg] \ 
,\\
q_\infty(\hat{r}) &= - \bigg(\frac{\hat{r}^2+\hat{a}^2}{\Delta}\bigg)^{\!2} U(\hat{r}) \ .
\end{align}
The functions $p_\infty(\hat{r})$ and $q_\infty(\hat{r})$ are analytic on 
the positive real axis, so the series
\begin{alignat*}{2}
p_\infty(\hat{r}) &= \displaystyle \sum_{n=0}^{\infty} \frac{1}{n!}  \frac{p_\infty^{(n)}}{\hat{r}^n} \ ,  \qquad 
q_\infty(\hat{r}) &= 
\displaystyle \sum_{n=0}^{\infty} \frac{1}{n!}  \frac{q_\infty^{(n)}}{\hat{r}^n} \ ,
\end{alignat*}
converge, with $p_\infty^{(n)}$ and $q_\infty^{(n)}$ being the $n$-th derivatives of 
the coefficients $p_\infty$ and $q_\infty$  with respect to $\hat{r}$. 
If at least one of $p_\infty^{(0)}$, $q_\infty^{(0)}$ or $q_\infty^{(1)}$ is 
nonzero, the formal solution is given by 
\begin{equation}
X_{\ell m\hat{\omega}} = e^{\gamma \hat{r}} \hat{r}^\xi \displaystyle\sum_{n=0}^{\infty} \frac{b_n}{\hat{r}^n} \ ,
\end{equation}
where $\gamma$ is one of the solutions of the characteristic equation
\begin{equation}
\gamma^2 + p^{(0)}_\infty \gamma + q^{(0)}_\infty =0 \ ,
\end{equation}
while
\begin{equation}
\xi= -\frac{p_\infty^{(1)}\gamma +q_\infty^{(1)}}{p_\infty^{(0)}+2 \gamma} \ .
\end{equation}
For the SN equation
\begin{align}
p_\infty^{(0)} &= 0= p_\infty^{(1)} \ , \qquad q_\infty^{(0)} = \omega^2 \ , \qquad  q_\infty^{(1)} = 4 \hat{\omega}^2 \ 
,\\
\gamma^2 &+ \hat{\omega}^2 =0  \ ,\qquad \xi = -\frac{q_\infty^{(1)}}{2\gamma} = \pm 2i  \hat{\omega} \ .
\end{align}
Therefore, we have two series solutions
\begin{equation}
X_{\ell m \hat{\omega}}=\exp \{\pm i \hat{\omega} [\hat{r} +2 \log(\hat{r})]\} \displaystyle 
\sum_{n=0}^{\infty}\frac{b_n}{\hat{r}^n} \,.
\label{eq:BCInf}
\end{equation}
The general recursion relation for the coefficients $b_n$ is (we set again 
$b_0=1$)
\begin{align}
(p_\infty^{(0)} + 2 \gamma) n b_n = (n - \xi)(n-1 -\xi)b_{n-1} +  \nonumber\\
+\displaystyle \sum_{k=1}^{n}\Big[\gamma p_\infty^{(k+1)}+q_\infty^{(k+1)} - (n-k-\xi)p_\infty^{(k)}\Big]b_{n-k}\ .
\end{align}
It can be proved that the series solutions constructed in this way diverge, and 
they have to be considered as asymptotic expansions. 
However, these solutions are unique and linearly independent. 
We computed the series coefficients up to $n=13$.

\subsubsection{Cross check of the boundary conditions with Ref.~\cite{Gralla:2015rpa}} 
We compared our boundary conditions with the ones used in Ref.~\cite{Gralla:2015rpa}, which are in form
\begin{align}
&e^{\pm i \hat{\kappa} \hat{r}^{\ast}} \displaystyle \sum_{n=0}^{\infty} a^\text{H}_n (\hat{r}-\hat{r}_+)^n \,,\\
&e^{\pm i \hat{\omega} \hat{r}^{\ast}} \displaystyle \sum_{n=0}^{\infty}a^\infty_n\frac{1}{(\hat{\omega} \hat{r})^n} \,.
\end{align}
First, we notice that the tortoise coordinate $\hat{r}^{\ast}(\hat{r})$ at the boundaries can be written as 
\begin{align}
\hat{r}^{\ast}(\hat{r}) &\sim \hat{r} +  2\ln(\hat{r}) - 2 \ln(2) \,, \\
\hat{r}^{\ast}(\hat{r}) &\sim \frac{2 \hat{r}_+}{\hat{r}_+ - \hat{r}_-}\ln(\hat{r} - \hat{r}_+)  + \delta r^*(r_+) \,,
\end{align}
at $\hat{r} \to \infty$ and $\hat{r} \to \hat{r}_+ $, respectively, and where we defined
\begin{align}
\delta \hat{r}^*(\hat{r}_+) &\equiv  - 2 \ln(2)  - \frac{2\hat{r}_-}{\hat{r}_+ - \hat{r}_-}\ln(\hat{r}_+ -\hat{r}_-) + 
\hat{r}_+\,.
\end{align}

If we multiply Eq.~\eqref{eq:BCHor} by the phase factor $\exp\{\pm i \hat{\kappa}\delta \hat{r}^*(\hat{r}_+)\}$  and Eq.~\eqref{eq:BCInf} by $\exp\{\pm i \hat{\omega}[-  2 \ln(2)]\}$,
our boundary conditions have the same modulus and phase as those in Ref.~\cite{Gralla:2015rpa} for all the 
values of the parameters space we have considered, up to numerical error. In the worst case,  for $\hat{a} = 0.995$ 
and $\ell = 20 $ at the ISCO, the fractional difference in both modulus and phase is at most of one part in $10^{10}$, 
and typically much smaller.

Since the solutions by means of series expansion of an ordinary differential equation are uniquely determined a part for 
a constant complex factor, the boundary conditions~\eqref{eq:BCHor} 
and~\eqref{eq:BCInf} are consistent with the ones of Ref.~\cite{Gralla:2015rpa}.
\section{Teukolsky source term}\label{app:Teu source term}

\subsection{Spinning particle on a general bound orbit}
The source term of the Teukolsky equation reads
\begin{equation}
\mathcal {T} _ {\ell m \hat{\omega}} =  4 \!\int\! \!\dd \hat{t} \dd\theta\sin\theta \dd\phi \frac{\left(B' _ 2 + {B' _ 
2}^*\right)}{\bar{\rho}\rho^5} \! \nswsh e^{-  i (m\phi+ \hat{\omega} \hat{t})}\ ,
\end{equation}
where the functions $B' _ 2$ and $ {B' _ 2}^*$ are defined as
\begin{align}
B' _ 2 &= - \frac{1}{2} \rho^8\bar {\rho}\mathcal {L} _{-1} 
\bigg[\frac{1}{\rho^4}\mathcal{L}_0\bigg[\frac{T_{nn}}{\rho^2\bar{\rho}} \bigg]\bigg]+ \nonumber\\
& -\frac{1}{2\sqrt{2}}\Delta^2 \rho^8\bar{\rho}\mathcal{L}_ {-1}\bigg[\frac{\bar{\rho}^2}{\rho^4} 
J_+\bigg[\frac{T_{\overline{m}n}}{ \Delta \rho^2\bar{\rho}^2} \bigg]\bigg] \ , \\
 {B' _ 2}^*&= - \frac {1} {4}\Delta^2 \rho^8\bar{\rho} J_+\bigg[\frac{1}{\rho^4}J_+ 
\bigg[\frac{\bar{\rho}}{\rho^2}T_{\overline{m}\overline{m}}\bigg] \bigg] +\nonumber  \\
&- \frac{1}{2\sqrt {2}}\Delta^2 \rho^8\bar{\rho} J_+ \bigg[\frac{\bar{\rho}^2}{\Delta \rho^4}\mathcal {L}_ 
{-1}\bigg[\frac{ T_ {\overline{m}n}}{\rho^2\bar {\rho}^2}\bigg] \bigg] \ ,
\end{align}
with $J_+ = \frac{\partial}{\partial \hat{r}}+\frac{i K}{\Delta}$ and
\begin{alignat}{2}
\rho &= \frac {1} {\hat{r} - i \hat{a}\cos (\theta)} \ , \qquad  \quad \bar{\rho} = \frac {1}{\hat{r} + i \hat{a}\cos 
(\theta)}&\ ,\\
\mathcal{L} _s &= \frac{\partial}{\partial\theta}+\frac{m}{\sin (\theta)} - \hat{a}\hat{\omega} \sin (\theta) + s \cot 
(\theta) &\ ,\\
\mathcal{L} _s^\dagger &= \frac{\partial}{\partial\theta}-\frac{m}{\sin (\theta)} + \hat{a}\hat{\omega} \sin\theta + s 
\cot (\theta)&\ .
\end{alignat}
The components $T_{nn}, T_ {\overline{m}n}$, and $T_{\overline{m}\overline{m}}$ 
are the projections of the stress-energy tensor with respect to the 
Newman-Penrose~(NP) tetrad:
\begin{alignat}{2}
l^\mu &= \sqrt{\frac{\Sigma}{\Delta}}\left(e^{\mu}_{(0)} + e^{\mu}_{(1)} \right)\ , \quad &n^\mu = 
\frac{1}{2}\sqrt{\frac{\Delta}{\Sigma}}\left(e^{\mu}_{(0)} - e^{\mu}_{(1)} \right) \ ,\\
m^\mu &= \bar{\rho} \sqrt{\frac{\Sigma}{2}}\left(e^{\mu}_{(2)} + i e^{\mu}_{(3)} \right),\; & \overline{m}^\mu 
=\rho\sqrt{\frac{\Sigma}{2}}\left(e^{\mu}_{(2)} - i e^{\mu}_{(3)} \right)\ ,
\end{alignat}
where, for example, $T_{nn} = n^\mu n^\nu T_{\mu\nu}$~\cite{Mino:1995fm}.
Henceforth we use the notation $\rswsh$ instead of $\nswsh$ for the spin-weighted spheroidal harmonics to reduce 
clutter in the notation.

All $\theta$-derivatives in $T_ {nn},T_{\overline{m}n}$ and 
$T_{\overline{m}\overline{m}} $ can be removed by repeated integrations 
by parts and by making use of the following identity
\begin{equation}
\int_ 0^\pi \! h (\theta)\mathcal {L}_s[g (\theta)]\sin(\theta)\dd \theta = -\int_ 0^\pi\! g (\theta) \mathcal 
{L}^\dagger_s [h (\theta)]\sin(\theta)\dd\theta \ ,
\end{equation}
with $h(\theta)$ and $g(\theta)$ regular functions. 
It is thus possible to write
\begin{equation}
\mathcal {T} _{\ell m\hat{\omega}} = \int\!\!\dd t \dd \theta \dd\phi \Delta^2 e^{ i (\hat{\omega} \hat{t} -  
m\phi)}\big(\mathcal{T}_{nn}+ \mathcal{T} _ {\overline{m}n} + \mathcal {T} _ {\overline{m}\overline{m}} \big) \ 
,\label{eq:Teulmomega}
\end{equation}
with
\begin{equation}
\mathcal{T}_{nn} = -\frac{2}{\Delta^2 \rho^2\bar{\rho}}\mathcal{L}^\dagger_ 
1\bigg[\frac{1}{\rho^4}\mathcal{L}^\dagger_ 2\big[\rho^3 \rswsh \big]\bigg] \! \sin(\theta) T_{nn} \ ,
\end{equation}
\begin{align}
\mathcal{T}_ {\overline{m}n} &= 
\frac{4}{\sqrt{2}}\frac{\overline{\rho}}{\rho^2}\mathcal{L}^\dagger_2 \big[\rswsh \rho 
\overline{\rho}\big] J_+ \bigg[\frac{T_ {\overline{m}n}}{\Delta \rho^2 \overline{\rho}^2}\bigg] \!\sin(\theta) + 
\nonumber \\ 
& +\frac{2}{\sqrt{2}} \frac{1}{\rho^2\bar{\rho}^2\Delta}\mathcal{L}^\dagger_2\bigg[ \rho^3 
\rswsh\totder{}{\hat{r}}\bigg(\frac{\bar{\rho}^2}{ \rho^4}\bigg)\bigg] \!\sin(\theta) T_ {\overline{m}n} \ ,
\end{align}
\begin{equation}
\mathcal{T}_{\overline{m}\overline{m}} = -\rho^3 \rswsh J_+\bigg[\frac{1}{\rho^4}J_+\bigg[\frac{\bar{\rho}}{\rho^2} 
T_{\overline{m}\overline{m}} \bigg] \bigg] \!\sin(\theta) \ .
\end{equation}
It is convenient to expand the previous terms in order to isolate the derivatives of the 
projected stress-energy tensor with respect to $\hat{r}$ and the derivative of $\rswsh$ with respect to $\theta$. After 
some algebra, we get
\begin{equation}
\mathcal{T}_{nn} = -\frac{2\sin(\theta)}{\Delta^2 \rho^3\bar{\rho}}\Big[\Big(\mathcal{L}^\dagger_ 
1 -2 i \hat{a} \rho\sin(\theta)\Big) \mathcal{L}^\dagger_ 2\rswsh\Big]T_{nn} \ ,
\label{eq:Teunn1}
\end{equation}
\begin{align}
\mathcal{T}_ {\overline{m}n} &=\frac{4\sin(\theta)}{\sqrt{2}}\bigg\{ \partial_{\hat{r}} \bigg[ 
\Big(\mathcal{L}^\dagger_2\rswsh + i \hat{a} \sin\theta(\bar{\rho}- \rho) \rswsh \Big)\frac{T_ {\overline{m}n}}{\rho^3 
\Delta}\bigg]\nn\\
&+ \bigg[\bigg( \frac{i K}{\Delta} + \rho +\bar{\rho}\bigg)\mathcal{L}^\dagger_2\rswsh \nn\\
&- \hat{a} 
\sin(\theta)\frac{K}{\Delta}(\bar{\rho}-\rho) \rswsh \bigg] \frac{T_ {\overline{m}n}}{\rho^3 \Delta}\bigg\}  \ 
,\label{eq:Teumn1} \\
\mathcal{T}_{\overline{m}\overline{m}} &= \left\{ - \partial^2_{\hat{r}}\!\left( \frac{\bar{\rho}}{\rho^3} T_ 
{\overline{m}\overline{m}} \right)\! - 2\partial_{\hat{r}} \! \left( \left( \frac{\bar{\rho}}{\rho^2} 
+\frac{\bar{\rho}}{\rho^3}\frac{i K}{\Delta} \right) T_ {\overline{m}\overline{m}} \right) \right.\nn\\
&\left.+ 
\frac{\bar{\rho}}{\rho^3}\!\left( \totder{}{\hat{r}}\!\left( \frac{i K}{\Delta} \right) \ - 2 \rho \frac{ i K}{\Delta} + 
\frac{K^2}{\Delta^2} \right)\! T_{\overline{m}\overline{m}} \right\}\!\sin(\theta) \rswsh \ . \label{eq:Teumm1}
\end{align}
The stress-energy tensor for a spinning object is given by~\cite{Tanaka:1996ht}
\begin{align}
T^{\mu \nu} &= q\! \displaystyle\int\! \dd\hat{\lambda} \Bigg[\frac{\delta^{(4)}_{x ,z(\lambda )} }{\sqrt{-g}} u^{(\mu} 
v^{\nu)} - 
\nabla_ {\sigma}\Bigg( S^{\sigma (\mu} v^{\nu)}\frac{\delta^{(4)}_{x ,z(\lambda)} }{\sqrt {-g}}\Bigg) \Bigg] \ ,
\label{eq:spinenergytensor}
\end{align}
where $\delta^{(4)}_{x ,z(\lambda )} \equiv \prod_{\nu=0}^{4} \delta\big(x^\nu - z^\nu(\hat{\lambda} )\big)$ and 
indices within parenthesis denote symmetrization. The tetrad components are~\cite{Tanaka:1996ht}
\begin{align}
T^{(a)(b)} = q\!\int \!\!\frac{\dd\hat{\lambda} }{\sqrt {-g}} &\Big[u^{((a)} v^{(b))}\delta^{(4)}_{x ,z(\lambda)}  \nn\\
& - {e^{\mathlarger{(}(a)}}_\nu\, 
{e^{(b)\mathlarger{)}}}_\rho\nabla _\sigma\!\left (S^{\sigma \nu} v^{\rho}\delta^{(4)}_{x ,z(\lambda)}  \right)\! \Big] 
 \,.
\end{align}
The above equation can be written as
\begin{align}
 T^{(a)(b)}&= q\! \int \! \frac{\dd \hat{\lambda} }{\sqrt {-g}}\Big[\delta^{(4)}_{x ,z(\lambda)}\! \left(u^{\mathlarger{(}(a)} 
v^{(b)\mathlarger{)}} +\right.\nn\\
&\left. + {\omega_ {(d)(c)}}^{\mathlarger{(}(a)} v^{(b)\mathlarger{)}} S^{(d)(c)} -{\omega_ 
{(d)(c)}}^{\mathlarger{(}(a)} S^{(b)\mathlarger{)}(d)} v^{(c)} \right) + \nn \\
&-\partial_\sigma\!\left(S^{\mathlarger{(}(a)} 
v^{(b)\mathlarger{)}}\delta^{(4)}_{x ,z(\lambda)} \right)\! \Big]\ .
\end{align}
For bound orbits, it is 
useful to rewrite the energy-momentum tensor as
 \begin{align}
T^{(a)(b)} &= \frac{1}{\sqrt {-g}}\delta^{(3)}_{\vett{x}, \vett{x}(\hat{t})}\Big(\mathcal{P}^{(a)(b)}-\mathcal{S}^{t 
(a)(b)}\partial _{\hat{t}} \Big) + \nonumber \\
 &+ \frac{1}{\sqrt {-g}}\partial_i\Big(\mathcal {S}^{i 
(a)(b)}\delta^{(3)}_{\vett{x}, \vett{x}(t)}\Big) \ ,
\end{align}
where $i = \{r, \theta, \phi\}$, $\delta^{(3)}_{\vett{x}, \vett{x}(t)} = \delta\big(\hat{r} - 
\hat{r}(\hat{t})\big)\delta\big(\theta - 
\theta(t)\big)\delta\big(\phi - \phi(\hat{t})\big)$, and we defined
\begin{align}
\mathcal{P}^{(a)(b)}& \coloneqq q \modulo{\totder{\hat{t}}{\hat{\lambda}}}^{-1}\!\Big(u^{\mathlarger{(}(a)} 
v^{(b)\mathlarger{)}} + {\omega_ {(d)(c)}}^{\mathlarger{(}(a)} v^{(b)\mathlarger{)}} S^{(d)(c)} \nonumber\\
& - \omega_ {(d)(c)}{}^{\mathlarger{(}(a)} S^{(b)\mathlarger{)}(d)} v^{(c)}\Big) \ , \label{eq:P}\\
\mathcal{S}^{\sigma (a)(b)} &\coloneqq - q \modulo{\totder{\hat{t}}{\hat{\lambda}}}^{-1}S^{\sigma\mathlarger{(}(a)} 
v^{(b)\mathlarger{)}} \ .\label{eq:S}
\end{align}
To rewrite the stress-energy tensor we used the well-known property of the derivative of a Dirac delta: 
\begin{equation}
 \int_{-\infty}^{\infty} \! \dd x h(x) \totder{}{x} \delta (x-x_0) = -\left.\totder{h}{x} \right \rvert_{x=x_0}\,.
\end{equation}
In this way, the stress-energy tensor can be interpreted as a linear differential operator that acts on the smooth 
functions inside of the Teukolsky source term. 

We now need to project $T^{ab}$  with respect to the NP null 
tetrad. In the following, we will employ a reduced version of the NP tetrad: 
\begin{alignat}{2}
\tilde l^\mu &=\left(e^{\mu}_{(0)} + e^{\mu}_{(1)} \right)\,, \quad & \tilde n^\mu &= \frac{1}{2}\left(e^{\mu}_{(0)} - 
e^{\mu}_{(1)} 
\right)\,, \\
\tilde m^\mu &= \frac{1}{\sqrt{2}} \left(e^{\mu}_{(2)} + i e^{\mu}_{(3)} \right)\,, \quad & \tilde k^\mu &= 
\frac{1}{\sqrt{2}} \left(e^{\mu}_{(2)} - i e^{\mu}_{(3)} \right)\,,
\end{alignat}
where $\tilde k^\mu$ is the complex conjugate of $\tilde m^\mu$.
Taking into account that the $\hat{t}$ and $\phi$ coordinates in the Teukolsky source term are only present in the 
exponential, and using the definitions $T_{nn}=  n^\mu n^\nu e_{\mu (a)} 
e_{\nu(b)}T^{(a)(b)}$ and so on, the projected components read
\begin{align}
T_{nn} &= \delta^{(3)}_{\vett{x}, \vett{x}(t)}\mathcal{D}_{\tilde{n} \tilde{n}}\!\left[N_{nn} \,\cdot\right] +\partial_{\hat{r}}\Big(\mathcal 
{S}^r_{\tilde{n}\tilde{n}}\delta^{(3)}_{\vett{x}, \vett{x}(t)}\Big) N_{nn} \,, \label{eq:Tnn1}\\
T_{\overline{m}n} &=\delta^{(3)}_{\vett{x}, \vett{x}(t)}\mathcal{D}_{\tilde{k}\tilde{n}}\!\left[ N_{\overline{m}n} 
\,\cdot\right] + \partial_{\hat{r}}\Big(\mathcal{S}^r_{\tilde{k}\tilde{n}}\delta^{(3)}_{\vett{x}, \vett{x}(t)}\Big) 
N_{\overline{m}n} \label{eq:Tmn1} \ ,\\
T_{\overline{m}\overline{m}} &=\delta^{(3)}_{\vett{x}, 
\vett{x}(t)}\mathcal{D}_{\tilde{k}\tilde{k}}\!\left[N_{\overline{m}\overline{m}} \,\cdot\right]+ 
\partial_{\hat{r}}\Big(\mathcal{S}^r_{\tilde{k}\tilde{k}}\delta^{(3)}_{\vett{x}, \vett{x}(t)}\Big) N_{\overline{m} 
\overline{m}}\label{eq:Tmm1} \ ,
\end{align}
with
\begin{align}
N_{nn}&= \frac{\Delta}{\sqrt {-g}\Sigma}\,,\quad N_{\overline{m}n}=\frac{\sqrt{\Delta}\rho}{\sqrt {-g}}\,,  \quad    
N_{\overline{m} \overline{m}}=\frac{\Sigma \rho^2}{\sqrt {-g}}\,,
\end{align}
and where we define the following linear operators acting on a generic smooth function $h(\hat{r},\theta)$:
\begin{align}
\mathcal{D}_{\tilde{n} \tilde{n}}\!\left[N_{nn} h(\hat{r},\theta) \right] & \equiv \left(\mathcal{P}_{\tilde{n} \tilde{n}}- i \hat{\omega}\mathcal{S}^{t}_{\tilde{n} \tilde{n}} + \right.  \nn \\
&\left. + i m \mathcal{S}^{\phi}_{\tilde{n} \tilde{n}} - \mathcal{S}^{\theta}_{\tilde{n} \tilde{n}}\partial_\theta \right)\!\left(\frac{\Delta}{\sqrt{-g}\Sigma} h(\hat{r},\theta) \right) \ , \\
\mathcal{D}_{\tilde{k}\tilde{n}}\!\left[ N_{\tilde{k}\tilde{n}} h(\hat{r},\theta) \right] &\equiv \left(\mathcal{P}_{\tilde{k}\tilde{n}}- i \hat{\omega}\mathcal{S}^t_{\tilde{k}\tilde{n}} + \right. \nn \\
&\left. + i m \mathcal{S}^{\phi}_{\tilde{k}\tilde{n}} - \mathcal{S}^{\theta}_{\tilde{k}\tilde{n}}\partial_\theta \right)\!\left(\frac{\sqrt{\Delta}\rho}{\sqrt {-g}} h(\hat{r},\theta) 
\right) \ , \\
\mathcal{D}_{\tilde{k}\tilde{k}}\!\left[ N_{\overline{m} \overline{m}} h(\hat{r},\theta) \right] &\equiv  
\left(\mathcal{P}_{\tilde{k}\tilde{k}}- i \omega\mathcal{S}^{t}_{\tilde{k}\tilde{k}} \right. \nn \\
&\left. + i m \mathcal{S}^{\phi}_{\tilde{k}\tilde{k}} - \mathcal{S}^{\theta}_{\tilde{k}\tilde{k}}\partial_\theta 
\right)\!\left(\frac{\Sigma \rho^2}{\sqrt {-g}} h(\hat{r},\theta) \right) \ .
\end{align}
Using the relations~\eqref{eq:Tnn1},~\eqref{eq:Tmn1} and~\eqref{eq:Tmm1}, we can now rewrite the terms 
$\mathcal{T}_{nn},\mathcal{T}_ {\overline{m}n}$ and $ \mathcal{T}_ {\overline{m}\overline{m}} $, obtaining
\begin{equation}
\mathcal{T}_{nn} = \left[\delta^{(3)}_{\vett{x}, \vett{x}(t)}\mathcal{D}_{\tilde{n} \tilde{n}} + \partial_{\hat{r}}
\!\left(\mathcal {S}^{r}_{\tilde{n} \tilde{n}}\delta^{(3)}_{\vett{x}, \vett{x}(t)}\right)\right]\!f_{nn}^{(0)} \ ,\label{eq:Teunn2}
\end{equation}
\begin{equation}
f_{nn}^{(0)} \coloneqq -\frac{2}{\Delta}\frac{\bar{\rho}}{\rho}\Big(\mathcal{L}^\dagger_ 1  -2 i \hat{a}
\rho \sin(\theta)\Big)\mathcal{L}^\dagger_ 2\rswsh \ , \label{eq:fnn0} 
\end{equation}
\begin{align}
&\mathcal{T}_{\overline{m}n} = \Big[\delta^{(3)}_{\vett{x}, \vett{x}(t)}\mathcal{D}_{\tilde{k}\tilde{n}}
+ \partial_{\hat{r}} \big(\mathcal{S}^r_{\tilde{k}\tilde{n}}\delta^{(3)}_{\vett{x}, \vett{x}(t)}\big) \Big]
f_{\overline{m}n}^{(0)} +\nonumber \\
&+ \partial_r \!\left[\Big(\delta^{(3)}_{\vett{x}, \vett{x}(t)}\mathcal{D}_{\tilde{k}\tilde{n}} + 
\partial_r \big(\mathcal{S}^r_{\tilde{k}\tilde{n}}\delta^{(3)}_{\vett{x}, \vett{x}(t)}\big)\Big) f_{\overline{m}n}^{(1)} 
\right] \ , \label{eq:Teumn2}
\end{align}
\begin{align}
f_{\overline{m}n}^{(0)} \coloneqq \frac{4}{\sqrt{2}} \frac{\bar{\rho}}{\rho\sqrt{\Delta}}\! &\left( 
 \left(\frac{i K}{\Delta} + \rho +\bar{\rho}\right) \mathcal{L}^\dagger_2\rswsh
\right.\nn\\
&\left. -  \hat{a}
\sin\theta\frac{K}{\Delta}(\bar{\rho}-\rho) \rswsh \right) \ ,\label{eq:fmn0}
\end{align}
\begin{align}
f_{\overline{m}n}^{(1)} &\coloneqq \frac{4}{\sqrt{2}} \frac{\bar{\rho}}{\rho 
\sqrt{\Delta}}\!\left(\mathcal{L}^\dagger_2\rswsh + i 	\hat{a} \sin(\theta)(\bar{\rho}- \rho) \right) \ ,
\label{eq:fmn1}
\end{align}
\begin{align}
&\mathcal{T}_{\overline{m}\overline{m}} =  \Big[\!\delta^{(3)}_{\vett{x}, 
\vett{x}(t)}\mathcal{D}_{\tilde{k}\tilde{k}} +\partial_{\hat{r}} 
\big(\mathcal{S}^r_{\tilde{k}\tilde{k}}\delta^{(3)}_{\vett{x}, 
\vett{x}(t)}\big)\!\Big]f_{\overline{m}\overline{m}}^{(0)} +\nonumber \\
&+ \partial_{\hat{r}}\!\left[ \Big( \delta_{r,r(t)}\mathcal{D}_{\tilde{k}\tilde{k}} + \partial_{\hat{r}} 
\big(\mathcal {S}^r_{\tilde{k}\tilde{k}}\big)\Big)f_{\overline{m}\overline{m}}^{(1)}\,\right]+  
\nonumber\\
&+ \partial_{\hat{r}}^2\!\left[\Big( \delta_{r,r(t)}\mathcal{D}_{\tilde{k}\tilde{k}}  +  
\partial_{\hat{r}}\big(\mathcal{S}^r_{\tilde{k}\tilde{k}}\delta_{r, r(t)}\big)\Big) 
f_{\overline{m}\overline{m}}^{(2)}\right] \ ,\label{eq:Teumm2}
\end{align}
\begin{align}
f_{\overline{m}\overline{m}}^{(0)} &\coloneqq \frac{\bar{\rho}}{\rho}\left( \totder{}{\hat{r}}\!\left(\frac{i K}{\Delta} 
\right) \ - 2 \rho \frac{ i K}{\Delta} + \frac{K^2}{\Delta^2} \right) \rswsh \ ,\label{eq:fmm0} \\
f_{\overline{m}\overline{m}}^{(1)} &\coloneqq - 2\frac{\bar{\rho}}{\rho}\left(\rho + \frac{i K}{\Delta}\right)\rswsh \ ,
\label{eq:fmm1}\\
f_{\overline{m}\overline{m}}^{(2)} &\coloneqq -\frac{\bar{\rho}}{\rho} \rswsh \ . \label{eq:fmm2}
\end{align}

We now have all the necessary ingredients to rewrite the inhomogeneous solutions of the Teukolsky equation in a form 
suitable to exploit the possible quasi-periodicities in the bound orbits. First of all, by plugging the 
terms~\eqref{eq:Teunn2},~\eqref{eq:Teumn2} and~\eqref{eq:Teumm2} into Eq.~\eqref{eq:Teulmomega}, integrating over the 
angles and using the $\delta(\theta-\theta(\hat{t}))\delta(\phi-\phi(\hat{t}))$ function, the Teukolsky source 
term becomes
\begin{align}
\mathcal{T}_{\ell m \hat{\omega}} &=\! \int\limits_{-\infty}^{\infty}\! \dd \hat{t}\, e^{i(\hat{\omega} \hat{t}- 
m\phi(\hat{t}))}\Delta^2\Big\{\mathcal{T}^{(0)}_{\mathcal{D}}\delta_{r,r(t)} + \nn\\
& + \partial_{\hat{r}}\!\left(\mathcal{T}^{(0)}_{\mathcal{D}}\delta_{r,r(t)}\right) 
+\partial_{\hat{r}}^2\!\left(\mathcal{T}^{(0)}_{\mathcal{D}}\delta_{r,r(t)}\right)+ \nn \\
& \left.+\mathcal{T}^{(0)}_{\mathcal{S}^r} + \partial_{\hat{r}}\mathcal{T}^{(1)}_{\mathcal{S}^r} +\partial_{\hat{r}}^2 \mathcal{T}^{(2)}_{\mathcal{S}^r} \Big\} 
\right\rvert_{\theta = \theta(\hat{t})} \ ,
\end{align}
when $\delta_{r,r(t)} \coloneqq \delta(\hat{r}-\hat{r}(\hat{t}))$, and we have rearranged the previous terms, defining
\begin{align}
\mathcal{T}^{(0)}_{\mathcal{D}} &= \mathcal{D}_{\tilde{n}\tilde{n}}f_{nn}^{(0)} +\mathcal{D}_{\tilde{k}\tilde{n}}f_{\overline{m}n}^{(0)} +\mathcal{D}_{\tilde{k} \tilde{k}} f_{{\overline{m} 
\overline{m}}}^{(0)} \ , \\
\mathcal{T}^{(1)}_{\mathcal{D}} &= \mathcal{D}_{\tilde{k}\tilde{n}}f_{\overline{m}n}^{(1)} 
+\mathcal{D}_{\tilde{k} \tilde{k}}f_{{\overline{m} \overline{m}}}^{(1)} \ , \\
\mathcal{T}^{(2)}_{\mathcal{D}} &=\mathcal{D}_{\tilde{k} \tilde{k}}f_{\overline{m}\overline{m}}^{(2)} \ ,
\end{align}
and
\begin{align}
 \mathcal{T}^{(0)}_{\mathcal{S}^r} &= \partial_{\hat{r}}\big[\mathcal{S}^r_{\tilde{n} \tilde{n}}\delta_{r,r(t)}\big]f_{n n}^{(0)} +  
\partial_{\hat{r}} \big[\mathcal{S}^r_{\tilde{k} \tilde{n}}\delta_{r,r(t)}\big]f_{\overline{m}n}^{(0)} + \nn
 \\
&+ \partial_{\hat{r}} \big[\mathcal{S}^r_{\tilde{k} \tilde{k}}\delta_{r,r(t)}\big]f_{\overline{m}\overline{m}}^{(0)} \ ,
\\
 \mathcal{T}^{(1)}_{\mathcal{S}^r} &= \partial_{\hat{r}} \big[\mathcal{S}^r_{\tilde{k} 
\tilde{n}}\delta_{r,r(t)}\big]f_{\overline{m}n}^{(1)} + \partial_{\hat{r}} \big[\mathcal{S}^r_{\tilde{k} 
\tilde{k}}\delta_{r,r(t)}\big]f_{\overline{m}\overline{m}}^{(1)} \ , \\
 \mathcal{T}^{(2)}_{\mathcal{S}^r} &= \partial_{\hat{r}} \big[\mathcal{S}^r_{\tilde{k} 
\tilde{k}}\delta_{r,r(t)}\big]f_{\overline{m}\overline{m}}^{(2)} \ .
\end{align}
To obtain the asymptotic fluxes, we need to calculate the amplitudes~\eqref{eq:infamp},~\eqref{eq:horamp}, namely
\begin{equation}
Z^{H,\infty}_{\ell m \hat{\omega}} = C^{H,\infty}_{\ell m \hat{\omega}}\int_{\hat{r}_+}^\infty\! \dd
\hat{r}'\frac{R^{\textup{in},\textup{up}}_{\ell m \hat{\omega}}(\hat{r}')}{\Delta^2} \mathcal{T}_{\ell m 
\hat{\omega}}(\hat{r}') \,.
\end{equation}
By changing the order of integration between $\hat{r}'$ and $\hat{t}$, we get
\begin{align}
&Z^{H,\infty}_{\ell m \hat{\omega}} = C^{H,\infty}_{\ell m \hat{\omega}}\int\limits_{-\infty}^{\infty} \left[ \left( 
\mathcal{T}^{(0)}_{\mathcal{D}} -\mathcal{T}^{(1)}_{\mathcal{D}}\totder{}{\hat{r}} +\mathcal{T}^{(2)}_{\mathcal{D}} 
\frac{\dd^2}{\dd \hat{r}^2}  \right) R^{\textup{in},\textup{up}}_{\ell m\hat{\omega}} \right. \nn\\
&+ \int\limits_{\hat{r}_+}^{\infty}\!\dd \hat{r} \! \left. \left( 
\mathcal{T}^{(0)}_{\mathcal{S}^r} +\partial_{\hat{r}}\mathcal{T}^{(1)}_{\mathcal{S}^r} +\partial_{\hat{r}}^2 
\mathcal{T}^{(2)}_{\mathcal{S}^r} \right) R^{\textup{in},\textup{up}}_{\ell m\hat{\omega}} \right. \Bigg]\!  
e^{i(\hat{\omega} \hat{t}- m\phi(\hat{t}))} \dd \hat{t} \ ,
\end{align}
which is calculated at $\theta=\theta(\hat{t})$. In the integral on the first line we have used the 
$\delta(\hat{r}-\hat{r}(\hat{t})$) function. The double integral on the second line can be simplified with multiple 
integrations by parts, obtaining the general expression 
\begin{align}
Z^{H,\infty}_{\ell m \hat{\omega}}&=C^{H,\infty}_{\ell m \hat{\omega}} \int\limits_{-\infty}^{\infty} \! \dd \hat{t} 
e^{i(\hat{\omega} \hat{t} - 
m\phi(\hat{t}))}\!\left. \left(A_0  - (A_1 + B_1) \totder{}{\hat{r}} \right.\right.\nn\\
& \left.\left.+ (A_2+B_2) \frac{\dd^2}{\dd \hat{r}^2}  - B_3 \frac{\dd^3}{\dd 
\hat{r}^3} 
\right)R^{\textup{in},\textup{up}}_{\ell m\hat{\omega}}  \right\rvert_{\theta = \theta(\hat{t}) , \hat{r} = \hat{r}(\hat{t})} \label{eq:genTeuamp}
\end{align}
where
\begin{align}
A_0 &\coloneqq O_{\tilde{n} \tilde{n}}f_{n n}^{(0)} + O_{\tilde{k}\tilde{n}}f_{\overline{m}n}^{(0)} + 
O_{\tilde{k}\tilde{k}}f_{\overline{m}\overline{m}}^{(0)} \ ,\label{eq:A0} \\
A_1 & \coloneqq O_{\tilde{k} \tilde{n}}f_{\overline{m}n}^{(1)} + 
O_{\tilde{k}\tilde{k}}f_{\overline{m}\overline{m}}^{(1)} \ , \label{eq:A1}\\
A_2 &\coloneqq O_{\tilde{k} \tilde{k}} f_{\overline{m}\overline{m}}^{(2)} \ ,\label{eq:A2} 
\end{align}
and
\begin{align}
B_1 &\coloneqq \mathcal{S}^r_{\tilde{n} \tilde{n}} f_{n n}^{(0)} + \mathcal{S}^r_{\tilde{k} \tilde{n}} f_{\overline{m}n}^{(0)} + 
\mathcal{S}^r_{\tilde{k} \tilde{k}}f_{\overline{m}\overline{m}}^{(0)}\ ,  \label{eq:B1}\\
B_2 & \coloneqq\mathcal{S}^r_{\tilde{k} \tilde{n}} f_{\overline{m}n}^{(1)} + \mathcal{S}^r_{\tilde{k} \tilde{k}} 
f_{\overline{m}\overline{m}}^{(1)} \ , \label{eq:B2}\\
B_3 &\coloneqq \mathcal{S}^r_{\tilde{k} \tilde{k}} f_{\overline{m}\overline{m}}^{(2)} \ ,\label{eq:B3}
\end{align}
with the operators $O_{\tilde{n} \tilde{n}}, O_{\tilde{k} \tilde{n}} , O_{\tilde{k} \tilde{k}}$ 
being defined as
\begin{align}
O_{\tilde{n} \tilde{n}}&\coloneqq \mathcal{P}_{\tilde{n} \tilde{n}} - i \hat{\omega}\mathcal{S}^t_{\tilde{n} \tilde{n}} +  i m \mathcal{S}^{\phi}_{\tilde{n} \tilde{n}}  - \mathcal{S}^{\theta}_{\tilde{n} \tilde{n}}\partial_\theta -\mathcal{S}^{r}_{\tilde{n} \tilde{n}}\partial_{\hat{r}} \ ,
\label{eq:Onn}\\
O_{\tilde{k} \tilde{n}}&\coloneqq \mathcal{P}_{\tilde{k} \tilde{n}} - i 
\hat{\omega} \mathcal{S}^t_{\tilde{k} \tilde{n}} +  i m \mathcal{S}^{\phi}_{\tilde{k} \tilde{n}} - 
\mathcal{S}^{\theta}_{\tilde{k} \tilde{n}}\partial_\theta -\mathcal{S}^{r}_{\tilde{k} \tilde{n}}\partial_{\hat{r}} \ ,
\label{eq:Okn}\\
O_{\tilde{k} \tilde{k}} &\coloneqq \mathcal{P}_{\tilde{k} \tilde{k}} - i 
\hat{\omega}\mathcal{S}^t_{\tilde{k} \tilde{k}} +  i m \mathcal{S}^{\phi}_{\tilde{k} \tilde{k}} - \mathcal{S}^{\theta}_{\tilde{k} \tilde{k}}\partial_\theta -\mathcal{S}^{r}_{\tilde{k} \tilde{k}}\partial_{\hat{r}} \ , \label{eq:Okk}
\end{align}
and $\mathcal{P}_{\tilde{n}\tilde{n}}=  \tilde{n}^\mu \tilde{n}^\nu e_{\mu (a)} 
e_{\nu(b)}\mathcal{P}^{(a)(b)}$, while $\mathcal{S}^\sigma_{\tilde{n}\tilde{n}}=  \tilde{n}^\mu \tilde{n}^\nu e_{\mu 
(a)} e_{\nu(b)}\mathcal{S}^{\sigma(a)(b)}$ and so on.
The terms $f_{n n}^{(i)} $, $f_{\overline{m} n}^{(i)}$, $f_{\overline{m} \overline{m}}^{(i)}$ (with $i=0,1,2$) are 
defined in Eqs.~~\eqref{eq:fnn0}--\eqref{eq:fmm2}.

We remark that Eq.~\eqref{eq:genTeuamp} is general: it 
is valid for any bound orbit for a spinning test particle in Kerr spacetime.

\subsection{Circular equatorial orbits} \label{app:Source term CEO}
On the equatorial plane, $\theta = \pi/2$, the Teukolsky source term drastically simplifies. First of all, some terms 
of the previous equations vanish, namely
\begin{equation}
\mathcal{S}^\theta_{\tilde{n} \tilde{n}} = \mathcal{S}^\theta_{\tilde{k} \tilde{n}} = 
\mathcal{S}^\theta_{\tilde{k}\tilde{k}} =0 \,,
\end{equation}
for $\theta = \pi/2$. Furthermore, we can write
\begin{align}
f_{nn}^{(0)} &= -4 \frac{\hat{S}(r)}{\Delta}\,,\\
f_{\overline{m}n}^{(0)} &= \frac{4}{\sqrt{2}} \frac{\tilde{S}}{\sqrt{\Delta}}\left(\frac{i K}{\Delta} + 
\frac{2}{\hat{r}}\right)\,, \\
f_{\overline{m}n}^{(1)} &= \frac{4}{\sqrt{2}}\frac{\tilde{S}}{\sqrt{\Delta}}\,,
\end{align}
where we applied the angular Teukolsky equation, with
\begin{align}
\tilde{S} &\coloneqq \left.\totder{\rswsh}{\theta}\right \rvert_{\theta = \pi/2} + (\hat{a}\hat{\omega} - m) 
\rswsh(\pi/2)\,, \\
\hat{S}(\hat{r}) &\coloneqq \left(\hat{a} \hat{\omega} - m - i \frac{\hat{a}}{\hat{r}}\right) \tilde{S} - 
\frac{\lambda_{\ell \hat{\omega}m}}{2}\rswsh(\pi/2)\,.
\end{align}
Moreover
\begin{align}
f_{\overline{m}\overline{m}}^{(0)} &=\left( \totder{}{\hat{r}}\!\left(\frac{i K}{\Delta} \right) - \frac{2}{\hat{r}} \frac{i 
K}{\Delta} + \frac{K^2}{\Delta^2} \right) \rswsh(\pi/2) \,, \\
f_{\overline{m}\overline{m}}^{(1)} &=- 2\left(\frac{1}{\hat{r}} + \frac{i K}{\Delta}\right)\rswsh(\pi/2) \,,\\
f_{\overline{m}\overline{m}}^{(2)} &= -\rswsh(\pi/2) \,.
\end{align}

Finally, for a circular equatorial orbit the projected components of $\mathcal{P}^{(a)(b)}$ and 
$\mathcal{S}^{\sigma(a)(b)}$ onto the reduced 
NP basis are
\begin{widetext}
\begin{align}
\mathcal{P}_{\tilde{n}\tilde{n}} &= -\frac{q}{4}\frac{ P_\sigma}{\Sigma_\sigma \Gamma_+}\! \left( \left(\hat{r}^3 + 2\sigma^2 
\right)\!\Delta\hat{x} \sigma - \hat{r} \Sigma_\sigma \! \left[2\hat{x} \sigma (\hat{r} -\hat{a}^2)  + P_\sigma 
(\hat{r}^2 - \sigma\hat{a}) \right] \right) \,,\\
\mathcal{P}_{\tilde{k}\tilde{n}} &= -\frac{ i q}{4 \sqrt{2}}\frac{\sqrt{\Delta}}{\Sigma_\sigma\Gamma_+} \!\left( -\hat{x} 
(\hat{r}^3 + 2 \sigma^2)\!\left[ \hat{x}\sigma (\hat{r} -\hat{a}^2) + P_\sigma(\hat{r}^2 + \hat{a} \sigma)\right]\! - 
\hat{r} P_\sigma \Sigma_\sigma \!\left[\hat{r}^2 \hat{x} +  \sigma(3 \hat{x} \hat{a} + P_\sigma) \right] 
\right)\,,\\
\mathcal{P}_{\tilde{k}\tilde{k}} &= \frac{q}{2}\frac{1}{\Sigma_\sigma \Gamma_-} \!\left\{ \hat{x} \Delta \left[ 
\sigma(P_\sigma + 2 \hat{x}\hat{a}) + \hat{x}\hat{r}^2 \right](\hat{r}^3+2\sigma^2) + \hat{a}\sigma 
\hat{r}\Sigma_\sigma  P_\sigma^2  \right\}\,,
\end{align}
with $\hat{x} \coloneqq \hat{J}_z- (\hat{a} + \sigma) \hat{E}$, $\Gamma_\pm \coloneqq 3 \hat{x} 
\hat{a}\sigma^2 \Delta  \pm \hat{r} \Sigma_\sigma \!\left[ P_\sigma (\hat{r}^2 
+\hat{a}^2) + \hat{x} \hat{a} \Delta \right]$, and 
\begin{align}
\mathcal{S}^\nu_{\tilde{n}\tilde{n}} &= \frac{1}{4}q \sigma\hat{r}^2 P_\sigma \!\left( \frac{\hat{a} P_\sigma + \hat{x} 
(\hat{r}^2+\hat{a}^2)}{\Gamma_+} , \; -\frac{\Delta \hat{x}}{\Gamma_-} , \; 0 , \; -\frac{\hat{a}\hat{x}+ P_\sigma 
}{\Gamma_-} \right) \,, \\
\mathcal{S}^\nu_{\tilde{k}\tilde{n}} &= \frac{i q 
\sigma}{4\sqrt{2}}\frac{\hat{r}\hat{x}\sqrt{\Delta}}{\Sigma_\sigma\Gamma_+} \Big((\hat{r}^3 + 2 
\sigma^2)\!\left[\hat{a} P_\sigma + \hat{x} (\hat{r}^2+\hat{a}^2) \right]\! ,\;\hat{x} \Delta (\hat{r}^3 +2 \sigma^2) + 
\frac{\hat{r}}{\hat{x}} \Sigma_\sigma P_\sigma^2 ,\; 0 ,\; (\hat{a}\hat{x}+ P_\sigma)(\hat{r}^3+2 \sigma^2) 
\Big) \,,\\
\mathcal{S}^\nu_{\tilde{k}\tilde{k}}  &= \frac{1}{2}q \sigma \frac{\hat{r} P_\sigma}{\Sigma_\sigma \Gamma_+} 
\!\left(0,\; \Delta \hat{x} (\hat{r}^3 + 2 \sigma^2),\; 0 ,\; 0\right)\,.
\end{align}
\end{widetext}
In Ref.~\cite{Tanaka:1996ht} the Teukolsky source was calculated at first order in the spin. Our results for the source 
term are general and, when truncated at ${\cal O}(\sigma)$, agree with those in Ref.~\cite{Tanaka:1996ht}, except for a 
factor $1/\sqrt{2}$ in their $\tilde{Z}^{\bar{m}\bar{m}}_{lm\omega}$ term. This is probably a typo in their source 
term, since with our source term we can reproduce previous results for the fluxes of a nonspinning particle (see also 
Appendix~\ref{app:comparison}).

\section{Comparisons of the GW fluxes with previous work}\label{app:comparison}

We have tested our code by comparing the GW fluxes against results already published in the
literature. In this section we provide a detailed comparison in order to assess the 
accuracy of our method. 

\subsection{Comparison with Harms {\it et al.}}

The GW fluxes at infinity for a spinning particle have been calculated in Ref.~\cite{Harms:2015ixa} by 
solving the Teukolsky equation in the time domain and assuming $q=1$, so that $\sigma=q\chi$ is not 
small when $\chi={\cal O}(1)$. To make the comparison, we also set $q=1$. We remark that we use the same spin 
supplementary conditions and the same orbital dynamics as in Ref.~\cite{Harms:2015ixa}.

Tables~\ref{tab:newtonfluxa0}--\ref{tab:newtonfluxa09pro} show the relative percentage 
difference between our results and those listed in Table II, III, and IV of Ref.~\cite{Harms:2015ixa} 
for the $\ell=2,3$ modes. The fluxes are normalized with respect to the leading  
Post-Newtonian order. Here the normalized fluxes are denoted as follows:
\begin{equation}
\mathcal{\hat{F}}^\infty_{\ell m} =\mathcal{F}^\infty_{\ell m} / k_{\ell m}\ ,\label{fractionalerror}
\end{equation}
where 
\begin{equation}
k_{22}=\frac{32}{5}\lvert\widehat{\Omega}\rvert^\frac{10}{3}\ ,\ 
k_{21}=\frac{8}{45}\lvert\widehat{\Omega}\rvert^\frac{12}{3}\ ,\ 
k_{33}=\frac{243}{28}\lvert\widehat{\Omega}\rvert^\frac{12}{3}
\end{equation}
and $\mathcal{F}^\infty_{\ell m}$ includes only the fluxes at infinity, assuming $q=1$, and therefore 
$\sigma = \chi$.  Moreover, we define
\begin{equation}
\Delta_{\ell m}= 100\modulo{1- \mathcal{\hat{F}}^\infty_{\ell m}/\hat{F}_{S\ell m}}\,,
\end{equation}
where $ \hat{F}_{S\ell m}$ given in~\cite{Harms:2015ixa}.
Note that Ref.~\cite{Harms:2015ixa} assumed $\hat{J}_z >0$, distinguishing prograde and retrograde 
orbits on the  base of the sign of $\hat{a}$. In our work we consider the opposite 
convention: we fix $\hat{a}\geq 0$, while $\hat{J}_z$ is positive (negative) for corotating (counter-rotating) orbits. 
Therefore, for retrograde orbits we compare our fluxes for $\sigma>0$ with the 
results $\sigma<0$ of Ref.~\cite{Harms:2015ixa} and vice versa. 

Tables~\ref{tab:newtonfluxa0}-\ref{tab:newtonfluxa09pro} show that our results are in good agreement
with those of Ref.~\cite{Harms:2015ixa}, with relative errors of the order of the percent or below for all the 
considered configurations. For the $\ell=m=2$ and $\ell=m=3$ modes the fractional difference is always less than 
$0.5\%$. 

This picture does not change for $\Delta_{21}$ except for fast spinning bodies with $\hat{a}=0.9$: 
in this case retrograde and prograde orbits lead to maximum discrepancies of $1.3\%$ and $16\%$, 
respectively. We believe that the last value may be given by numerical rounding, since the
corresponding flux is given in Ref.~\cite{Harms:2015ixa} with only one significant figure.

\begin{table}[!htpb]
\begin{equation*}
\begin{array}{*{8}{c}}
\multicolumn{8}{c}{\hat{a}=0} \\
\hline
\hline
\hat{r}  & \sigma &\mathcal{\hat{F}}^\infty_{22} &\Delta_{22}[\%]  &\mathcal{\hat{F}}^\infty_{21} &\Delta_{21}[\%]  
&\mathcal{\hat{F}}^\infty_{33} &\Delta_{33}[\%]   \\
\hline
4  &-0.9 &  2.2135   & 0.2 &  2.1607  & 0.4 &  2.4238  & 0.3 \\
    &-0.5 &  1.7954   & 0.2 &  2.3052  & 0.4 &  1.8302  & 0.3 \\ 
    & 0.5 &  1.0422   & 0.3 &  2.1033  & 0.5 &  0.8709  & 0.4 \\ 
    & 0.9 &  0.8538   & 0.3 &  2.0157  & 0.5 &  0.6549  & 0.4 \\
\hline
5  &-0.9 &  1.2143   & 0.2 &  0.9541  & 0.5 &  1.2187  & 0.3 \\
    &-0.5 &  1.1143   & 0.2 &  1.2514  & 0.5 &  1.0605  & 0.3 \\ 
    & 0.5 &  0.8703   & 0.2 &  1.7777  & 0.5 &  0.7181  & 0.3 \\ 
    & 0.9 &  0.7849   & 0.2 &  1.9312  & 0.5 &  0.6110  & 0.4 \\
\hline
6  &-0.9 &  1.0137   & 0.2 &  0.7042  & 0.5 &  0.9780  & 0.3 \\
    &-0.5 &  0.9610   & 0.2 &  0.9837  & 0.5 &  0.8881  & 0.3 \\ 
    & 0.5 &  0.8249   & 0.2 &  1.6424  & 0.5 &  0.6837  & 0.3 \\ 
    & 0.9 &  0.7727   & 0.2 &  1.8835  & 0.5 &  0.6132  & 0.3 \\
\hline
8  &-0.9 &  0.9042   & 0.2 &  0.5629  & 0.5 &  0.8430  & 0.3 \\
    &-0.5 &  0.8778   & 0.2 &  0.8124  & 0.5 &  0.7955  & 0.3 \\ 
    & 0.5 &  0.8093   & 0.2 &  1.5136  & 0.5 &  0.6837  & 0.3 \\ 
    & 0.9 &  0.7818   & 0.2  &  1.8115 & 0.5 &  0.6424  & 0.3 \\
\hline
10 &-0.9 &  0.8779  & 0.2 &  0.5292  & 0.5 &  0.8110  & 0.3 \\
    &-0.5 &  0.8608   & 0.2 &  0.7602  & 0.5 &  0.7792  & 0.3 \\ 
    & 0.5 &  0.8166   & 0.2 &  1.4464  & 0.5 &  0.7030  & 0.3 \\ 
    & 0.9 &  0.7987   & 0.2 &  1.7537  & 0.5 &  0.6741  & 0.3\\
\hline
20 &-0.9 &  0.8875  & 0.2 &  0.5560  & 0.4 &  0.8290  & 0.3\\
    &-0.5 &  0.8820   & 0.2 &  0.7426  & 0.4 &  0.8179  & 0.3 \\ 
    & 0.5 &  0.8680   & 0.2 &  1.3100  & 0.4 &  0.7907  & 0.3 \\ 
    & 0.9 &  0.8623   & 0.2 &  1.5745  & 0.4 &  0.7799  & 0.3 \\
    \hline
    \hline
\bottomrule
\end{array}
\end{equation*}
\caption{Normalized fluxes and fractional differences [Eq.~\eqref{fractionalerror}] between our results and 
those obtained in Table~II of~Ref.~\cite{Harms:2015ixa} for $\hat{a}=0$, and different values of $\hat{r}$. Note that 
we set $q=1$ to agree with Ref.~\cite{Harms:2015ixa}.}
\label{tab:newtonfluxa0}
\end{table}

\begin{table}[!htpb]
\begin{equation*}
\begin{array}{*{8}{c}}
\multicolumn{8}{c}{\hat{a}=0.9 \quad \text{retrograde orbits}}\\
\hline
\hline
\hat{r}  & \sigma &\mathcal{\hat{F}}^\infty_{22} &\Delta_{22}[\%]  &\mathcal{\hat{F}}^\infty_{21} &\Delta_{21}[\%]  
&\mathcal{\hat{F}}^\infty_{33} &\Delta_{33}[\%]   \\
\hline
5  &-0.9 &  1.2361  & 0.2 &  5.6616  & 0.4 & 1.0827 & 0.3  \\
    &-0.5 &  1.6251  & 0.2 &  6.6959  & 0.4 & 1.5729 & 0.3  \\ 
    & 0.5 &  3.3150  & 0.2 &  10.789    & 0.3 &  3.9783 & 0.3  \\ 
    & 0.9 &  4.4462  & 0.2 &  13.255    & 0.3 &  5.7567 & 0.3  \\
 \hline   
6  &-0.9 &  1.0335  & 0.2 &  4.6842  & 0.4 &  0.8937 & 0.3  \\
    &-0.5 &  1.2023  & 0.2 &  4.8148 & 0.4 &  1.1143 & 0.3  \\ 
    & 0.5 &  1.7181  & 0.2 &  4.8963  & 0.4 &  1.8635 & 0.3  \\ 
    & 0.9 &  1.9563  & 0.2 &  4.7277  & 0.3 &  2.2404 & 0.3 \\
\hline
8  &-0.9 &  0.9123  & 0.2 &  3.7900  & 0.4 &  0.7911 & 0.3  \\
    &-0.5 &  0.9784  & 0.2 &  3.5167  & 0.4 &  0.8842 & 0.3 \\ 
    & 0.5 &  1.1510  & 0.2 &  2.6978  & 0.4 &  1.1499 & 0.3  \\ 
    & 0.9 &  1.2208  & 0.2 &  2.3159  & 0.3 &  1.2679 & 0.3  \\
\hline
10 &-0.9 &  0.8816 & 0.2 &  3.3399  & 0.4 &  0.7727 & 0.3  \\
    &-0.5 &  0.9193  & 0.2 &  2.9873  & 0.4 &  0.8286 & 0.3  \\ 
    & 0.5 &  1.0142  & 0.2 &  2.0862  & 0.4 &  0.9799 & 0.3   \\ 
    & 0.9 &  1.0519  & 0.2 &  1.7269  & 0.3 &  1.0446  & 0.3  \\
\hline
20 &-0.9 &  0.8875  & 0.3 &  2.4826  & 0.7 &  0.8130 & 1.2  \\
    &-0.5 &  0.8969   & 0.1 &  2.1581 & 0.6  &  0.8290 & 0.4  \\ 
    & 0.5 &  0.9202   & 0.2 &  1.4249  & 0.3 &  0.8699  & 0.0  \\ 
    & 0.9 &  0.9294   & 0.2 &  1.1662  & 1.3 &  0.8866  & 0.3  \\
    \hline
    \hline
\bottomrule
\end{array}
\end{equation*}
\caption{Normalized fluxes and fractional differences with the fluxes in Table~III of Ref.~\cite{Harms:2015ixa} in the 
case 
$\hat{a}=0.9$, retrogade orbits. The fluxes $\mathcal{\hat{F}}^\infty_{\ell m}$ with $\sigma <0$ have to be compared with 
the fluxes $ \hat{F}_{S\ell m}$ with $\sigma>0$ and vice versa.}
\label{tab:newtonfluxa09ret}
\end{table}

\begin{figure}[!htpb]
\includegraphics[width=0.48\textwidth]{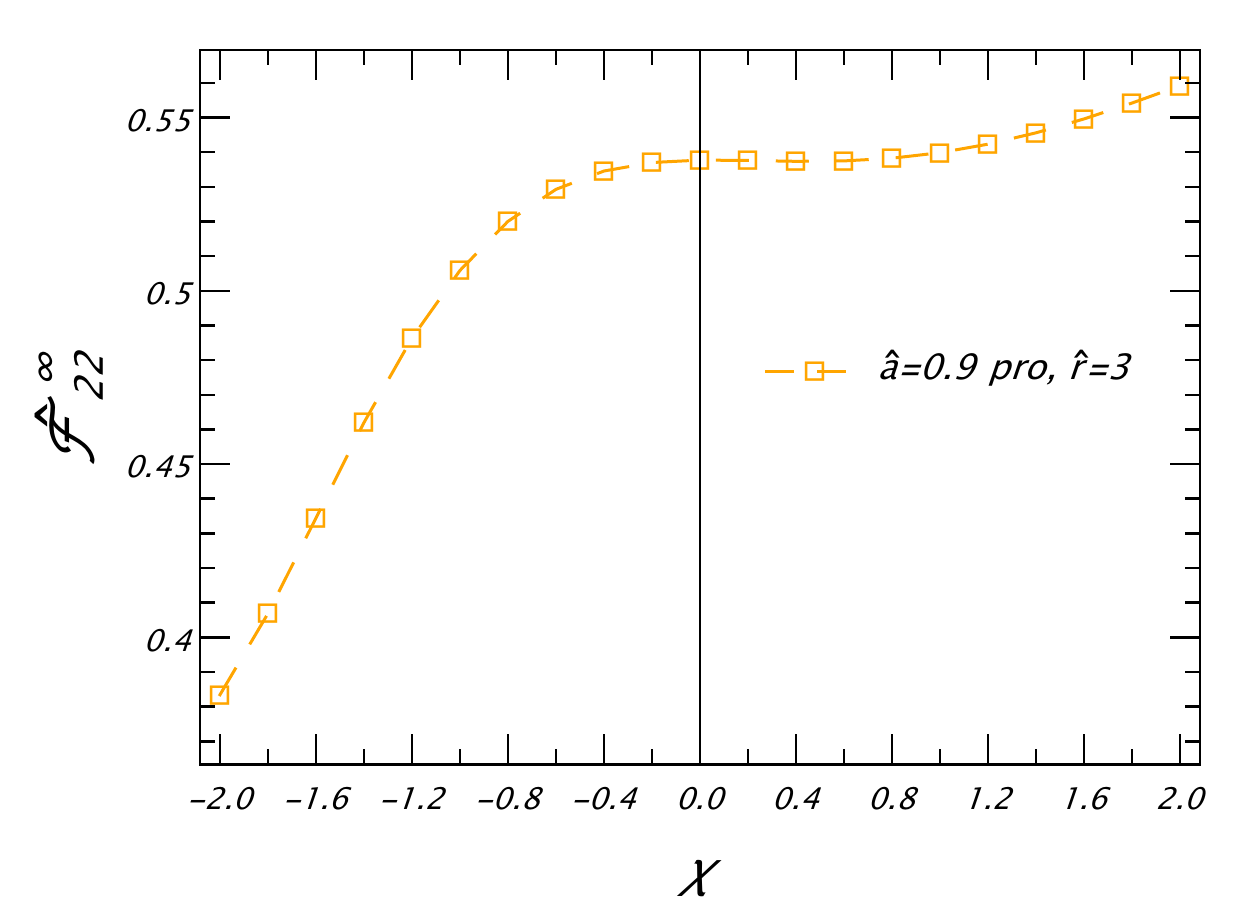}
\caption{Fluxes $\mathcal{\hat{F}}^\infty_{22}$ for the $\ell=m=2$ modes as a function of $\sigma$ for 
$\hat{a}=0.9$, prograde orbits and $\hat{r}=3$. Notice the nonlinear dependence of the fluxes on $\sigma$ for the 
extreme case $q=1$.}
 \label{fig:newtonflux} 
\end{figure}

\begin{table}[!htpb]
\begin{equation*}
\begin{array}{*{8}{c}}
\multicolumn{8}{c}{\hat{a}=0.9 \quad \text{prograde orbits}}\\
\hline
\hline
\hat{r}  & \sigma &\mathcal{\hat{F}}^\infty_{22} &\Delta_{22}[\%]  &\mathcal{\hat{F}}_{21} &\Delta_{21}[\%]  
&\mathcal{\hat{F}}_{33} &\Delta_{33}[\%]   \\
\hline
4  &-0.9 & 0.6037 & 0.2 & 3.3 \times 10^{-4} & 16 & 0.5052 & 0.3  \\
    &-0.5 & 0.6077 & 0.2 & 0.0315 & 1.3 & 0.4888 & 0.3 \\ 
    & 0.5 & 0.6038 & 0.2 & 0.3081 & 0.7 & 0.4458 & 0.3 \\ 
    & 0.9 & 0.6015 & 0.2 & 0.4651 & 0.7 & 0.4314 & 0.3 \\
\hline 
6  &-0.9 & 0.6900 & 0.2 & 0.0093 &\ast & 0.5826 & 0.3  \\
    &-0.5 & 0.6880 & 0.2 & 0.0737 & \ast & 0.5671  & 0.3\\ 
    & 0.5 & 0.6792 & 0.2 & 0.4314 & \ast & 0.5294 & 0.3 \\ 
    & 0.9 & 0.6750 & 0.2 & 0.6330 & 0.7 & 0.5154 & 0.3 \\
\hline
8  &-0.9 & 0.7384 & 0.2 & 0.0324 & 1.2 & 0.6357 & 0.3 \\
    &-0.5 & 0.7354 & 0.2 & 0.1164 &\ast & 0.6223 & 0.3  \\ 
    & 0.5 & 0.7261 & 0.2 & 0.5092 & \ast & 05899 & 0.3   \\ 
    & 0.9 & 0.7221 & 0.2 & 0.7264 & 0.7 & 0.5776 & 0.3  \\
\hline
10 &-0.9 & 0.7716 & 0.2 & 0.0596 & 1.1 & 0.6755  & 0.3\\
    &-0.5 & 0.7685 & 0.2 & 0.1558  & \ast & 0.6640 & 0.3 \\ 
    & 0.5 & 0.7598 & 0.2 & 0.5633  & \ast & 0.6361 & 0.3  \\ 
    & 0.9 & 0.7560 & 0.2 & 0.7842  & 0.6 & 0.6253 & 0.3  \\
\hline
20 &-0.9 & 0.8558 & 0.1 & 0.1848 & 1.0 & 0.7862 & 0.00 \\
    &-0.5 &  0.8537 & 0.2 & 0.2998 & 1.7 &  0.7800 & 0.01  \\ 
    & 0.5 &  0.8481 & 0.2 & 0.6982 & 0.3 & 0.7646 & 0.2 \\ 
    & 0.9 &  0.8458 & 0.2 & 0.8998 & 1.9 &  0.7586  & 0.1 \\
    \hline
    \hline
\bottomrule
\end{array}
\end{equation*}
\caption{Normalized fluxes compared against the fluxes shown in 
Table~IV of Ref.~\cite{Harms:2015ixa} for $\hat{a}=0.9$ and prograde orbits. 
The $\ast$ indicates fluxes not calculated in Ref.~\cite{Harms:2015ixa}.}
\label{tab:newtonfluxa09pro}
\end{table}
Finally, in Fig.~\ref{fig:newtonflux} we plot $\mathcal{\hat{F}}_{22}$  for prograde orbits with 
$\hat{a}=0.9$ and $\hat{r}=3$ as a function of $\chi$. Owing to the fact that $q=1$ (and therefore $\sigma$ is not 
small), the fluxes depend on the spin of the secondary in a nonlinear fashion when $\chi={\cal O}(1)$. 

\subsection{Comparison with Akcay {\it et al.}}

Recently, a new flux  balance law relating the local changes of 
energy of a spinning particle in Kerr spacetime with the asymptotic fluxes of energy and
angular momentum was obtained in Ref.~\cite{Akcay:2019bvk}. This procedure has been applied 
to particles with spin perpendicular to the orbital plane on circular orbits
in the Schwarzschild spacetime, computing the linear spin corrections to the fluxes. 
Table~\ref{tab:fluxcorrschwarz} provides our spin corrections to the flux and the fractional difference with respect 
to the sum of the spin's contributions at horizon and infinity given in Table~I of Ref.~\cite{Akcay:2019bvk}.
The errors show a very good agreement between the two results.

\begin{table}[!htpb]
\begin{equation*}
\begin{array}{*{3}{c}}
\hline
\hline
\hat{r} &\delta\mathcal{F}^\sigma &\Delta^\text{rel} (\delta\mathcal{F}^\sigma )  \\
\hline
10 & -1.35324081460517 \times 10^{-5}  & 3.0 \times 10^{-14}\\
8  & -6.28540371972  \times 10^{-5} & 1.9 \times 10^{-13} \\
6  &  -5.074933017 \times 10^{-4}  & 2.5 \times 10^{-11}  \\
\hline
\hline
\bottomrule
\end{array}
\end{equation*}
\caption{Linear spin correction to the GW flux $\delta\mathcal{F}^\sigma$ and fractional differences 
$\Delta^\text{rel} (\delta\mathcal{F}^\sigma )$ with respect to the fluxes shown in Table I 
of Ref.~\cite{Akcay:2019bvk} for $\hat{a}=0$. }
\label{tab:fluxcorrschwarz}
\end{table}

\subsection{Comparison with Taracchini {\it et al.}}

Reference~\cite{Taracchini:2013wfa} computed high-precision GW fluxes for nonspinning particles orbiting 
around Schwarzschild and Kerr BHs solving the Teukolsky equation in the frequency domain. 
We have checked our code against both their set-up. The relative errors are shown in 
Tables~\ref{tab:fluxISCO}-\ref{tab:fluxa9} for the values of the GW fluxes computed at the ISCO and 
at a different orbital separations $\hat{r}$, as a function of the primary spin. 
Note that in Ref.~\cite{Taracchini:2013wfa} the sum over the harmonic index 
$\ell$ was truncated at a certain value $\ell_\text{max}$ such that the fractional error between 
the flux at $\ell_{\text{max}}$ and $\ell_{\text{max}}-1$ was less than $10^{-14}$. To achieve this accuracy the 
required 
$\ell_\text{max}$ is in general very large: at the ISCO, for example, $\ell_\text{max}=30$ for $\hat{a}=0$,  and 
$\ell_\text{max}=66$  for $\hat{a}=0.99$. In our calculations we fixed $\ell_\text{max}=20$.
Nonetheless, the agreement between our results and those computed in Ref.~\cite{Taracchini:2013wfa} is 
extremely good. Even for the fastest spinning BH considered (with $\hat{a}=0.9$), we find 
a relative difference smaller than $10^{-5}$.

\begin{table}[!htpb]
\begin{equation*}
\begin{array}{c|c|c|c}
\hline
\hline
\hat{a} & \text{ISCO} & \mathcal{F}^0 & \Delta^\text{rel} (\mathcal{F}^0)  \\
\hline 
0.1 & 5.669 & 1.203797640  \times 10^{-3}  & 8.5 \times 10^{-11} \\
0.3 & 4.979 & 2.10037308  \times 10^{-3} & 1.4 \times 10^{-9} \\
0.5 & 4.233 & 4.11717449 \times 10^{-3} & 6.9 \times 10^{-10} \\
0.8 & 2.907 & 1.71190 \times 10^{-2} & 4.4 \times 10^{-7} \\
0.9 & 2.321 & 3.5223 \times 10^{-2} & 5.4 \times 10^{-6}\\
\hline
\hline
\end{array}
\end{equation*}
\caption{Fluxes for a nonspinning objects around Kerr BHs $\mathcal{F}^0$ at the ISCO and 
fractional difference $\Delta^\text{rel} (\mathcal{F}^0)$ compared to the results of Ref.~\cite{Taracchini:2013wfa}. }
\label{tab:fluxISCO}
\end{table}
\begin{table*}[!htpb]
\begin{equation*}
\begin{array}{c|ll|ll|ll}
\multicolumn{1}{c}{}&\multicolumn{2}{c}{\hat{a}=0}  & \multicolumn{2}{c}{\hat{a}=0.3}  &\multicolumn{2}{c}{\hat{a}=0.5} 
\\
\hline
\hline
\hat{r} &\mathcal{F}^0 &\Delta^\text{rel} (\mathcal{F}^0)&\mathcal{F}^0  &\Delta^\text{rel} (\mathcal{F}^0) 
&\mathcal{F}^0  &\Delta^\text{rel} (\mathcal{F}^0) \\
\hline
10 & 6.15163167846 \times 10^{-5}  & 1.8 \times 10^{-13} &  5.72185605812 \times 10^{-5}  & 1.1 \times 10^{-12} &  
5.4706016232 \times 10^{-5}  & 3.0 \times 10^{-12} \\
8 & 1.9610454858336 \times 10^{-4} & 1.6 \times 10^{-14}  & 1.757401400491 \times 10^{-4}  & 2.4 \times 10^{-14} & 
1.64390512713 \times 10^{-4}  & 7.2 \times 10^{-13}  \\
6 &  9.40339356  \times 10^{-4}  & 3.8 \times 10^{-11}  & 7.7105423521 \times 10^{-4}  & 1.2 \times 10^{-11} & 
6.8651481394 \times 10^{-4}  & 7.1 \times 10^{-12} \\
\hline
\hline
\bottomrule
\end{array}
\end{equation*}
\caption{Same as Table~\ref{tab:fluxISCO} but for generic orbital separation different from the 
ISCO, and focusing on $\hat{a}=(0,0.3,0.5)$.} 
\label{tab:fluxa3a8}
\end{table*}
\begin{table*}[!htpb]
\begin{equation*}
\begin{array}{c|ll|ll}
\multicolumn{1}{c}{}  & \multicolumn{2}{c}{\hat{a}=0.8}  &\multicolumn{2}{c}{\hat{a}=0.9} \\
\hline
\hline
\hat{r} &\mathcal{F}  &\Delta^\text{rel} (\mathcal{F}) &\mathcal{F}  &\Delta^\text{rel} (\mathcal{F}) \\
\hline
10 &  5.13763911701 \times 10^{-5}  & 4.3 \times 10^{-13} &  5.0368602531 \times 10^{-5}  & 1.4 \times 10^{-12}    \\
8   & 1.49973726131 \times 10^{-4}  & 2.6 \times 10^{-13}   &  1.4574909234 \times 10^{-4}  & 9.5 \times 10^{-13}  \\
6   & 5.8851295900 \times 10^{-4}  & 2.7 \times 10^{-12} &  5.6168859157 \times 10^{-4}  & 1.5 \times 10^{-12}  \\
4  & 3.9084751 \times 10^{-3}  & 2.2 \times 10^{-9} & 3.53976293 \times 10^{-3}  & 1.4 \times 10^{-9}  \\
\hline
\hline
\bottomrule
\end{array}
\end{equation*}
\caption{Fluxes for a non spinning object $\mathcal{F}^0 $ and fractional difference $\Delta^\text{rel} 
(\mathcal{F}^0)$ 
with respect to the fluxes listed in~\cite{Taracchini:2013wfa} for fast rotating BHs with $\hat{a}=(0.8,09)$. }
\label{tab:fluxa9}
\end{table*}
\begin{table*}[!htpb]
\begin{equation*}
\begin{array}{c|ll|ll}
\multicolumn{1}{c}{} & \multicolumn{2}{c}{\hat{a}=0.990}  &\multicolumn{2}{c}{\hat{a}=0.995} \\
\hline
\hline
\hat{r}&\mathcal{F}  &\Delta^\text{rel} (\mathcal{F}) &\mathcal{F}  &\Delta^\text{rel} (\mathcal{F}) \\
\hline
10 &  4.9500572776 \times 10^{-5}  & 2.7 \times 10^{-12} & 4.9453383948 \times 10^{-5}  & 3.4 \times 10^{-12}   \\
8  & 1.4216152170 \times 10^{-4}  & 1.5 \times 10^{-11} & 1.419678387 \times 10^{-4}  & 1.4\times 10^{-11}   \\
6  & 5.395577551 \times 10^{-4}  & 6.6\times 10^{-11} & 5.38379633 \times 10^{-4}  & 6.6 \times 10^{-11}   \\
4  & 3.26013974  \times 10^{-3} & 1.3 \times 10^{-9} & 3.24583765 \times 10^{-3}  & 1.3 \times 10^{-9}   \\
2  & 4.301  \times 10^{-2} & 1.1 \times 10^{-5} & 4.221 \times 10^{-2}  & 1.0 \times 10^{-5}   \\
\text{ISCO} & 9.17 \times 10^{-2} & 5.0 \times 10^{-4} & 9.5 \times 10^{-2} &  1.0 \times 10^{-3} \\
\hline
\hline
\bottomrule
\end{array}
\end{equation*}
\caption{Fluxes for a nonspinning object $\mathcal{F}^0 $ and fractional difference $\Delta^\text{rel} (\mathcal{F}^0)$ 
with respect to the fluxes listed in~\cite{BHPToolkit}. The ISCO is at $\hat{r}= 1.454$ and $\hat{r}= 1.341$  for $\hat{a}=0.990$ and $\hat{a}=0.995$ 
respectively.}
\label{tab:fluxa9a995b}
\end{table*}

\subsection{Comparison with Gralla {\it et al.}}
%
Finally, we tested our code in the case of a nonspinning secondary and fast spinning primary 
BHs with $\hat{a}>0.9$. In this case we use the data obtained in Ref.~\cite{Gralla:2016qfw} using 
the Teukolsky formalism in the frequency domain and assuming $\ell_\text{max}= 30$~\cite{Gralla:2016qfw}. The comparison 
is shown in Table~\ref{tab:fluxa9a995b} for $\hat{a}=0.99$ 
and $\hat{a}=0.995$ for orbital radii equal to and larger than the ISCO.
The discrepancy between our results and those of Ref.~\cite{Gralla:2016qfw} increases for larger spins 
and smaller orbital separation. However, in the worst case scenario, the fluxes differ at most 
by one part over~$10^3$.

\FloatBarrier

\bibliographystyle{utphys}
\bibliography{Ref}

\providecommand{\href}[2]{#2}\begingroup\raggedright\begin{thebibliography}{100}

\bibitem{Audley:2017drz}
{\bfseries LISA} Collaboration, P.~Amaro-Seoane {\em et~al.}, ``{Laser
  Interferometer Space Antenna},''
\href{http://arxiv.org/abs/1702.00786}{{\ttfamily arXiv:1702.00786
  [astro-ph.IM]}}.

\bibitem{Baibhav:2019rsa}
V.~Baibhav {\em et~al.}, ``{Probing the Nature of Black Holes: Deep in the mHz
  Gravitational-Wave Sky},''
\href{http://arxiv.org/abs/1908.11390}{{\ttfamily arXiv:1908.11390
  [astro-ph.HE]}}.

\bibitem{Babak:2017tow}
S.~Babak, J.~Gair, A.~Sesana, E.~Barausse, C.~F. Sopuerta, C.~P.~L. Berry,
  E.~Berti, P.~Amaro-Seoane, A.~Petiteau, and A.~Klein, ``{Science with the
  space-based interferometer LISA. V: Extreme mass-ratio inspirals},''
  \href{http://dx.doi.org/10.1103/PhysRevD.95.103012}{{\em Phys. Rev.}
  {\bfseries D95} no.~10, (2017) 103012},
\href{http://arxiv.org/abs/1703.09722}{{\ttfamily arXiv:1703.09722 [gr-qc]}}.

\bibitem{Chua:2019wgs}
A.~J.~K. Chua, N.~Korsakova, C.~J. Moore, J.~R. Gair, and S.~Babak, ``{Gaussian
  processes for the interpolation and marginalization of waveform error in
  extreme-mass-ratio-inspiral parameter estimation},''
  \href{http://dx.doi.org/10.1103/PhysRevD.101.044027}{{\em Phys. Rev.}
  {\bfseries D101} no.~4, (2020) 044027},
\href{http://arxiv.org/abs/1912.11543}{{\ttfamily arXiv:1912.11543
  [astro-ph.IM]}}.

\bibitem{Barack:2018yly}
L.~Barack {\em et~al.}, ``{Black holes, gravitational waves and fundamental
  physics: a roadmap},'' \href{http://dx.doi.org/10.1088/1361-6382/ab0587}{{\em
  Class. Quant. Grav.} {\bfseries 36} no.~14, (2019) 143001},
\href{http://arxiv.org/abs/1806.05195}{{\ttfamily arXiv:1806.05195 [gr-qc]}}.

\bibitem{Barausse:2020rsu}
E.~Barausse {\em et~al.}, ``{Prospects for Fundamental Physics with LISA},''
\href{http://arxiv.org/abs/2001.09793}{{\ttfamily arXiv:2001.09793 [gr-qc]}}.

\bibitem{Sopuerta:2009iy}
C.~F. Sopuerta and N.~Yunes, ``{Extreme and Intermediate-Mass Ratio Inspirals
  in Dynamical Chern-Simons Modified Gravity},''
  \href{http://dx.doi.org/10.1103/PhysRevD.80.064006}{{\em Phys. Rev.}
  {\bfseries D80} (2009) 064006},
\href{http://arxiv.org/abs/0904.4501}{{\ttfamily arXiv:0904.4501 [gr-qc]}}.

\bibitem{Yunes:2011aa}
N.~Yunes, P.~Pani, and V.~Cardoso, ``{Gravitational Waves from Quasicircular
  Extreme Mass-Ratio Inspirals as Probes of Scalar-Tensor Theories},''
  \href{http://dx.doi.org/10.1103/PhysRevD.85.102003}{{\em Phys. Rev.}
  {\bfseries D85} (2012) 102003},
\href{http://arxiv.org/abs/1112.3351}{{\ttfamily arXiv:1112.3351 [gr-qc]}}.

\bibitem{Pani:2011xj}
P.~Pani, V.~Cardoso, and L.~Gualtieri, ``{Gravitational waves from extreme
  mass-ratio inspirals in Dynamical Chern-Simons gravity},''
  \href{http://dx.doi.org/10.1103/PhysRevD.83.104048}{{\em Phys. Rev.}
  {\bfseries D83} (2011) 104048},
\href{http://arxiv.org/abs/1104.1183}{{\ttfamily arXiv:1104.1183 [gr-qc]}}.

\bibitem{Barausse:2016eii}
E.~Barausse, N.~Yunes, and K.~Chamberlain, ``{Theory-Agnostic Constraints on
  Black-Hole Dipole Radiation with Multiband Gravitational-Wave
  Astrophysics},'' \href{http://dx.doi.org/10.1103/PhysRevLett.116.241104}{{\em
  Phys. Rev. Lett.} {\bfseries 116} no.~24, (2016) 241104},
\href{http://arxiv.org/abs/1603.04075}{{\ttfamily arXiv:1603.04075 [gr-qc]}}.

\bibitem{Chamberlain:2017fjl}
K.~Chamberlain and N.~Yunes, ``{Theoretical Physics Implications of
  Gravitational Wave Observation with Future Detectors},''
  \href{http://dx.doi.org/10.1103/PhysRevD.96.084039}{{\em Phys. Rev.}
  {\bfseries D96} no.~8, (2017) 084039},
\href{http://arxiv.org/abs/1704.08268}{{\ttfamily arXiv:1704.08268 [gr-qc]}}.

\bibitem{Cardoso:2018zhm}
V.~Cardoso, G.~Castro, and A.~Maselli, ``{Gravitational waves in massive
  gravity theories: waveforms, fluxes and constraints from extreme-mass-ratio
  mergers},'' \href{http://dx.doi.org/10.1103/PhysRevLett.121.251103}{{\em
  Phys. Rev. Lett.} {\bfseries 121} no.~25, (2018) 251103},
\href{http://arxiv.org/abs/1809.00673}{{\ttfamily arXiv:1809.00673 [gr-qc]}}.

\bibitem{Barack:2006pq}
L.~Barack and C.~Cutler, ``{Using LISA EMRI sources to test off-Kerr deviations
  in the geometry of massive black holes},''
  \href{http://dx.doi.org/10.1103/PhysRevD.75.042003}{{\em Phys. Rev.}
  {\bfseries D75} (2007) 042003},
\href{http://arxiv.org/abs/gr-qc/0612029}{{\ttfamily arXiv:gr-qc/0612029
  [gr-qc]}}.

\bibitem{Pani:2010em}
P.~Pani, E.~Berti, V.~Cardoso, Y.~Chen, and R.~Norte, ``{Gravitational-wave
  signatures of the absence of an event horizon. II. Extreme mass ratio
  inspirals in the spacetime of a thin-shell gravastar},''
  \href{http://dx.doi.org/10.1103/PhysRevD.81.084011}{{\em Phys. Rev.}
  {\bfseries D81} (2010) 084011},
\href{http://arxiv.org/abs/1001.3031}{{\ttfamily arXiv:1001.3031 [gr-qc]}}.

\bibitem{Pani:2019cyc}
P.~Pani and A.~Maselli, ``{Love in Extrema Ratio},''
  \href{http://dx.doi.org/10.1142/S0218271819440012}{{\em Int. J. Mod. Phys.}
  {\bfseries D28} no.~14, (2019) 1944001},
\href{http://arxiv.org/abs/1905.03947}{{\ttfamily arXiv:1905.03947 [gr-qc]}}.

\bibitem{Datta:2019epe}
S.~Datta, R.~Brito, S.~Bose, P.~Pani, and S.~A. Hughes, ``{Tidal heating as a
  discriminator for horizons in extreme mass ratio inspirals},''
  \href{http://dx.doi.org/10.1103/PhysRevD.101.044004}{{\em Phys. Rev.}
  {\bfseries D101} no.~4, (2020) 044004},
\href{http://arxiv.org/abs/1910.07841}{{\ttfamily arXiv:1910.07841 [gr-qc]}}.

\bibitem{Pound:2015tma}
A.~Pound, ``{Motion of small objects in curved spacetimes: An introduction to
  gravitational self-force},''
  \href{http://dx.doi.org/10.1007/978-3-319-18335-0_13}{{\em Fund. Theor.
  Phys.} {\bfseries 179} (2015) 399--486},
\href{http://arxiv.org/abs/1506.06245}{{\ttfamily arXiv:1506.06245 [gr-qc]}}.

\bibitem{Barack:2018yvs}
L.~Barack and A.~Pound, ``{Self-force and radiation reaction in general
  relativity},'' \href{http://dx.doi.org/10.1088/1361-6633/aae552}{{\em Rept.
  Prog. Phys.} {\bfseries 82} no.~1, (2019) 016904},
\href{http://arxiv.org/abs/1805.10385}{{\ttfamily arXiv:1805.10385 [gr-qc]}}.

\bibitem{Dolan:2013roa}
S.~R. Dolan, N.~Warburton, A.~I. Harte, A.~Le~Tiec, B.~Wardell, and L.~Barack,
  ``{Gravitational self-torque and spin precession in compact binaries},''
  \href{http://dx.doi.org/10.1103/PhysRevD.89.064011}{{\em Phys. Rev.}
  {\bfseries D89} no.~6, (2014) 064011},
\href{http://arxiv.org/abs/1312.0775}{{\ttfamily arXiv:1312.0775 [gr-qc]}}.

\bibitem{Burko:2015sqa}
L.~M. Burko and G.~Khanna, ``{Self-force gravitational waveforms for extreme
  and intermediate mass ratio inspirals. III: Spin-orbit coupling revisited},''
  \href{http://dx.doi.org/10.1103/PhysRevD.91.104017}{{\em Phys. Rev.}
  {\bfseries D91} no.~10, (2015) 104017},
\href{http://arxiv.org/abs/1503.05097}{{\ttfamily arXiv:1503.05097 [gr-qc]}}.

\bibitem{Warburton:2017sxk}
N.~Warburton, T.~Osburn, and C.~R. Evans, ``{Evolution of small-mass-ratio
  binaries with a spinning secondary},''
  \href{http://dx.doi.org/10.1103/PhysRevD.96.084057}{{\em Phys. Rev.}
  {\bfseries D96} no.~8, (2017) 084057},
\href{http://arxiv.org/abs/1708.03720}{{\ttfamily arXiv:1708.03720 [gr-qc]}}.

\bibitem{Akcay:2019bvk}
S.~Akcay, S.~R. Dolan, C.~Kavanagh, J.~Moxon, N.~Warburton, and B.~Wardell,
  ``{Dissipation in extreme-mass ratio binaries with a spinning secondary},''
\href{http://arxiv.org/abs/1912.09461}{{\ttfamily arXiv:1912.09461 [gr-qc]}}.

\bibitem{Mino:1995fm}
Y.~Mino, M.~Shibata, and T.~Tanaka, ``{Gravitational waves induced by a
  spinning particle falling into a rotating black hole},''
  \href{http://dx.doi.org/10.1103/PhysRevD.53.622,
  10.1103/PhysRevD.59.047502}{{\em Phys. Rev.} {\bfseries D53} (1996)
  622--634}.
[Erratum: Phys. Rev.D59,047502(1999)].

\bibitem{Saijo:1998mn}
M.~Saijo, K.-i. Maeda, M.~Shibata, and Y.~Mino, ``{Gravitational waves from a
  spinning particle plunging into a Kerr black hole},''
\href{http://dx.doi.org/10.1103/PhysRevD.58.064005}{{\em Phys. Rev.} {\bfseries
  D58} (1998) 064005}.

\bibitem{Tominaga:2000cs}
K.~Tominaga, M.~Saijo, and K.-i. Maeda, ``{Gravitational waves from a spinning
  particle scattered by a relativistic star: Axial mode case},''
  \href{http://dx.doi.org/10.1103/PhysRevD.63.124012}{{\em Phys. Rev.}
  {\bfseries D63} (2001) 124012},
\href{http://arxiv.org/abs/gr-qc/0009055}{{\ttfamily arXiv:gr-qc/0009055
  [gr-qc]}}.

\bibitem{Tanaka:1996ht}
T.~Tanaka, Y.~Mino, M.~Sasaki, and M.~Shibata, ``{Gravitational waves from a
  spinning particle in circular orbits around a rotating black hole},''
  \href{http://dx.doi.org/10.1103/PhysRevD.54.3762}{{\em Phys. Rev.} {\bfseries
  D54} (1996) 3762--3777},
\href{http://arxiv.org/abs/gr-qc/9602038}{{\ttfamily arXiv:gr-qc/9602038
  [gr-qc]}}.

\bibitem{Nagar:2019wrt}
A.~Nagar, F.~Messina, C.~Kavanagh, G.~Lukes-Gerakopoulos, N.~Warburton,
  S.~Bernuzzi, and E.~Harms, ``{Factorization and resummation: A new paradigm
  to improve gravitational wave amplitudes. III: the spinning test-body
  terms},'' \href{http://dx.doi.org/10.1103/PhysRevD.100.104056}{{\em Phys.
  Rev.} {\bfseries D100} no.~10, (2019) 104056},
\href{http://arxiv.org/abs/1907.12233}{{\ttfamily arXiv:1907.12233 [gr-qc]}}.

\bibitem{Burko:2003rv}
L.~M. Burko, ``{Orbital evolution of a particle around a black hole. 2.
  Comparison of contributions of spin orbit coupling and the selfforce},''
  \href{http://dx.doi.org/10.1103/PhysRevD.69.044011}{{\em Phys. Rev. D}
  {\bfseries 69} (2004) 044011},
  \href{http://arxiv.org/abs/gr-qc/0308003}{{\ttfamily arXiv:gr-qc/0308003}}.

\bibitem{Harms:2015ixa}
E.~Harms, G.~Lukes-Gerakopoulos, S.~Bernuzzi, and A.~Nagar, ``{Asymptotic
  gravitational wave fluxes from a spinning particle in circular equatorial
  orbits around a rotating black hole},''
  \href{http://dx.doi.org/10.1103/PhysRevD.100.129901,
  10.1103/PhysRevD.93.044015}{{\em Phys. Rev.} {\bfseries D93} no.~4, (2016)
  044015}, \href{http://arxiv.org/abs/1510.05548}{{\ttfamily arXiv:1510.05548
  [gr-qc]}}.
[Addendum: Phys. Rev.D100,no.12,129901(2019)].

\bibitem{Harms:2016ctx}
E.~Harms, G.~Lukes-Gerakopoulos, S.~Bernuzzi, and A.~Nagar, ``{Spinning test
  body orbiting around a Schwarzschild black hole: Circular dynamics and
  gravitational-wave fluxes},''
  \href{http://dx.doi.org/10.1103/PhysRevD.100.129902,
  10.1103/PhysRevD.94.104010}{{\em Phys. Rev.} {\bfseries D94} no.~10, (2016)
  104010},
\href{http://arxiv.org/abs/1609.00356}{{\ttfamily arXiv:1609.00356 [gr-qc]}}.

\bibitem{Lukes-Gerakopoulos:2017vkj}
G.~Lukes-Gerakopoulos, E.~Harms, S.~Bernuzzi, and A.~Nagar, ``{Spinning
  test-body orbiting around a Kerr black hole: circular dynamics and
  gravitational-wave fluxes},''
  \href{http://dx.doi.org/10.1103/PhysRevD.96.064051}{{\em Phys. Rev.}
  {\bfseries D96} no.~6, (2017) 064051},
\href{http://arxiv.org/abs/1707.07537}{{\ttfamily arXiv:1707.07537 [gr-qc]}}.

\bibitem{Yunes:2010zj}
N.~Yunes, A.~Buonanno, S.~A. Hughes, Y.~Pan, E.~Barausse, M.~Miller, and
  W.~Throwe, ``{Extreme Mass-Ratio Inspirals in the Effective-One-Body
  Approach: Quasi-Circular, Equatorial Orbits around a Spinning Black Hole},''
  \href{http://dx.doi.org/10.1103/PhysRevD.83.044044}{{\em Phys. Rev. D}
  {\bfseries 83} (2011) 044044},
  \href{http://arxiv.org/abs/1009.6013}{{\ttfamily arXiv:1009.6013 [gr-qc]}}.
  [Erratum: Phys.Rev.D 88, 109904 (2013)].

\bibitem{Chen:2019hac}
B.~Chen, G.~Compère, Y.~Liu, J.~Long, and X.~Zhang, ``{Spin and Quadrupole
  Couplings for High Spin Equatorial Intermediate Mass-ratio Coalescences},''
  \href{http://dx.doi.org/10.1088/1361-6382/ab4fb0}{{\em Class. Quant. Grav.}
  {\bfseries 36} no.~24, (2019) 245011},
  \href{http://arxiv.org/abs/1901.05370}{{\ttfamily arXiv:1901.05370 [gr-qc]}}.

\bibitem{Piovano:2020ooe}
G.~A. Piovano, A.~Maselli, and P.~Pani, ``{Model independent tests of the Kerr
  bound with extreme mass ratio inspirals},''
\href{http://arxiv.org/abs/2003.08448}{{\ttfamily arXiv:2003.08448 [gr-qc]}}.

\bibitem{xAct}
``{xAct: Efficient tensor computer algebra for the Wolfram Language}.''
  (\href{http://www.xact.es/}{xact.es}).

\bibitem{Tulczyjew:1959}
W.~Tulczyjew, ``Motion of multipole particles in general relativity theory,''
  {\em Acta Phys. Pol.} {\bfseries 18} (1959) 393.

\bibitem{Dixon:1964NCim}
W.~Dixon, ``{A covariant multipole formalism for extended test bodies in
  general relativity},'' \href{http://dx.doi.org/10.1007/BF02734579}{{\em Il
  Nuovo Cimento} {\bfseries 34} no.~2, (Oct, 1964) 317--339}.

\bibitem{Dixon:1970I}
W.~G. Dixon, ``{Dynamics of extended bodies in general relativity. I. Momentum
  and angular momentum},''
\href{http://dx.doi.org/10.1098/rspa.1970.0020}{{\em Proc. Roy. Soc. Lond.}
  {\bfseries A314} (1970) 499--527}.

\bibitem{Dixon:1970II}
W.~G. Dixon, ``{Dynamics of extended bodies in general relativity. II. Moments
  of the charge-current vector},''
\href{http://dx.doi.org/10.1098/rspa.1970.0191}{{\em Proc. Roy. Soc. Lond.}
  {\bfseries A319} (1970) 509--547}.

\bibitem{Kyrian:2007zz}
K.~Kyrian and O.~Semerak, ``{Spinning test particles in a Kerr field},''
\href{http://dx.doi.org/10.1111/j.1365-2966.2007.12502.x}{{\em Mon. Not. Roy.
  Astron. Soc.} {\bfseries 382} (2007) 1922}.

\bibitem{Dixon:1978}
W.~Dixon, ``Extended bodies in general relativity; their description and
  motion,'' in {\em Isolated Gravitating Systems in General Relativity -
  Proceedings of the International School of Physics "Enrico Fermi"}.
\newblock 1978.

\bibitem{Mathisson:1937zz}
M.~Mathisson, ``{Neue mechanik materieller systemes},''
{\em Acta Phys. Polon.} {\bfseries 6} (1937) 163--2900.

\bibitem{Papapetrou:1951pa}
A.~Papapetrou, ``{Spinning test particles in general relativity. 1.},''
\href{http://dx.doi.org/10.1098/rspa.1951.0200}{{\em Proc. Roy. Soc. Lond.}
  {\bfseries A209} (1951) 248--258}.

\bibitem{Corinaldesi:1951pb}
E.~Corinaldesi and A.~Papapetrou, ``{Spinning test particles in general
  relativity. 2.},''
\href{http://dx.doi.org/10.1098/rspa.1951.0201}{{\em Proc. Roy. Soc. Lond.}
  {\bfseries A209} (1951) 259--268}.

\bibitem{Steinhoff:2009tk}
J.~Steinhoff and D.~Puetzfeld, ``{Multipolar equations of motion for extended
  test bodies in General Relativity},''
  \href{http://dx.doi.org/10.1103/PhysRevD.81.044019}{{\em Phys. Rev.}
  {\bfseries D81} (2010) 044019},
\href{http://arxiv.org/abs/0909.3756}{{\ttfamily arXiv:0909.3756 [gr-qc]}}.

\bibitem{Semerak:1999qc}
O.~Semerak, ``{Spinning test particles in a Kerr field. 1.},''
\href{http://dx.doi.org/10.1046/j.1365-8711.1999.02754.x}{{\em Mon. Not. Roy.
  Astron. Soc.} {\bfseries 308} (1999) 863--875}.

\bibitem{Costa:2011zn}
F.~Costa, C.~A.~R. Herdeiro, J.~Natario, and M.~Zilhao, ``{Mathisson's helical
  motions for a spinning particle: Are they unphysical?},''
  \href{http://dx.doi.org/10.1103/PhysRevD.85.024001}{{\em Phys. Rev.}
  {\bfseries D85} (2012) 024001},
\href{http://arxiv.org/abs/1109.1019}{{\ttfamily arXiv:1109.1019 [gr-qc]}}.

\bibitem{Costa:2014nta}
L.~F.~O. Costa and J.~Natário, ``{Center of mass, spin supplementary
  conditions, and the momentum of spinning particles},''
  \href{http://dx.doi.org/10.1007/978-3-319-18335-0_6}{{\em Fund. Theor. Phys.}
  {\bfseries 179} (2015) 215--258},
\href{http://arxiv.org/abs/1410.6443}{{\ttfamily arXiv:1410.6443 [gr-qc]}}.

\bibitem{Ehlers:1977}
J.~Ehlers and E.~Rudolph, ``{Dynamics of extended bodies in general relativity
  center-of-mass description and quasirigidity},''
  \href{http://dx.doi.org/10.1007/BF00763547}{{\em General Relativity and
  Gravitation} {\bfseries 8} no.~3, (Mar, 1977) 197--217}.

\bibitem{Lukes-Gerakopoulos:2017cru}
G.~Lukes-Gerakopoulos, ``{Time parameterizations and spin supplementary
  conditions of the Mathisson-Papapetrou-Dixon equations},''
  \href{http://dx.doi.org/10.1103/PhysRevD.96.104023}{{\em Phys. Rev.}
  {\bfseries D96} no.~10, (2017) 104023},
\href{http://arxiv.org/abs/1709.08942}{{\ttfamily arXiv:1709.08942 [gr-qc]}}.

\bibitem{Witzany:2018ahb}
V.~Witzany, J.~Steinhoff, and G.~Lukes-Gerakopoulos, ``{Hamiltonians and
  canonical coordinates for spinning particles in curved space-time},''
  \href{http://dx.doi.org/10.1088/1361-6382/ab002f}{{\em Class. Quant. Grav.}
  {\bfseries 36} no.~7, (2019) 075003},
\href{http://arxiv.org/abs/1808.06582}{{\ttfamily arXiv:1808.06582 [gr-qc]}}.

\bibitem{Moller:1949}
C.~M\o\~ller, ``Sur la dynamique des systèmes ayant un moment angulaire
  interne,'' {\em Annales de l'institut Henri Poincaré} {\bfseries 11} no.~5,
  (1949) 251--278. \url{http://eudml.org/doc/79030}.

\bibitem{Steinhoff:2012rw}
J.~Steinhoff and D.~Puetzfeld, ``{Influence of internal structure on the motion
  of test bodies in extreme mass ratio situations},''
  \href{http://dx.doi.org/10.1103/PhysRevD.86.044033}{{\em Phys. Rev.}
  {\bfseries D86} (2012) 044033},
\href{http://arxiv.org/abs/1205.3926}{{\ttfamily arXiv:1205.3926 [gr-qc]}}.

\bibitem{Jefremov:2015gza}
P.~I. Jefremov, O.~{\relax Yu}. Tsupko, and G.~S. Bisnovatyi-Kogan,
  ``{Innermost stable circular orbits of spinning test particles in
  Schwarzschild and Kerr space-times},''
  \href{http://dx.doi.org/10.1103/PhysRevD.91.124030}{{\em Phys. Rev.}
  {\bfseries D91} no.~12, (2015) 124030},
\href{http://arxiv.org/abs/1503.07060}{{\ttfamily arXiv:1503.07060 [gr-qc]}}.

\bibitem{Suzuki:1997by}
S.~Suzuki and K.-i. Maeda, ``{Innermost stable circular orbit of a spinning
  particle in Kerr space-time},''
  \href{http://dx.doi.org/10.1103/PhysRevD.58.023005}{{\em Phys. Rev.}
  {\bfseries D58} (1998) 023005},
\href{http://arxiv.org/abs/gr-qc/9712095}{{\ttfamily arXiv:gr-qc/9712095
  [gr-qc]}}.

\bibitem{1972ApJ...178..347B}
J.~M. {Bardeen}, W.~H. {Press}, and S.~A. {Teukolsky}, ``{Rotating Black Holes:
  Locally Nonrotating Frames, Energy Extraction, and Scalar Synchrotron
  Radiation},'' \href{http://dx.doi.org/10.1086/151796}{{\em The Astrophysical
  Journal} {\bfseries 178} (Dec, 1972) 347--370}.

\bibitem{Hinderer:2008dm}
T.~Hinderer and E.~E. Flanagan, ``{Two timescale analysis of extreme mass ratio
  inspirals in Kerr. I. Orbital Motion},''
  \href{http://dx.doi.org/10.1103/PhysRevD.78.064028}{{\em Phys. Rev.}
  {\bfseries D78} (2008) 064028},
\href{http://arxiv.org/abs/0805.3337}{{\ttfamily arXiv:0805.3337 [gr-qc]}}.

\bibitem{Hughes:2018qxz}
S.~A. Hughes, ``{Bound orbits of a slowly evolving black hole},''
  \href{http://dx.doi.org/10.1103/PhysRevD.100.064001}{{\em Phys. Rev.}
  {\bfseries D100} no.~6, (2019) 064001},
\href{http://arxiv.org/abs/1806.09022}{{\ttfamily arXiv:1806.09022 [gr-qc]}}.

\bibitem{Ori:2000zn}
A.~Ori and K.~S. Thorne, ``{The Transition from inspiral to plunge for a
  compact body in a circular equatorial orbit around a massive, spinning black
  hole},'' \href{http://dx.doi.org/10.1103/PhysRevD.62.124022}{{\em Phys. Rev.
  D} {\bfseries 62} (2000) 124022},
  \href{http://arxiv.org/abs/gr-qc/0003032}{{\ttfamily arXiv:gr-qc/0003032}}.

\bibitem{Burke:2019yek}
O.~Burke, J.~R. Gair, and J.~Simón, ``{Transition from Inspiral to Plunge: A
  Complete Near-Extremal Trajectory and Associated Waveform},''
  \href{http://dx.doi.org/10.1103/PhysRevD.101.064026}{{\em Phys. Rev. D}
  {\bfseries 101} no.~6, (2020) 064026},
  \href{http://arxiv.org/abs/1909.12846}{{\ttfamily arXiv:1909.12846 [gr-qc]}}.

\bibitem{Compere:2019cqe}
G.~Compère, K.~Fransen, and C.~Jonas, ``{Transition from inspiral to plunge
  into a highly spinning black hole},''
  \href{http://dx.doi.org/10.1088/1361-6382/ab79d3}{{\em Class. Quant. Grav.}
  {\bfseries 37} no.~9, (2020) 095013},
  \href{http://arxiv.org/abs/1909.12848}{{\ttfamily arXiv:1909.12848 [gr-qc]}}.

\bibitem{Kennefick:1998ab}
D.~Kennefick, ``{Stability under radiation reaction of circular equatorial
  orbits around Kerr black holes},''
  \href{http://dx.doi.org/10.1103/PhysRevD.58.064012}{{\em Phys. Rev. D}
  {\bfseries 58} (1998) 064012},
  \href{http://arxiv.org/abs/gr-qc/9805102}{{\ttfamily arXiv:gr-qc/9805102}}.

\bibitem{Kennefick:1995za}
D.~Kennefick and A.~Ori, ``{Radiation reaction induced evolution of circular
  orbits of particles around Kerr black holes},''
  \href{http://dx.doi.org/10.1103/PhysRevD.53.4319}{{\em Phys. Rev. D}
  {\bfseries 53} (1996) 4319--4326},
  \href{http://arxiv.org/abs/gr-qc/9512018}{{\ttfamily arXiv:gr-qc/9512018}}.

\bibitem{BHPToolkit}
``{Black Hole Perturbation Toolkit}.''
  (\href{http://bhptoolkit.org/}{bhptoolkit.org}).

\bibitem{Mino:1997bx}
Y.~Mino, M.~Sasaki, M.~Shibata, H.~Tagoshi, and T.~Tanaka, ``{Black hole
  perturbation: Chapter 1},'' \href{http://dx.doi.org/10.1143/PTPS.128.1}{{\em
  Prog. Theor. Phys. Suppl.} {\bfseries 128} (1997) 1--121},
\href{http://arxiv.org/abs/gr-qc/9712057}{{\ttfamily arXiv:gr-qc/9712057
  [gr-qc]}}.

\bibitem{Mano:1996vt}
S.~Mano, H.~Suzuki, and E.~Takasugi, ``{Analytic solutions of the Teukolsky
  equation and their low frequency expansions},''
  \href{http://dx.doi.org/10.1143/PTP.95.1079}{{\em Prog. Theor. Phys.}
  {\bfseries 95} (1996) 1079--1096},
\href{http://arxiv.org/abs/gr-qc/9603020}{{\ttfamily arXiv:gr-qc/9603020
  [gr-qc]}}.

\bibitem{Fujita:2004rb}
R.~Fujita and H.~Tagoshi, ``{New numerical methods to evaluate homogeneous
  solutions of the Teukolsky equation},''
  \href{http://dx.doi.org/10.1143/PTP.112.415}{{\em Prog. Theor. Phys.}
  {\bfseries 112} (2004) 415--450},
\href{http://arxiv.org/abs/gr-qc/0410018}{{\ttfamily arXiv:gr-qc/0410018
  [gr-qc]}}.

\bibitem{Fujita:2009us}
R.~Fujita, W.~Hikida, and H.~Tagoshi, ``{An Efficient Numerical Method for
  Computing Gravitational Waves Induced by a Particle Moving on Eccentric
  Inclined Orbits around a Kerr Black Hole},''
  \href{http://dx.doi.org/10.1143/PTP.121.843}{{\em Prog. Theor. Phys.}
  {\bfseries 121} (2009) 843--874},
\href{http://arxiv.org/abs/0904.3810}{{\ttfamily arXiv:0904.3810 [gr-qc]}}.

\bibitem{Hughes:1999bq}
S.~A. Hughes, ``{The Evolution of circular, nonequatorial orbits of Kerr black
  holes due to gravitational wave emission},''
  \href{http://dx.doi.org/10.1103/PhysRevD.65.069902,
  10.1103/PhysRevD.90.109904, 10.1103/PhysRevD.61.084004,
  10.1103/PhysRevD.63.049902, 10.1103/PhysRevD.67.089901}{{\em Phys. Rev.}
  {\bfseries D61} no.~8, (2000) 084004},
  \href{http://arxiv.org/abs/gr-qc/9910091}{{\ttfamily arXiv:gr-qc/9910091
  [gr-qc]}}.
[Erratum: Phys. Rev.D63,no.4,049902(2001); Erratum: Phys.
  Rev.D65,no.6,069902(2002); Erratum: Phys. Rev.D67,no.8,089901(2003); Erratum:
  Phys. Rev.D78,no.10,109902(2008); Erratum: Phys. Rev.D90,no.10,109904(2014)].

\bibitem{webpage}
Data and relevant codes are publicly available at
  \url{https://web.uniroma1.it/gmunu}.

\bibitem{Gralla:2016qfw}
S.~E. Gralla, S.~A. Hughes, and N.~Warburton, ``{Inspiral into Gargantua},''
  \href{http://dx.doi.org/10.1088/0264-9381/33/15/155002}{{\em Class. Quant.
  Grav.} {\bfseries 33} no.~15, (2016) 155002},
\href{http://arxiv.org/abs/1603.01221}{{\ttfamily arXiv:1603.01221 [gr-qc]}}.

\bibitem{Huerta:2011kt}
E.~Huerta and J.~R. Gair, ``{Importance of including small body spin effects in
  the modelling of extreme and intermediate mass-ratio inspirals},''
  \href{http://dx.doi.org/10.1103/PhysRevD.84.064023}{{\em Phys. Rev. D}
  {\bfseries 84} (2011) 064023},
  \href{http://arxiv.org/abs/1105.3567}{{\ttfamily arXiv:1105.3567 [gr-qc]}}.

\bibitem{LISADataChallenge}
LISA Data Challenge Working Group. LISA Data Challenges, 2019.
  \url{https://lisa-ldc.lal.in2p3.fr}.

\bibitem{Lindblom:2008cm}
L.~Lindblom, B.~J. Owen, and D.~A. Brown, ``{Model Waveform Accuracy Standards
  for Gravitational Wave Data Analysis},''
  \href{http://dx.doi.org/10.1103/PhysRevD.78.124020}{{\em Phys. Rev.}
  {\bfseries D78} (2008) 124020},
\href{http://arxiv.org/abs/0809.3844}{{\ttfamily arXiv:0809.3844 [gr-qc]}}.

\bibitem{Flanagan:1997kp}
E.~E. Flanagan and S.~A. Hughes, ``{Measuring gravitational waves from binary
  black hole coalescences: 2. The Waves' information and its extraction, with
  and without templates},''
  \href{http://dx.doi.org/10.1103/PhysRevD.57.4566}{{\em Phys. Rev.} {\bfseries
  D57} (1998) 4566--4587},
\href{http://arxiv.org/abs/gr-qc/9710129}{{\ttfamily arXiv:gr-qc/9710129
  [gr-qc]}}.

\bibitem{Hartl:2002ig}
M.~D. Hartl, ``{Dynamics of spinning test particles in Kerr space-time},''
  \href{http://dx.doi.org/10.1103/PhysRevD.67.024005}{{\em Phys. Rev.}
  {\bfseries D67} (2003) 024005},
\href{http://arxiv.org/abs/gr-qc/0210042}{{\ttfamily arXiv:gr-qc/0210042
  [gr-qc]}}.

\bibitem{Gimon:2007ur}
E.~G. Gimon and P.~Horava, ``{Astrophysical violations of the Kerr bound as a
  possible signature of string theory},''
  \href{http://dx.doi.org/10.1016/j.physletb.2009.01.026}{{\em Phys. Lett.}
  {\bfseries B672} (2009) 299--302},
\href{http://arxiv.org/abs/0706.2873}{{\ttfamily arXiv:0706.2873 [hep-th]}}.

\bibitem{Pani:2010jz}
P.~Pani, E.~Barausse, E.~Berti, and V.~Cardoso, ``{Gravitational instabilities
  of superspinars},'' \href{http://dx.doi.org/10.1103/PhysRevD.82.044009}{{\em
  Phys. Rev.} {\bfseries D82} (2010) 044009},
\href{http://arxiv.org/abs/1006.1863}{{\ttfamily arXiv:1006.1863 [gr-qc]}}.

\bibitem{Maggio:2017ivp}
E.~Maggio, P.~Pani, and V.~Ferrari, ``{Exotic Compact Objects and How to Quench
  their Ergoregion Instability},''
  \href{http://dx.doi.org/10.1103/PhysRevD.96.104047}{{\em Phys. Rev.}
  {\bfseries D96} no.~10, (2017) 104047},
\href{http://arxiv.org/abs/1703.03696}{{\ttfamily arXiv:1703.03696 [gr-qc]}}.

\bibitem{Maggio:2018ivz}
E.~Maggio, V.~Cardoso, S.~R. Dolan, and P.~Pani, ``{Ergoregion instability of
  exotic compact objects: electromagnetic and gravitational perturbations and
  the role of absorption},''
  \href{http://dx.doi.org/10.1103/PhysRevD.99.064007}{{\em Phys.\ Rev.\ D}
  {\bfseries 99} no.~6, (2019) 064007},
  \href{http://arxiv.org/abs/1807.08840}{{\ttfamily arXiv:1807.08840 [gr-qc]}}.

\bibitem{Roy:2019uuy}
R.~Roy, P.~Kocherlakota, and P.~S. Joshi, ``{Mode stability of a near-extremal
  Kerr superspinar},''
\href{http://arxiv.org/abs/1911.06169}{{\ttfamily arXiv:1911.06169 [gr-qc]}}.

\bibitem{1997A&A...317..815B}
B.~C. {Bisscheroux}, O.~R. {Pols}, P.~{Kahabka}, T.~{Belloni}, and E.~P.~J.
  {van den Heuvel}, ``{The nature of the bright subdwarf HD 49798 and its X-ray
  pulsating companion.},'' {\em \aap} {\bfseries 317} (Feb., 1997) 815--822.

\bibitem{Hessels:2006ze}
J.~W. Hessels, S.~M. Ransom, I.~H. Stairs, P.~C.~C. Freire, V.~M. Kaspi, and
  F.~Camilo, ``{A radio pulsar spinning at 716-hz},''
  \href{http://dx.doi.org/10.1126/science.1123430}{{\em Science} {\bfseries
  311} (2006) 1901--1904},
  \href{http://arxiv.org/abs/astro-ph/0601337}{{\ttfamily
  arXiv:astro-ph/0601337}}.

\bibitem{Manchester:2004bp}
R.~N. Manchester, G.~B. Hobbs, A.~Teoh, and M.~Hobbs, ``{The Australia
  Telescope National Facility pulsar catalogue},''
  \href{http://dx.doi.org/10.1086/428488}{{\em Astron. J.} {\bfseries 129}
  (2005) 1993}, \href{http://arxiv.org/abs/astro-ph/0412641}{{\ttfamily
  arXiv:astro-ph/0412641}}.

\bibitem{Cardoso:2019rvt}
V.~Cardoso and P.~Pani, ``{Testing the nature of dark compact objects: a status
  report},'' \href{http://dx.doi.org/10.1007/s41114-019-0020-4}{{\em Living
  Rev. Rel.} {\bfseries 22} no.~1, (2019) 4},
\href{http://arxiv.org/abs/1904.05363}{{\ttfamily arXiv:1904.05363 [gr-qc]}}.

\bibitem{Berti:2015itd}
E.~Berti {\em et~al.}, ``{Testing General Relativity with Present and Future
  Astrophysical Observations},''
  \href{http://dx.doi.org/10.1088/0264-9381/32/24/243001}{{\em Class. Quant.
  Grav.} {\bfseries 32} (2015) 243001},
\href{http://arxiv.org/abs/1501.07274}{{\ttfamily arXiv:1501.07274 [gr-qc]}}.

\bibitem{Barausse:2014tra}
E.~Barausse, V.~Cardoso, and P.~Pani, ``{Can environmental effects spoil
  precision gravitational-wave astrophysics?},''
  \href{http://dx.doi.org/10.1103/PhysRevD.89.104059}{{\em Phys.\ Rev.\ D}
  {\bfseries 89} no.~10, (2014) 104059},
  \href{http://arxiv.org/abs/1404.7149}{{\ttfamily arXiv:1404.7149 [gr-qc]}}.

\bibitem{Maselli:2020zgv}
A.~Maselli, N.~Franchini, L.~Gualtieri, and T.~P. Sotiriou, ``{Detecting scalar
  fields with Extreme Mass Ratio Inspirals},''
  \href{http://arxiv.org/abs/2004.11895}{{\ttfamily arXiv:2004.11895 [gr-qc]}}.

\bibitem{Witzany:2019nml}
V.~Witzany, ``{Hamilton-Jacobi equation for spinning particles near black
  holes},'' \href{http://dx.doi.org/10.1103/PhysRevD.100.104030}{{\em Phys.
  Rev.} {\bfseries D100} no.~10, (2019) 104030},
\href{http://arxiv.org/abs/1903.03651}{{\ttfamily arXiv:1903.03651 [gr-qc]}}.

\bibitem{Bini:2006pc}
D.~Bini, A.~Geralico, R.~T. Jantzen, and F.~de~Felice, ``{Spin precession along
  circular orbits in the Kerr spacetime: the Frenet-Serret description},''
  \href{http://dx.doi.org/10.1088/0264-9381/23/10/003}{{\em Class. Quant.
  Grav.} {\bfseries 23} (2006) 3287--3304},
\href{http://arxiv.org/abs/1408.4278}{{\ttfamily arXiv:1408.4278 [gr-qc]}}.

\bibitem{Ruangsri:2015cvg}
U.~Ruangsri, S.~J. Vigeland, and S.~A. Hughes, ``{Gyroscopes orbiting black
  holes: A frequency-domain approach to precession and spin-curvature coupling
  for spinning bodies on generic Kerr orbits},''
  \href{http://dx.doi.org/10.1103/PhysRevD.94.044008}{{\em Phys. Rev.}
  {\bfseries D94} no.~4, (2016) 044008},
\href{http://arxiv.org/abs/1512.00376}{{\ttfamily arXiv:1512.00376 [gr-qc]}}.

\bibitem{Hinderer:2013uwa}
T.~Hinderer {\em et~al.}, ``{Periastron advance in spinning black hole
  binaries: comparing effective-one-body and Numerical Relativity},''
  \href{http://dx.doi.org/10.1103/PhysRevD.88.084005}{{\em Phys. Rev.}
  {\bfseries D88} no.~8, (2013) 084005},
\href{http://arxiv.org/abs/1309.0544}{{\ttfamily arXiv:1309.0544 [gr-qc]}}.

\bibitem{Bini:2014xyr}
D.~Bini and A.~Geralico, ``{Extended bodies in a Kerr spacetime: exploring the
  role of a general quadrupole tensor},''
  \href{http://dx.doi.org/10.1088/0264-9381/31/7/075024}{{\em Class. Quant.
  Grav.} {\bfseries 31} (2014) 075024},
\href{http://arxiv.org/abs/1408.5484}{{\ttfamily arXiv:1408.5484 [gr-qc]}}.

\bibitem{Zelenka:2019nyp}
O.~Zelenka, G.~Lukes-Gerakopoulos, V.~Witzany, and O.~Kop\'{a}\u{c}ek,
  ``{Growth of resonances and chaos for a spinning test particle in the
  Schwarzschild background},''
  \href{http://dx.doi.org/10.1103/PhysRevD.101.024037}{{\em Phys. Rev.}
  {\bfseries D101} no.~2, (2020) 024037},
\href{http://arxiv.org/abs/1911.00414}{{\ttfamily arXiv:1911.00414 [gr-qc]}}.

\bibitem{Lukes-Gerakopoulos:2016udm}
G.~Lukes-Gerakopoulos,
  \href{http://dx.doi.org/10.1142/9789813226609_0209}{``{Spinning particles
  moving around black holes: integrability and chaos},''} in {\em {Proceedings,
  14th Marcel Grossmann Meeting on Recent Developments in Theoretical and
  Experimental General Relativity, Astrophysics, and Relativistic Field
  Theories (MG14) (In 4 Volumes): Rome, Italy, July 12-18, 2015}}, vol.~2,
  pp.~1960--1965.
\newblock 2017.
\newblock
\href{http://arxiv.org/abs/1606.09430}{{\ttfamily arXiv:1606.09430 [gr-qc]}}.
\newblock

\bibitem{Sasaki:1981sx}
M.~Sasaki and T.~Nakamura, ``{Gravitational Radiation From a Kerr Black Hole.
  1. Formulation and a Method for Numerical Analysis},''
\href{http://dx.doi.org/10.1143/PTP.67.1788}{{\em Prog. Theor. Phys.}
  {\bfseries 67} (1982) 1788}.

\bibitem{Olver:1994:AEC}
F.~W.~J. Olver, ``Asymptotic expansions of the coefficients in asymptotic
  series solutions of linear differential equations,'' {\em Methods Appl.
  Anal.} {\bfseries 1} no.~1, (1994) 1–13.

\bibitem{Olver:1997:ASL}
F.~W.~J. Olver, ``Asymptotic solutions of linear ordinary differential
  equations at an irregular singularity of rank unity,'' {\em Methods Appl.
  Anal.} {\bfseries 4} no.~4, (1997) 375–403.

\bibitem{Olver:1974asymptotics}
F.~Olver, {\em Asymptotics and Special Functions}.
\newblock Computer science and applied mathematics : a series of monographs and
  textbooks. Academic Press, 1974.

\bibitem{Gralla:2015rpa}
S.~E. Gralla, A.~P. Porfyriadis, and N.~Warburton, ``{Particle on the Innermost
  Stable Circular Orbit of a Rapidly Spinning Black Hole},''
  \href{http://dx.doi.org/10.1103/PhysRevD.92.064029}{{\em Phys. Rev.}
  {\bfseries D92} no.~6, (2015) 064029},
\href{http://arxiv.org/abs/1506.08496}{{\ttfamily arXiv:1506.08496 [gr-qc]}}.

\bibitem{Taracchini:2013wfa}
A.~Taracchini, A.~Buonanno, S.~A. Hughes, and G.~Khanna, ``{Modeling the
  horizon-absorbed gravitational flux for equatorial-circular orbits in Kerr
  spacetime},'' \href{http://dx.doi.org/10.1103/PhysRevD.88.109903,
  10.1103/PhysRevD.88.044001}{{\em Phys. Rev.} {\bfseries D88} (2013) 044001},
  \href{http://arxiv.org/abs/1305.2184}{{\ttfamily arXiv:1305.2184 [gr-qc]}}.
[Erratum: Phys. Rev.D88,no.10,109903(2013)].

\end{thebibliography}\endgroup

\end{document}